\documentclass{amsart}

\usepackage[margin=4.5cm]{geometry}
\usepackage[utf8]{inputenc}
\usepackage[T1]{fontenc}
\usepackage{microtype}
\usepackage{scrbase}
\usepackage[shortlabels]{enumitem}
\usepackage{etoolbox}
\usepackage{graphicx}

\usepackage{biblatex}

\usepackage{amsmath}
\usepackage{amssymb}
\usepackage{amsthm}
\usepackage{thmtools}
\usepackage{mathtools}
\usepackage{braket}

\usepackage{float}
\usepackage{booktabs}
\usepackage{tikz}
\usepackage{tikz-cd}
\usepackage{pgfmath}
\usepackage{pgfplots}
\usepackage{ifthen}
\usetikzlibrary{
  positioning,
  arrows.meta,
  fit,
  shapes,
  calc,
  through,
  intersections
}

\tikzset{
vertex/.style={fill,circle,inner sep=2pt},
edge/.style={line width=1},
wire/.style={color=gray,line width=.5},
every loop/.style={}
}

\usepackage{lipsum}

\usepackage{xcolor} 
\definecolor{rwth-blue}{cmyk}{1,.5,0,0}\colorlet{rwth-lblue}{rwth-blue!50}\colorlet{rwth-llblue}{rwth-blue!25}
\definecolor{rwth-violet}{cmyk}{.6,.6,0,0}\colorlet{rwth-lviolet}{rwth-violet!50}\colorlet{rwth-llviolet}{rwth-violet!25}
\definecolor{rwth-purple}{cmyk}{.7,1,.35,.15}\colorlet{rwth-lpurple}{rwth-purple!50}\colorlet{rwth-llpurple}{rwth-purple!25}
\definecolor{rwth-carmine}{cmyk}{.25,1,.7,.2}\colorlet{rwth-lcarmine}{rwth-carmine!50}\colorlet{rwth-llcarmine}{rwth-carmine!25}
\definecolor{rwth-red}{cmyk}{.15,1,1,0}\colorlet{rwth-lred}{rwth-red!50}\colorlet{rwth-llred}{rwth-red!25}
\definecolor{rwth-magenta}{cmyk}{0,1,.25,0}\colorlet{rwth-lmagenta}{rwth-magenta!50}\colorlet{rwth-llmagenta}{rwth-magenta!25}
\definecolor{rwth-orange}{cmyk}{0,.4,1,0}\colorlet{rwth-lorange}{rwth-orange!50}\colorlet{rwth-llorange}{rwth-orange!25}
\definecolor{rwth-yellow}{cmyk}{0,0,1,0}\colorlet{rwth-lyellow}{rwth-yellow!50}\colorlet{rwth-llyellow}{rwth-yellow!25}
\definecolor{rwth-grass}{cmyk}{.35,0,1,0}\colorlet{rwth-lgrass}{rwth-grass!50}\colorlet{rwth-llgrass}{rwth-grass!25}
\definecolor{rwth-green}{cmyk}{.7,0,1,0}\colorlet{rwth-lgreen}{rwth-green!50}\colorlet{rwth-llgreen}{rwth-green!25}
\definecolor{rwth-cyan}{cmyk}{1,0,.4,0}\colorlet{rwth-lcyan}{rwth-cyan!50}\colorlet{rwth-llcyan}{rwth-cyan!25}
\definecolor{rwth-teal}{cmyk}{1,.3,.5,.3}\colorlet{rwth-lteal}{rwth-teal!50}\colorlet{rwth-llteal}{rwth-teal!25}
\definecolor{rwth-gold}{cmyk}{.35,.46,.7,.35}
\definecolor{rwth-silver}{cmyk}{.39,.31,.32,.14}

\usepackage{array}
\newcolumntype{C}{>{$}c<{$}}

\usepackage{hyperref}
\usepackage[capitalise,noabbrev]{cleveref}
\hypersetup{%
	pdftoolbar=false,
	pdfmenubar=false,
	colorlinks,
	urlcolor={rwth-magenta},
	linkcolor={black},
	citecolor={rwth-green}
}

\theoremstyle{plain}
\newtheorem{theorem}{Theorem}
\newtheorem{lemma}[theorem]{Lemma}
\newtheorem{proposition}[theorem]{Proposition}
\newtheorem{corollary}[theorem]{Corollary}

\newtheorem{observation}[theorem]{Observation}

\newtheorem*{convention}{Convention}
\newtheorem*{conjecture*}{Conjecture}
\newtheorem{claim}{Claim}
\AtEndEnvironment{lemma}{\setcounter{claim}{0}}

\theoremstyle{definition}
\newtheorem{definition}[theorem]{Definition}
\newtheorem{remark}[theorem]{Remark}
\newtheorem{examplex}[theorem]{Example}
\newenvironment{example}
  {\pushQED{\qed}\examplex}
  {\popQED\endexamplex}
  
\newenvironment{subproof}[1][\proofname]{%
  \begin{proof}[#1]%
}{%
  \end{proof}%
}

\newcommand\numberthis{\refstepcounter{equation}\tag{\theequation}} 

\theoremstyle{plain}
\usepackage{thm-restate}

\renewcommand{\epsilon}{\varepsilon}
\renewcommand{\theta}{\vartheta}

\newcommand{\C}{\ensuremath{\mathbb{C}}}
\newcommand{\R}{\ensuremath{\mathbb{R}}}
\newcommand{\N}{\ensuremath{\mathbb{N}}}
\newcommand{\Z}{\ensuremath{\mathbb{Z}}}

\DeclareMathOperator{\Aut}{Aut}
\DeclareMathOperator{\Qut}{Qut}
\DeclareMathOperator{\GL}{GL}
\DeclareMathOperator{\Rep}{Rep}
\DeclareMathOperator{\Mor}{Mor}
\newcommand{\gF}{\mathcal{F}}
\renewcommand{\P}{\mathbb{P}}
\newcommand{\gbG}{\boldsymbol{\mathcal{G}}} 
\newcommand{\gbP}{\boldsymbol{\mathcal{P}}} 

\newcommand{\U}{\mathcal{U}}
\newcommand{\V}{\mathcal{V}}
\newcommand{\Utp}[1]{\U^{\otimes #1}}

\newcommand{\bM}{\mathbf{M}}
\newcommand{\bI}{\mathbf{I}}
\newcommand{\bS}{\mathbf{S}}

\DeclareMathOperator{\id}{id}

\DeclareMathOperator{\lspan}{span}
\DeclareMathOperator{\soe}{soe}

\newcommand{\tup}[1]{\mathbf{#1}}
\newcommand{\transpose}{^\mathsf{T}}
\newcommand{\inverse}{^{-1}}
\newcommand{\identity}{I}
\newcommand{\1}{\mathbf 1}

\renewcommand{\iff}{\Leftrightarrow}

\renewcommand{\implies}{\Rightarrow}
\newcommand{\abs}[1]{\lvert #1 \rvert}
\newcommand{\norm}[1]{\lVert #1 \rVert}
\newcommand{\Cstar}{C\ensuremath{^*}\nobreakdash}

\newcommand{\ccdot}{\makebox[1ex]{\textbf{$\cdot$}}}

\newcommand{\argp}{\ccdot}

\newcommand{\lrangle}[1]{\langle #1 \rangle}
\newcommand{\intertwclosure}[1]{\ensuremath{\langle{} #1 \rangle{}_{\circ, \otimes, *}}}

\DeclareMathOperator{\img}{img}
\DeclareMathOperator{\HOM}{Hom}

\makeatletter
\let\amsmath@bigm\bigm

\renewcommand{\bigm}[1]{%
  \ifcsname fenced@\string#1\endcsname
    \expandafter\@firstoftwo
  \else
    \expandafter\@secondoftwo
  \fi
  {\expandafter\amsmath@bigm\csname fenced@\string#1\endcsname}%
  {\amsmath@bigm#1}%
}

\newcommand{\DeclareFence}[2]{\@namedef{fenced@\string#1}{#2}}
\makeatother

\DeclareFence{\mid}{|} 

\makeatletter
\newcommand{\vast}{\bBigg@{3}}
\newcommand{\Vast}{\bBigg@{4}}
\makeatother

\newcommand{\partition}[3]{\begin{tikzpicture}[scale=.5,line width=1]
    \def\nk{#1}
    \def\nl{#2}

    \foreach \x in {1,...,\nk} {
      \coordinate (u\x) at (\x, 3);
      \coordinate (ua\x) at (\x, 2);
    }

    \foreach \x in {2,...,\nk} {
      \pgfmathsetmacro{\xmh}{\x - 0.5}
      \pgfmathsetmacro{\xmo}{int(\x - 1)}
      \coordinate (um\xmo) at (\xmh, 2);
    }

    \foreach \x in {1,...,\nl} {
      \coordinate (l\x) at (\x, 0);
      \coordinate (la\x) at (\x, 1);
    }

    \foreach \x in {2,...,\nl} {
      \pgfmathsetmacro{\xmh}{\x - 0.5}
      \pgfmathsetmacro{\xmo}{int(\x - 1)}
      \coordinate (lm\xmo) at (\xmh, 2);
    }

  \draw #3;
\end{tikzpicture}}

\newcommand{\lblPartition}[4]{\begin{tikzpicture}[scale=.5, line width=1]
  
    \def\nk{#1}
    \def\nl{#2}
    \pgfmathsetmacro{\cappos}{ifthenelse(\nk > \nl, \nk/2, \nl/2)}%

    \foreach \x in {1,...,\nk} {
      \coordinate (u\x) at (\x, 3);
      \coordinate (ua\x) at (\x, 2);
    }

    \foreach \x in {1,...,\nl} {
      \coordinate (l\x) at (\x, 0);
      \coordinate (la\x) at (\x, 1);
    }

  \draw #3;
     
  \node[align=center] at (\cappos, -2) {#4};
\end{tikzpicture}}

\newcommand{\pa}[1][false]{\begin{tikzpicture}[yscale=.8, ultra thick]
    \ifthenelse{\equal{#1}{true}}
    {\begin{scope}[yscale=-1]}{}
    \useasboundingbox (-.25, 0) rectangle (1.25,2);

    \coordinate (a) at (0,0) {};
    \coordinate (b) at (1,0) {};
    \coordinate (c) at (0,2) {};
    \coordinate (d) at (1,2) {};

    \draw (a) -- (d)
    (b) -- (c);
    \ifthenelse{\equal{#1}{true}}{\end{scope}}{}
\end{tikzpicture}}

\newcommand{\pb}[1][false]{\begin{tikzpicture}[yscale=.8, ultra thick]
    \ifthenelse{\equal{#1}{true}}
    {\begin{scope}[yscale=-1]}{}
    \useasboundingbox (-.6, 0) rectangle (.5,2);
    \coordinate (a) at (0,0) {};
    \coordinate (b) at (0,2) {};

    \draw[-{To[scale=3]}] (a) -- (b);
    \ifthenelse{\equal{#1}{true}}{\end{scope}}{}
\end{tikzpicture}}

\newcommand{\pc}[1][false]{\begin{tikzpicture}[yscale=.8, ultra thick]
    \ifthenelse{\equal{#1}{true}}
    {\begin{scope}[yscale=-1]}{}
    \useasboundingbox (-.25, 0) rectangle (1.25,2);
    \coordinate (a) at (0,0) {};
    \coordinate (b) at (1,0) {};
    \coordinate (a1) at (0, .6) {};
    \coordinate (b1) at (1, .6) {};
    \coordinate (ab) at (.5, .6) {};
    
    \coordinate (c) at (0,2) {};
    \coordinate (d) at (1,2) {};
    \coordinate (c1) at (0, 1.4) {};
    \coordinate (d1) at (1, 1.4) {};
    \coordinate (cd) at (.5, 1.4) {};

    \draw 
    (a) -- (a1)
    (b) -- (b1)
    (a1) -- (b1);

    \draw
    (c) -- (c1)
    (d) -- (d1)
    (c1) -- (d1);

    \draw (ab) -- (cd);
    \ifthenelse{\equal{#1}{true}}{\end{scope}}{}
\end{tikzpicture}}

\newcommand{\pd}[1][false]{\begin{tikzpicture}[yscale=.8, ultra thick]
    \ifthenelse{\equal{#1}{true}}
    {\begin{scope}[yscale=-1]}{}
    \useasboundingbox (-.5, 0) rectangle (.5,2);
    \coordinate (a) at (0,0) {};
    \coordinate (b) at (0,2) {};

    \draw (a) -- (b);
    \ifthenelse{\equal{#1}{true}}{\end{scope}}{}
\end{tikzpicture}}

\newcommand{\pe}[1][false]{\begin{tikzpicture}[yscale=.8, ultra thick]
    \ifthenelse{\equal{#1}{true}}
    {\begin{scope}[yscale=-1]}{}
    \useasboundingbox (-.25, 0) rectangle (1.25,2);
    \coordinate (a) at (0,0) {};
    \coordinate (a1) at (0,1.15) {};
    \coordinate (b) at (1,0) {};
    \coordinate (b1) at (1,1.15) {};

    \draw (a) -- (a1)
    (a1) -- (b1)
    (b1) -- (b);
    \ifthenelse{\equal{#1}{true}}{\end{scope}}{}
\end{tikzpicture}}

\newcommand{\pf}[1][false]{\begin{tikzpicture}[yscale=.8, ultra thick]
    \ifthenelse{\equal{#1}{true}}{\begin{scope}[yscale=-1]}{}
      \useasboundingbox (-.25, 0) rectangle (1.25,2);
      \coordinate (a) at (0,0) {};
      \coordinate (a1) at (0,1) {};
      \coordinate (b) at (1,0) {};
      \coordinate (b1) at (1,1) {};
      \coordinate (c) at (0.5, 1.75);
      \coordinate (c1) at (0.5, 1);

      \draw (a) -- (a1)
        (a1) -- (b1)
        (b1) -- (b)
        (c1) -- (c);
    \ifthenelse{\equal{#1}{true}}{\end{scope}}{}
\end{tikzpicture}}

\newcommand{\pg}[1][false]{\begin{tikzpicture}[yscale=.8, ultra thick]
    \ifthenelse{\equal{#1}{true}}{\begin{scope}[yscale=-1]}{}
      \useasboundingbox (-.25, 0) rectangle (1.75,2);
      \coordinate (a) at (0,0);
      \coordinate (a1) at (.75,0);
      \coordinate (a2) at (1.5, 0);
      \coordinate (b) at (0,2);
      \coordinate (b1) at (.75,2);
      \coordinate (b2) at (1.5, 2);

      \draw (a) -- (b2)
        (a1) -- (b1)
        (a2) -- (b);
    \ifthenelse{\equal{#1}{true}}{\end{scope}}{}
\end{tikzpicture}}

\newcommand{\ph}[1][false]{\begin{tikzpicture}[yscale=.8, ultra thick]
    \ifthenelse{\equal{#1}{true}}{\begin{scope}[yscale=-1]}{}
      \useasboundingbox (-.25, 0) rectangle (1.75,2);
      \coordinate (a) at (0,2);
      \coordinate (b) at (1.5,0);

      \coordinate (c1u) at (0.75, 2);
      \coordinate (c2u) at (1.5, 2);
      \coordinate (c1l) at (0.75, 1.5);
      \coordinate (c2l) at (1.5, 1.5);

      \coordinate (d1u) at (0.75, 0);
      \coordinate (d2u) at (0, 0);
      \coordinate (d1l) at (0.75, .5);
      \coordinate (d2l) at (0, .5);

      \coordinate (cm) at (1.125, 1.5);
      \coordinate (dm) at (0.375, .5);

      \draw (a) -- (b);
      \draw (c1u) -- (c1l)
            (c2u) -- (c2l)
            (c1l) -- (c2l);
      \draw (d1u) -- (d1l)
            (d2u) -- (d2l)
            (d1l) -- (d2l);
      \draw (cm) -- (dm);
    \ifthenelse{\equal{#1}{true}}{\end{scope}}{}
\end{tikzpicture}}

\newcommand{\pj}[1][false]{\begin{tikzpicture}[yscale=.8, ultra thick]
    \ifthenelse{\equal{#1}{true}}
    {\begin{scope}[yscale=-1]}{}
    \useasboundingbox (-.25, 0) rectangle (1.25,2);

    \coordinate (a) at (0,0) {};
    \coordinate (a1) at (0, 0.75);
    \coordinate (b) at (1,0) {};
    \coordinate (c) at (0,2) {};
    \coordinate (d) at (1,2) {};
    \coordinate (d1) at (1,1.25) {};

    \draw (a) -- (a1)
          (d) -- (d1)
          (b) -- (c);

    \ifthenelse{\equal{#1}{true}}{\end{scope}}{}
\end{tikzpicture}}

\newcommand{\spa}{\leavevmode\lower.1cm\hbox{\resizebox{!}{.4cm}{\pa}}}
\newcommand{\spai}{\leavevmode\lower.1cm\hbox{\resizebox{!}{.4cm}{\pa[true]}}}
\newcommand{\spb}{\leavevmode\ensuremath{\uparrow}}
\newcommand{\spbi}{\leavevmode\ensuremath{\downarrow}}
\newcommand{\spc}{\leavevmode\lower.1cm\hbox{\resizebox{!}{.4cm}{\pc}}}
\newcommand{\spci}{\leavevmode\lower.1cm\hbox{\resizebox{!}{.4cm}{\pc[true]}}}
\newcommand{\spd}{\leavevmode\lower.1cm\hbox{\resizebox{!}{.4cm}{\pd}}}
\newcommand{\spdi}{\leavevmode\lower.1cm\hbox{\resizebox{!}{.4cm}{\pd[true]}}}
\newcommand{\spe}{\leavevmode\lower.1cm\hbox{\resizebox{!}{.4cm}{\pe}}}
\newcommand{\spei}{\leavevmode\lower.1cm\hbox{\resizebox{!}{.4cm}{\pe[true]}}}
\newcommand{\spf}{\leavevmode\lower.1cm\hbox{\resizebox{!}{.4cm}{\pf}}}
\newcommand{\spfi}{\leavevmode\lower.1cm\hbox{\resizebox{!}{.4cm}{\pf[true]}}}
\newcommand{\spg}{\leavevmode\lower.1cm\hbox{\resizebox{!}{.4cm}{\pg}}}
\newcommand{\spgi}{\leavevmode\lower.1cm\hbox{\resizebox{!}{.4cm}{\pg[true]}}}
\newcommand{\sph}{\leavevmode\lower.1cm\hbox{\resizebox{!}{.4cm}{\ph}}}
\newcommand{\sphi}{\leavevmode\lower.1cm\hbox{\resizebox{!}{.4cm}{\ph[true]}}}
\newcommand{\spj}{\leavevmode\lower.1cm\hbox{\resizebox{!}{.4cm}{\pj}}}
\newcommand{\spji}{\leavevmode\lower.1cm\hbox{\resizebox{!}{.4cm}{\pj[true]}}}

\title{Homomorphism Indistinguishability Relations induced by Quantum Groups}

\author{Tim Seppelt}
\address[Author 1]{IT-Universitetet i København, Rued Langgaards Vej 7, 2300, København S, Denmark}
\email{tise@itu.dk}

\author{Gian Luca Spitzer}
\address[Author 2]{Univ. Bordeaux, CNRS, Bordeaux INP, LaBRI, UMR-5800, F-33400 Talence, France}
\address[Author 2]{Laboratoire de Physique Th\'eorique, Universit\'e de Toulouse, CNRS, UPS, France}
\email{gian-luca.spitzer@u-bordeaux.fr}

\begin{document}
  \begin{abstract}
    Homomorphism indistinguishability is a way of characterising many natural equivalence relations on graphs. Two graphs $G$ and $H$ are called homomorphism indistinguishable over a graph class $\gF$ if for each $F \in \gF$, the number of homomorphisms from $F$ to $G$ equals the number of homomorphisms from $F$ to $H$. Examples of such equivalence relations include isomorphism and cospectrality, as well as equivalence with respect to many formal logics.
    Quantum groups are a generalisation of topological groups that describe ``non-commutative symmetries'' and, inter alia, have applications in quantum information theory.
    An important subclass are the easy quantum groups, which enjoy a combinatorial characterisation and have been fully classified by Raum and Weber.

    A recent connection between these seemingly distant concepts was made by Man\v{c}inska and Roberson, who showed that quantum isomorphism, a relaxation of classical isomorphism that can be phrased in terms of the quantum symmetric group, is equivalent to homomorphism indistinguishability over the class of planar graphs. 
    We generalise Man\v{c}inska and Roberson's result to all orthogonal easy quantum groups. We obtain for each orthogonal easy quantum group a graph isomorphism relaxation $\approx$ and a graph class $\gF$, such that homomorphism indistinguishability over $\gF$ coincides with $\approx$.
    Our results include a full classification of the $(0, 0)$-intertwiners of the graph-theoretic quantum group obtained by adding the adjacency matrix of a graph to the intertwiners of an orthogonal easy quantum group. 
  \end{abstract}

  \maketitle

  \section{Introduction}
  Homomorphism counts capture graph structure. More precisely, two graphs $G$ and $H$ are said to be \emph{homomorphism indistinguishable} over a class of graphs $\gF$ if for all graphs $F \in \gF$, the number of homomorphisms from $F$ to $G$ equals the number of homomorphisms from $F$ to $H$. In 1967, Lovász \cite{lovasz1967operations} first showed that isomorphism is equivalent to homomorphism indistinguishability over all graphs. Since then, a wide spectrum of equivalence relations have been characterised as homomorphism indistinguishability relations. Examples include natural equivalence relations such as cospectrality and having the same degree sequence, as well as
equivalences in different formal logics \cite{dvorak_recognizing_2010,grohe_counting_2020,fluck2024going}, results in category theory \cite{dawar2021lovasz,montacute2022pebble}, and characterisations of different linear programming relaxations \cite{dell2018lov,grohe_homomorphism_2025,roberson2024lasserre}, cf.\ the monograph \cite{seppelt_homomorphism_2024}.

Particularly interesting is a recent result by Man\v{c}inska and Roberson \cite{manvcinska2020quantum}, who showed that \emph{quantum isomorphism} is equivalent to homomorphism indistinguishability over the class of planar graphs. Quantum isomorphism was originally defined in \cite{atserias2019quantum} in terms of the \emph{graph isomorphism game}. This is a non-local game in which two players have to convince a verifier that two graphs are isomorphic. They are not allowed to communicate, but may agree on a strategy beforehand. Classically, the players have a winning strategy if and only if the graphs are in fact isomorphic. On the other hand, we call the two graphs \emph{quantum isomorphic} if the players have a winning strategy when being allowed to coordinate their answers by measuring a shared entangled quantum state. 

Results by Lupini, Man\v{c}inska, and Roberson \cite{lupini2020nonlocal} show that classical and quantum and isomorphism share deep similarities. Classical isomorphism of graphs $G$ and $H$ can be equivalently phrased as the existence of a permutation matrix $P$ such that $P A_G = A_H P$, where $A_G, A_H$ are the adjacency matrices of $G$ and $H$ respectively. Analogously, quantum isomorphism turns out to be equivalent to the existence of a so-called \emph{quantum permutation matrix} $\U$ such that $\U A_G = A_H \U$. Classically, if $G$ and $H$ are connected, then they are isomorphic if and only if there exists an orbit of the automorphism group of the disjoint union $G \cup H$, that intersects both $G$ and $H$ non-trivially. As such, classical isomorphism is closely related to the symmetric group $S_n$ and automorphism groups. Again, an analogous picture turns out to hold for quantum isomorphism: Two connected graphs are quantum isomorphic if and only if there exists an orbit of the \emph{quantum automorphism group} of $G \cup H$ that intersects $G$ and $H$ non-trivially. Quantum isomorphism is thus closely related to the \emph{quantum symmetric group} $S^+_n$ and quantum automorphism groups $\Qut(\argp)$.

Quantum groups such as the quantum symmetric group generalise classical groups and can be seen as describing ``non-commutative symmetries''. They have their origins in mathematical physics and have been extensively studied. Man\v{c}inska and Roberson showed that the intertwiners of a certain subclass of quantum groups, among which the \emph{orthogonal easy quantum groups}, can be seen as \emph{homomorphism tensors}, which are matrices that keep track of homomorphism counts from bilabelled graphs. Through a Tannaka-Krein-type duality originally due to Woronowicz \cite{woronowicz_tannakakrein_1988}, the bilabelled graphs in question then entirely determine the quantum group. For the quantum symmetric group, these graphs are shown to be planar bilabelled graphs, which eventually yields the equivalence of quantum isomorphism and homomorphism indistinguishability over planar graphs.

In light of these connections, a natural question is whether similar homomorphism indistinguishability results can be obtained for other orthogonal easy quantum groups, which have, in fact, been fully classified by Raum and Weber \cite{raum_full_2016}. We answer this question affirmatively. We introduce \emph{graph instantiations} of easy quantum groups, which generalise automorphism and quantum automorphism groups of graphs, and eventually prove the following theorem.

{
  \renewcommand{\thetheorem}{\ref{thm:mainresult}}
  \begin{theorem}[Informal]
    Let $Q$ be an orthogonal easy quantum group and let $\gF$ be the set of its intertwiner graphs. Then for all graphs $G$ and $H$, the following are equivalent.
    \begin{enumerate}
      \item $G$ and $H$ are homomorphism indistinguishable over the unlabelled graphs in $\gF$.
      \item There exists a matrix $\U$ with entries in some \Cstar-algebra, such that $\U A_G = A_H \U$ and $\Utp{k}M = M\Utp{\ell}$ for all intertwiners $M$ of $Q$ and suitable $k, \ell$.
    \end{enumerate}
  \end{theorem}
  \addtocounter{theorem}{-1}
}
 
We will see that the constraints imposed on $\U$ by the intertwiners of $Q$ correspond to very natural properties of matrices, such as orthogonality or restrictions on row and column sums.
For example, if we take $Q$ to be the quantum symmetric group,
then $\mathcal{F}$ is the class of all planar graphs.
In this case, we recover the result of Man\v{c}inska and Roberson \cite{manvcinska2020quantum} that $G$ and $H$ are homomorphism indistinguishable over all planar graphs if and only if there is a quantum permutation matrix $\mathcal{U}$ such that $\mathcal{U} A_G = A_H \mathcal{U}$.
Notably, this argument does not rely on quantum orbits.

Based on Raum and Weber's classification, we finally identify the classes of unlabelled intertwiner graphs of all easy quantum groups.
We find that for every orthogonal easy quantum group, these graph classes are either all graphs, all planar graphs, all paths and cycles, or all cycles  (\cref{thm:graphclassclassification}). 
This yields several concrete corollaries of our theorem and partially answers a question by Man\v{c}inska and Roberson.

  \section{Background}
  \label{sec:background}
  \subsection{Graphs and Homomorphisms}

A \emph{graph} $G$ is a set $V(G)$ of vertices equipped with a binary relation $E(G) \subseteq V(G)^2$ defining the edges. Unless otherwise specified, we assume all graphs to be \emph{simple}: they are undirected and do not contain loops, that is, $E(G)$ is irreflexive and symmetric. We usually write $uv$ or $vu$ to denote the edge $(u, v) = (v, u) \in E(G)$. 
We can define a matrix $A \in \{0, 1\}^{V(G) \times V(G)}$ with $A_{uv} = 1 \iff uv \in E(G)$, which is called the \emph{adjacency matrix} of $G$. The class of all graphs is denoted by $\mathcal{G}$. 

We call $H$ a \emph{subgraph} of $G$ if $H$ can be obtained from $G$ by removing vertices and edges. A graph is \emph{connected} if there exists a path between any two vertices. The \emph{disjoint union} $G \cup H$ of $G$ and $H$ is the graph given by $(V(G) \sqcup V(H), E(G) \sqcup E(H))$, where $\sqcup$ denotes the disjoint union of sets. We \emph{subdivide} an edge $uv \in E(G)$ by removing it, adding a new vertex $w$ and adding the edges $uw$ and $wv$. If $v_1, \dots, v_k$ is a path in $G$ such that $v_i$ has degree $2$ for all $1 < i < k$, we \emph{unsubdivide} it by removing the inner vertices and adding the edge $v_1v_k$. We \emph{contract} an edge $uv \in E(G)$ by removing it, identifying $u$ and $v$, and suppressing any resulting multiple edges.

A \emph{planar embedding} of a graph $G$ consists of an injective function $f \colon V(G) \to \R^2$ and a set of curves $h_{uv}\colon[0, 1] \to \R^2$ for all $uv \in E(G)$ such that $h_{uv}(0) = f(u)$ and $h_{uv}(1) = f(v)$, and for distinct edges $e, e' \in E(G)$ we have $h_e(x) \neq h_{e'}(y)$ except for $x, y \in \{0, 1\}$. A graph is called \emph{planar} if it has a planar embedding, or informally if it can be drawn on the plane without any edges crossing. The \emph{faces} of planar embedding are the connected regions of $\R^2 \setminus \bigcup_{e \in E(G)} \img h_e$. A \emph{facial cycle} is a cycle in $G$ that is the boundary of a face in some planar embedding.

\begin{convention}
  For simplicity, we will assume that $V(G) = [\abs{V(G)}]$. This means that the class of all graphs is a set, and that the adjacency matrix of a graph uniquely corresponds to a matrix $\C^{\abs{V(G)} \times \abs{V(G)}}$. Note that this is done without loss of generality, since every graph is isomorphic to a graph of this form and all our results will be isomorphism-invariant. 
\end{convention}

A \emph{graph homomorphism} from a graph $G$ to a graph $H$ is a mapping $h\colon V(G) \to V(H)$ that preserves adjacency, that is, $uv \in E(G)$ implies $h(u)h(v) \in E(H)$. For any graph $F$, we denote by $\HOM(F, G)$ the set of homomorphisms from $F$ to $G$, and let $\hom(F, G) \coloneqq \abs{\HOM(F, G)}$. We then say that $G$ and $H$ are \emph{homomorphism indistinguishable} over a class of graphs $\gF \subseteq \mathcal{G}$ if $\hom(F, G) = \hom(F, H)$ for all $F \in \gF$.
See \cite{seppelt_homomorphism_2024} for background on homomorphism indistinguishability.

\subsection{Easy Quantum Groups}
\label{ssec:easyqgroups}
Our main objects of study will be easy quantum groups. We give a short overview here, for a more in-depth treatment see for example the monograph \cite{maassen2021representation}. Easy quantum groups are a subclass of compact quantum groups, which in turn generalise compact classical groups, that is, topological groups whose underlying space $X$ is compact and Hausdorff. Through Gelfand duality, we may consider instead of $X$ the commutative \Cstar-algebra $C(X)$ of complex-valued continuous functions on $X$. Now a group multiplication $m\colon X \times X \to X$ induces a \emph{comultiplication} $\Delta\colon C(X) \to C(X) \otimes C(X) \cong C(X \times X)$ as
\begin{equation*}
  \Delta(f)(x, y) = f(xy),
\end{equation*}
that is coassociative, that is, $(\id \circ \Delta) \circ \Delta = (\Delta \circ \id) \circ \Delta$, and that satisfies that the linear spans of $\Delta(C(X))(\mathbf{1} \otimes C(X))$ and $\Delta(C(X))(C(X) \otimes \mathbf{1})$ are dense in $C(X) \otimes C(X)$. 
Importantly, one can show that, from every $*$-homomomorphism $C(X) \to C(X) \otimes C(X)$ satisfying these properties, one can uniquely recover a group multiplication $m \colon X \times X \to X$. Consequently, one may view compact groups equivalently as commutative \Cstar-algebras equipped with such a $*$\nobreakdash-homomorphism. We define compact quantum groups by dropping the commutativity requirement.

\begin{definition}\label{def:cqg}
  A \emph{compact quantum group} $G$ is given by a tuple $(C(G), \Delta)$ consisting of a \Cstar-algebra $C(G)$ and a $^*$-homomorphism $\Delta \colon C(G) \to C(G) \otimes C(G)$, called comultiplication, such that
  \begin{enumerate}
    \item $(\id \otimes \Delta) \circ \Delta = (\Delta \otimes \id) \circ \Delta$.
    \item $\Delta(C(G))(\mathbf{1} \otimes C(G))$ and $\Delta(C(G))(C(G) \otimes \mathbf{1})$ are linearly dense in $C(G) \otimes C(G)$.
  \end{enumerate}
\end{definition}

\begin{remark}
  The name ``quantum group'' may be justified by noting that quantum mechanics can be derived from classical mechanics in essentially the same way, see for example \cite{strocchi2008introduction}. In particular, non-commutativity is \emph{the} defining feature of quantum mechanics.
\end{remark}

Along similar lines, we may generalise compact matrix groups, that is, compact groups $G$ admitting a faithful matrix representation $\chi\colon G \to \GL_n(\C)$. Their corresponding function algebras turn out to be generated by the coordinate functions
\begin{equation*}
  u_{ij}\colon G \to \C, x \mapsto \chi(x)_{ij},
\end{equation*}
for which the comultiplication will be of the form
\begin{equation*}
  \Delta(u_{ij}) = \sum_{k = 1}^n u_{ik} \otimes u_{kj}.
\end{equation*}
A few additional properties of the $u_{ij}$ allow recovering the original matrix representation. Compact matrix quantum groups are then defined by dropping the commutativity requirement.

\begin{definition}\label{def:cqmg}
  A \emph{compact matrix quantum group} $G$ is given by a tuple $(C(G), \U, n)$, consisting of a \Cstar-algebra $C(G)$, an integer $n \in \N$, and a matrix $\U = (u_{ij}) \in C(G)^{n \times n}$, called the \emph{fundamental representation} of $G$, such that
  \begin{enumerate}
    \item The elements $\{u_{ij} \mid i, j \in [n]\}$ generate $C(G)$,
    \item The matrix $\U$ is unitary and $\U\transpose$ is invertible,
    \item The map $\Delta \colon C(G) \to C(G) \otimes C(G), u_{ij} \mapsto \sum_{k = 1}^n u_{ik} \otimes u_{kj}$ is a $^*$-homo\-morphism, called comultiplication.
  \end{enumerate}
\end{definition}

\begin{remark}
  The definition of the $u_{ij}$ suggests to arrange them as a matrix, that is, to view them as an element $\U \in C(G)^{n \times n}$. We can then define the standard matrix terminology as one would expect. The transpose of $\U = (u_{ij})$ is $\U\transpose = (u_{ji})$, the adjoint $\U^*$ is the conjugate transpose, and the identity matrix in $C(G)^{n \times n}$ is $\mathbf{1}I$.
  We may equivalently see $\U$ as an element of $C(G) \otimes B(\C^n)$.
\end{remark}

\begin{example}
  Wang \cite{wang1995free} defines the \emph{quantum orthogonal group} as the compact matrix quantum group $O_n^+$, whose underlying \Cstar\nobreakdash-algebra is the universal \Cstar-algebra generated by the $n^2$ entries of $\U = (u_{ij})$, subject to the conditions 
  \begin{equation}
    \label{eq:orthrep}
    u_{ij}^* = u_{ij} \qquad \text{and} \qquad \U\U\transpose = \U\transpose\U = \mathbf{1}\U. \qedhere
  \end{equation}
\end{example}

\begin{example}
  Another compact matrix quantum group introduced by Wang \cite{wang1998quantum} is the \emph{quantum symmetric group} $S_n^+$, whose underlying \Cstar-algebra is the universal \Cstar-algebra generated by the $n^2$ entries of $\U = (u_{ij})$, such that 
  \begin{equation}
    \label{eq:symrep}
    u_{ij}^2 = u_{ij}^* = u_{ij} \qquad \text{and} \qquad\sum_{k} u_{ik} = \mathbf{1} = \sum_k u_{kj} .
  \end{equation}
  Note that these are exactly the conditions that define permutation matrices in the case where $u_{ij} \in \C$.
\end{example}

In this paper, we will mostly be interested in quantum subgroups of the quantum orthogonal group, that is, compact matrix quantum groups whose fundamental representation satisfies \cref{eq:orthrep}. These \emph{orthogonal} compact matrix quantum groups turn out to have a particularly accessable representation theory.

\begin{definition}
  A \emph{finite-dimensional representation} of a compact matrix quantum group $G$ is a matrix $\mathcal{V} = (v_{ij}) \in C(G)^{m \times m}$ for some $m \in \N$, satisfying
  \begin{equation*}
    \Delta(v_{ij}) = \sum_k v_{ik} \otimes v_{kj}.
  \end{equation*}
\end{definition}

\begin{definition}
  An \emph{intertwiner} is an equivariant map between two finite-dim-ensional representations $\U \in C(G)^{m \times m}$, $\mathcal{V} \in C(G)^{m' \times m'}$, that is, a linear map $T \colon \C^m \to \C^{m'}$ such that $T\U = \mathcal{V} T$.
\end{definition}

We can arrange the finite-dimensional representations of a compact matrix quantum group $G$ into a category, called the \emph{representation category} $\Rep(G)$.

\begin{quote}
\begin{tabular}{ l l }
  Objects: & finite-dimensional unitary representations of $G$,\\
  Morphisms: & intertwiners,\\
  Composition: & natural composition.
\end{tabular}
\end{quote}

The following result shows that $\Rep(G)$ is completely determined by the fundamental representation of $G$.

\begin{theorem}
  Let $G = (C(G), \U, n)$ be an orthogonal compact matrix quantum group. Then 
  \begin{enumerate}
    \item Every invertible finite-dimensional representation $G$ is equivalent to a unitary representation of $G$.
    \item Every unitary representation of $G$ decomposes into a direct sum of irreducible representations of $G$.
    \item Every unitary irreducible representation of $G$ is either the trivial representation $(\1)$ or equivalent to a direct summand of a tensor power of $\U$.
  \end{enumerate}
\end{theorem}

This suggests to define the \emph{reduced representation category} $\Rep_0(G)$ as the full subcategory of $\Rep(G)$ consisting of the tensor powers of $\U$. One can show that both $\Rep(G)$ and $\Rep_0(G)$ are rigid monoidal categories where the monoid product is the tensor product, the monoidal unit is the trivial representation $(\1)$, and the dual of a representation $\V = (v_{ij})$ is given by $\overline{\V} = (v_{ij}^*)$. One may further show that the morphisms of $\Rep_0(G)$ form a so-called \emph{concrete tensor category with duals}.

\begin{definition}\label{def:ctensorcat}
  Let $n \in \N$. A \emph{concrete tensor category with duals} $C$ is a collection of $\C$-vector spaces $C(\ell, k) \subseteq \{T\colon (\C^n)^{\otimes k} \to (\C^n)^{\otimes \ell} ~\text{linear}\}$ for all $k, \ell \in \N$, such that
  \begin{enumerate}
    \item if $T \in C(\ell, k)$ and $T' \in C(r, s)$, then $T \otimes T' \in C(\ell + r, k + s)$,
    \item if $T \in C(\ell, k)$ and $T' \in C(k, r)$, then $TT' \in C(\ell, r)$,
    \item if $T \in C(\ell, k)$, then $T^* \in C(k, \ell)$,
    \item $\identity \in C(1,1)$,
    \item $\xi \coloneqq \sum_i e_i \otimes e_i \in C(2, 0)$.
  \end{enumerate}
\end{definition}

This allows us to state a central result in the theory of compact matrix quantum groups: a Tannaka-Krein-type duality due to Woronowicz \cite{woronowicz_tannakakrein_1988}, which allows to recover a compact matrix quantum group from its finite-dimensional representations.

\begin{theorem}[Woronowicz-Tannaka-Krein]\label{thm:tannakakrein}
  Let $n \in \N$ and $C$ be a concrete tensor category with duals. Then there exists a unique compact matrix quantum group $(C(G), \U, n)$ with 
  \begin{equation*}
    \operatorname{Mor}_{\Rep_0(G)}(\U^{\otimes k}, \U^{\otimes \ell}) = C(\ell, k)
  \end{equation*}
  for all $k, \ell \in \N$. 
\end{theorem}

In particular, we can reason about compact matrix quantum groups entirely through their intertwiners. We therefore introduce the following simplified notation.

\begin{definition}
  Let $(C(G), \U, n)$ be a compact matrix quantum group. We denote by $C^G$ the concrete tensor category with duals formed by the morphisms of $\Rep_0(G)$ and refer to it as the \emph{intertwiners of $G$}. Similarly, we refer to the sets $C^G(\ell, k)$ as the \emph{$(\ell, k)$-intertwiners of $G$}.
\end{definition}

Finally, we can introduce \emph{easy quantum groups}. These are compact matrix quantum groups whose intertwiners can be described combinatorically. Again, we are interested in the orthogonal case.

\begin{definition}
A \emph{$(\ell, k)$-partition} is a partition of the set $\{1_L, \dots, \ell_L, 1_U, \dots, k_U\}$, where we think of $1_L, \dots, \ell_L$ as the \emph{lower points}, and of $1_U, \dots, k_U$ as the \emph{upper points} of the partition. We denote the set of these partitions by $\mathbb{P}(\ell, k)$.
\end{definition}

\begin{figure}[h]
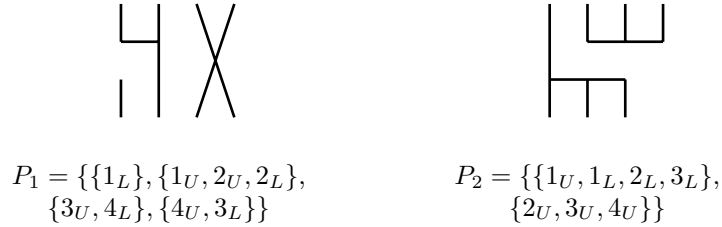

  \centering
  \lblPartition{4}{4}{
    (u3) -- (l4)
    (u4) -- (l3)
    (u2) -- (l2)
    (u1) -- (ua1)
    (ua1) -- (ua2)
    (l1) -- (la1)
  }
  {$P_1 = \{\{1_L\}, \{1_U, 2_U, 2_L\}$,\\ $\{3_U, 4_L\}, \{4_U, 3_L\}\}$}
  \hspace{1.5cm}
  \lblPartition{4}{3}{
    (u1) -- (l1)
    (la1) -- (la3)
    (l2) -- (la2)
    (l3) -- (la3)
    (u2) -- (ua2)
    (ua2) -- (ua4)
    (u3) -- (ua3)
    (u4) -- (ua4)
  }
  {$P_2 = \{\{1_U, 1_L, 2_L, 3_L\}$,\\ $\{2_U, 3_U, 4_U\}\}$}
  \caption{Drawing partitions}
  \label{fig:drawingparts}
\end{figure}

We can define such a partition graphically by drawing the upper points above the lower points and connecting points that are in the same block of the partition, as shown in \cref{fig:drawingparts}.

An important subclass are \emph{non-crossing} partitions, which are the partitions that can be drawn without any lines crossing.

\begin{definition}
  A partition $P$ of an ordered set $V$ is called \emph{non-crossing} if whenever $a < b < c < d$, and $a, c$ are in the same part and $b, d$ are in the same part, then the two parts coincide. We define non-crossing partitions of $\{1_L, \dots, \ell_L, 1_U, \dots, k_U\}$ by letting $1_L < \dots < \ell_L < 1_U < \dots < k_U$.
\end{definition}

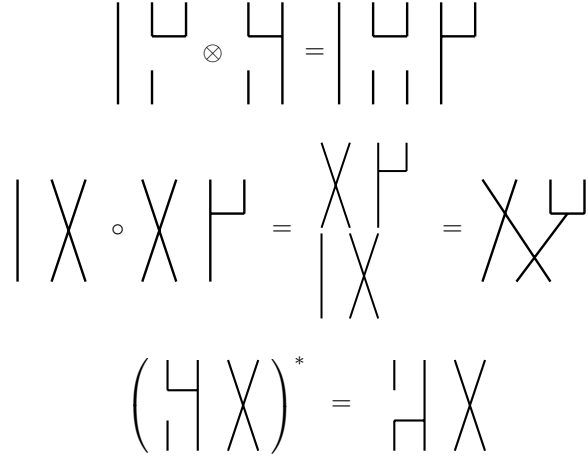
\begin{figure}[h]
  \centering
  \begin{tikzpicture}[scale=.75]
    \node[scale=.9,align=left] (p) at (0, 0) {
        \partition{3}{2}{
          (u1) -- (l1)
          (l2) -- (la2)
          (u2) -- (ua2)
          (u3) -- (ua3)
          (ua2) -- (ua3)
        }
      };
    \node (tp) at (1, 0) {$\otimes$};
    \node[scale=.9,align=left] (pp) at (2, 0) {
        \partition{2}{2}{
          (l1) -- (la1)
          (l2) -- (u2)
          (u1) -- (ua1)
          (ua1) -- (ua2)
        }
      };
    \node (eq) at (2.8, 0) {$=$};
    \node[scale=.9,align=left] (ptpp) at (4.5, 0) {
        \partition{5}{4}{
          (u1) -- (l1)
          (l2) -- (la2)
          (u2) -- (ua2)
          (u3) -- (ua3)
          (ua2) -- (ua3)
          (l3) -- (la3)
          (ua4) -- (ua5)
          (u4) -- (l4)
          (u5) -- (ua5)
        }
      };
  \end{tikzpicture}
  \vspace{.25cm}
  \\
  \begin{tikzpicture}[scale=.75]
    \node[scale=.9,align=left] (p) at (0, 0) {
        \partition{3}{3}{
          (u1) -- (l1)
          (u2) -- (l3)
          (u3) -- (l2)
        }
      };
    \node (tp) at (1.1, 0) {$\circ$};
    \node[scale=.9,align=left] (pp) at (2.5, 0) {
        \partition{4}{3}{
          (u1) -- (l2)
          (u2) -- (l1)
          (u3) -- (l3)
          (ua3) -- (ua4)
          (u4) -- (ua4)
        }
      };
    \node (eq) at (4, 0) {$=$};
    \node[scale=.75,align=left] (pp) at (5.5, .8) {
        \partition{4}{3}{
          (u1) -- (l2)
          (u2) -- (l1)
          (u3) -- (l3)
          (ua3) -- (ua4)
          (u4) -- (ua4)
        }
      };
    \node[scale=.75,align=left] (ptpp) at (5.25, -.8) {
        \partition{3}{3}{
          (u1) -- (l1)
          (u2) -- (l3)
          (u3) -- (l2)
        }
      };
    \node (eq) at (7, 0) {$=$};
    \node[scale=.9,align=left] (ptpp) at (8.5, 0) {
        \partition{4}{3}{
          (u1) -- (l3)
          (u2) -- (l1)
          (u3) -- (ua3)
          (u4) -- (ua4)
          (ua3) -- (ua4)
          (um3) -- (l2)
        }
      };
  \end{tikzpicture}
  \\
  \vspace{.25cm}
  \begin{tikzpicture}
    \node (pl) at (-1, 0) {$\vast($};
    \node[scale=.8] (p) at (0, 0) {
        \partition{4}{4}{
          (u3) -- (l4)
          (u4) -- (l3)
          (u2) -- (l2)
          (u1) -- (ua1)
          (ua1) -- (ua2)
          (l1) -- (la1)}};
    \node (pl) at (1, 0) {$\vast)^*$};
    \node (eq) at (1.7, 0) {$=$};
    \node[scale=.8] (p) at (3, 0) {
        \partition{4}{4}{
          (u3) -- (l4)
          (u4) -- (l3)
          (l2) -- (u2)
          (l1) -- (la1)
          (la1) -- (la2)
          (u1) -- (ua1)}};
  \end{tikzpicture}
  \caption{Operations on partitions}
  \label{fig:partops}
\end{figure}

It is possible to define certain operations on partitions. Let $P \in \P(\ell, k)$ and $P' \in \P(\ell', k')$. Then the tensor product $P \otimes P' \in P(\ell + \ell', k + k')$ is the partition obtained by drawing $P$ and $P'$ next to each other horizontally, as shown in \cref{fig:partops}. If $\ell' = k$, then the composition $P' \circ P \in \P(k', \ell)$ is obtained by drawing $P'$ below $P$ and identifying the lower row of $P$ and the upper row of $P'$. Note that this operation may create \emph{loops}, i.e.\ blocks that are no longer connected to the upper or lower row of the resulting partition. These become \emph{empty blocks} of $P' \circ P$. Finally, the \emph{adjoint} $P^* \in \P(k, \ell)$ is obtained by reflecting $P$ across the horizontal axis.

\begin{figure}
  \centering
  \begin{tikzpicture}[scale=.75,every edge/.style = {draw, ->},line width=1]
    \def\nk{4}
    \def\nl{3}
    
    \pgfmathsetmacro{\dist}{1.5}
    \pgfmathsetmacro{\disth}{\dist/2}

    \foreach \x in {1,...,\nk} {
      \pgfmathsetmacro{\xp}{\x*\dist}
      \coordinate (u\x) at (\xp, 3);
      \coordinate (ua\x) at (\xp, 2);
    }

    \foreach \x in {2,...,\nk} {
      \pgfmathsetmacro{\xp}{\x*\dist}
      \pgfmathsetmacro{\xmh}{\xp - \disth}
      \pgfmathsetmacro{\xmo}{int(\x - 1)}
      \coordinate (um\xmo) at (\xmh, 2);
    }

    \foreach \x in {1,...,\nl} {
      \pgfmathsetmacro{\xp}{\x*\dist}
      \coordinate (l\x) at (\xp, 0);
      \coordinate (la\x) at (\xp, 1);
    }

    \foreach \x in {2,...,\nl} {
      \pgfmathsetmacro{\xp}{\x*\dist}
      \pgfmathsetmacro{\xmh}{\xp - \disth}
      \pgfmathsetmacro{\xmo}{int(\x - 1)}
      \coordinate (lm\xmo) at (\xmh, 2);
    }

  \node[yshift=.5cm] at (u1) {$e_1$};
  \node[yshift=1.25cm] at (um1) {$\otimes$};
  \node[yshift=.5cm] at (u2) {$e_2$};
  \node[yshift=1.25cm] at (um2) {$\otimes$};
  \node[yshift=.5cm] at (u3) {$e_3$};
  \node[yshift=1.25cm] at (um3) {$\otimes$};
  \node[yshift=.5cm] at (u4) {$e_3$};

  \node[yshift=-.5cm] at (l1) {$e_2$};
  \node[yshift=-2cm] at (lm1) {$\otimes$};
  \node[yshift=-.5cm] at (l2) {$e_3$};
  \node[yshift=-2cm] at (lm2) {$\otimes$};
  \node[yshift=-.5cm] at (l3) {$e_1$};

  \draw
    (u1) edge (l3)
    (u2) edge (l1)
    (u3) -- (ua3)
    (u4) -- (ua4)
    (ua3) -- (ua4)
    (um3) edge (l2);
 \end{tikzpicture} 
 \hspace{1cm}
  \begin{tikzpicture}[scale=.75,every edge/.style = {draw, ->},line width=1]
    \def\nk{4}
    \def\nl{4}
    
    \pgfmathsetmacro{\dist}{1.5}
    \pgfmathsetmacro{\disth}{\dist/2}

    \foreach \x in {1,...,\nk} {
      \pgfmathsetmacro{\xp}{\x*\dist}
      \coordinate (u\x) at (\xp, 3);
      \coordinate (ua\x) at (\xp, 2);
    }

    \foreach \x in {2,...,\nk} {
      \pgfmathsetmacro{\xp}{\x*\dist}
      \pgfmathsetmacro{\xmh}{\xp - \disth}
      \pgfmathsetmacro{\xmo}{int(\x - 1)}
      \coordinate (um\xmo) at (\xmh, 2);
    }

    \foreach \x in {1,...,\nl} {
      \pgfmathsetmacro{\xp}{\x*\dist}
      \coordinate (l\x) at (\xp, 0);
      \coordinate (la\x) at (\xp, 1);
    }

    \foreach \x in {2,...,\nl} {
      \pgfmathsetmacro{\xp}{\x*\dist}
      \pgfmathsetmacro{\xmh}{\xp - \disth}
      \pgfmathsetmacro{\xmo}{int(\x - 1)}
      \coordinate (lm\xmo) at (\xmh, 2);
    }

  \node[yshift=.5cm] at (u1) {$e_1$};
  \node[yshift=1.25cm] at (um1) {$\otimes$};
  \node[yshift=.5cm] at (u2) {$e_1$};
  \node[yshift=1.25cm] at (um2) {$\otimes$};
  \node[yshift=.5cm] at (u3) {$e_2$};
  \node[yshift=1.25cm] at (um3) {$\otimes$};
  \node[yshift=.5cm] at (u4) {$e_3$};

  \node[yshift=-.5cm,xshift=-.3cm] at (l1) {$(\sum_k e_k)$};
  \node[yshift=-2cm] at (lm1) {$\otimes$};
  \node[yshift=-.5cm] at (l2) {$e_1$};
  \node[yshift=-2cm] at (lm2) {$\otimes$};
  \node[yshift=-.5cm] at (l3) {$e_3$};
  \node[yshift=-2cm] at (lm3) {$\otimes$};
  \node[yshift=-.5cm] at (l4) {$e_2$};

  \draw
    (u3) edge (l4)
    (u4) edge (l3)
    (u2) edge (l2)
    (u1) -- (ua1)
    (ua1) -- (ua2)
    (l1) -- (la1);
 \end{tikzpicture} 
 \caption{Partitions as linear maps}
 \label{fig:partmaps}
\end{figure}
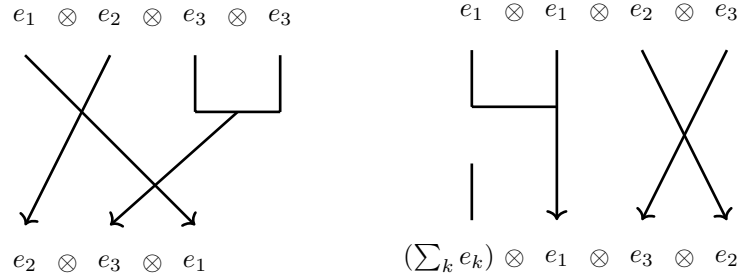

To any partition $P \in \P(\ell, k)$ and $n \geq 1$, one can associate a linear map $T_P^{(n)} \colon (\C^n)^{\otimes k} \to (\C^n)^{\otimes \ell}$ as follows. For $\tup{a} \in [n]^\ell, \tup{b} \in [n]^k$, we let $\delta(\tup{a}, \tup{b}) = 1$ if the assignments $i_U \mapsto b_i$ and $i_L \mapsto a_i$ are constant on the blocks of $P$, and $\delta(\tup{a}, \tup{b}) = 0$ otherwise.
Then $T_P^{(n)}$ is defined on basis vectors $e_{b_1} \otimes \dots \otimes e_{b_k}\in (\C^n)^{\otimes k}$ as
\begin{equation*}
  T_P^{(n)}(e_{b_1} \otimes \dots \otimes e_{b_k}) = n^{e(P)} \sum_{\tup{a} \in [n]^\ell} \delta(\tup{a}, \tup{b})(e_{a_1} \otimes \dots \otimes e_{a_\ell}),
\end{equation*}
where $e(P)$ denotes the number of empty blocks in $P$. This definition has a nice graphical interpretation, where we simply shuffle the input vectors according to the blocks, see \cref{fig:partmaps}. This suggest to write the matrix form of $T_P^{(n)}$ in terms of outer products: Consider, for instance, the partition $P_1$ in \cref{fig:drawingparts}. The corresponding matrix can be seen to be equal to
\begin{equation*}
  T_{P_1}^{(n)} = \sum_{ijk\ell} (e_\ell \otimes e_i \otimes e_k \otimes e_j) \cdot (e_i \otimes e_i \otimes e_j \otimes e_k)\transpose.
\end{equation*}

\begin{example}
 The linear maps corresponding to the single block partitions $\{\{1_L, \dots, \ell_L, 1_U, \dots k_U\}\} \in \P(\ell, k)$ map $k$ copies of a vector to $\ell$ copies. We call these maps the \emph{generalised multiplication maps} and denote them by $M^{(n), \ell, k}$. Two important special cases are $M^{(n), 1, 1}= \identity_n$ and $M^{(n), 2, 0}= \sum_i e_i \otimes e_i = \xi$. 
\end{example}

\begin{example}
  Another important partition is the \emph{swap} partition $\spa = \{\{1_U, 2_L\},$\\ $\{2_U, 1_L\}\}$. Its corresponding linear map is $S^{(n)}\colon e_i \otimes e_j \mapsto e_j \otimes e_i$.
\end{example}

\begin{example}
  We represent the single point partitions $\{\{1_L\}\}$ and $\{\{1_U\}\}$ graphically as $\uparrow$ and $\downarrow$, to distinguish them from each other and from the identity partition $\spd$.
\end{example}

Banica and Speicher \cite{banica2009liberation} showed that the operations of tensor product, composition, and adjoint on partitions coincide with the corresponding operations on linear maps. 

\begin{proposition}\label{prop:partmapcorrespondence}
  Let $P, P' \in \P$. Then we have $T_{P \otimes P'} = T_P \otimes T_{P'}$, $T_{P^*} = T_P^*$, and if $P$ and $P'$ are compatible, $T_{P \circ P'} = T_P \circ T_{P'}$.
\end{proposition}

This allows us to define \emph{partition categories} that give rise to concrete tensor categories with duals.

\begin{definition}
  A \emph{partition category} $\mathcal{C}$ is a set of partitions, such that
  \begin{enumerate}
    \item if $P \in \mathcal{C}(\ell, k)$ and $P' \in \mathcal{C}(r, s)$, then $P \otimes P' \in \mathcal{C}(\ell + r, k + s)$,
    \item if $P \in \mathcal{C}(\ell, k)$ and $P' \in \mathcal{C}(k, r)$, then $P \circ P' \in \mathcal{C}(\ell, r)$,
    \item if $P \in \mathcal{C}(\ell, k)$, then $P^* \in \mathcal{C}(\ell, r)$,
    \item $\{\{1_U, 1_L\}\} = \spd \in \mathcal{C}(1, 1)$,
    \item $\{\{1_L, 2_L\}\} = \spe \in \mathcal{C}(2, 0)$,
  \end{enumerate}
  where $\mathcal{C}(\ell, k) = \mathcal{C} \cap \P(\ell, k)$.
\end{definition}

For partitions $P_1, \dots, P_m \in \P$, we write $\lrangle{P_1, \dots, P_m}$ for the partition category generated by the $P_i$, that is, the closure of $\{P_1, \dots, P_m, \spe, \spd\}$ under tensor products, composition, and taking adjoints.
Now it is easy to see that for every partition category $\mathcal{C}$, the collection of vector spaces 
\begin{equation*}
  C(\ell, k) = \lspan \{T_P \mid P \in \mathcal{C}(\ell, k)\}
\end{equation*}
forms a concrete tensor category with duals. \cref{thm:tannakakrein} thus implies the following result.

\begin{proposition}
  Let $\mathcal{C}$ be a partition category. Then for each $n \geq 1$ there exists a unique compact matrix quantum group $G_n(\mathcal{C}) = (A, \U, n)$ such that
  \begin{equation*}
    \U^{\otimes k}T_P = T_P\U^{\otimes \ell}
  \end{equation*}
  for all $P \in \mathcal{C}(k, \ell)$.
\end{proposition}

Compact matrix quantum groups of this form are called \emph{orthogonal easy quantum groups}. Indeed, it follows from \cite[Remark 3.2.15]{maassen2021representation} that $\U$ is always orthogonal. Moreover, one can prove that $\U^{\otimes 2}S = S\U^{\otimes 2}$ if and only if the entries of $\U$ commute. Hence, an easy quantum group $G_n(\mathcal{C})$ is classical if and only if $\spa \in \mathcal{C}$.

\subsection{Classification of Orthogonal Easy Quantum Groups}
\label{ssec:fullclassification}
The orthogonal easy quantum groups have been fully classified by Raum and Weber \cite{raum_full_2016}. Since we will work extensively with this classification, we restate it here for the reader's convenience.

\begin{theorem}[Classification of Partition Categories]\label{thm:classifyingpartitions}
  Every category of partition falls into one of the following cases.
  \begin{enumerate}
    \item The partition categories $\mathcal{C}$ with $\spa \in \mathcal{C}$ are exactly 
      \begin{itemize}
        \item $P = \langle \spa, \spb, \spf \rangle$, all partitions,
        \item $P_2 \coloneqq \langle \spa \rangle$, the partitions with block size two,
        \item $P_{even} \coloneqq \langle \spa, \spc \rangle$, the partitions with even block size,
        \item $P' \coloneqq \langle \spa, \spb \otimes \spb, \spc \rangle$, the partitions with an even number of blocks of odd size,
        \item $P_b \coloneqq \langle \spa, \spb \rangle$, the partitions with block size one or two,
        \item $P_b' \coloneqq \langle \spa, \spb \otimes \spb \rangle = \langle \spa, \spj \rangle$, the partitions with an arbitrary number of blocks of size two and an even number of blocks of size one.
    \end{itemize}
  \item The partition categories that contain only non-crossing partitions are exactly 
      \begin{itemize}
        \item $NC = \langle \spb, \spf \rangle$, all non-crossing partitions,
        \item $NC_2 \coloneqq \langle \varnothing \rangle = \langle \spd, \spe \rangle$, the non-crossing partitions with block size two,
        \item $NC_{even} \coloneqq \langle \spc \rangle$, the non-crossing partitions with even block size,
        \item $NC' \coloneqq \langle \spb \otimes \spb, \spc \rangle$, the non-crossing partitions with an even number of blocks of odd size,
        \item $NC_b \coloneqq \langle \spb \rangle$, the non-crossing partitions with block size one or two,
        \item $NC_b' \coloneqq \langle \spj \rangle$, the non-crossing partitions with an arbitrary number of blocks of size two and an even number of blocks of size one.
        \item $NC_b^{\#} \coloneqq \langle \spb \otimes \spb \rangle$, the partitions with an arbitrary number of blocks of size two, each connecting an odd and an even point, and an even number of blocks of size one.
      \end{itemize}
    \item The partition categories $\mathcal{C}$ with $\spa \not\in \mathcal{C}$ and $\spg \in \mathcal{C}$ are exactly
      \begin{itemize}
        \item $E_o \coloneqq \langle \spg \rangle$, the partitions with blocks of size two, each connecting an odd and an even point,
        \item $E'_b \coloneqq \langle \spg, \spb \otimes \spb \rangle$, the partitions with an arbitrary number of blocks of size two---each connecting an odd and an even point, and an even number of blocks of size one,
        \item $E_h \coloneqq \langle \spg, \spc \rangle$, the partitions with each block having the same number of odd and even points,
        \item $E_h^s \coloneqq \langle \spg, \spc, h_s\rangle$, where $h_s = \{\{1, 3, \dots, (2s - 1)\}, \{2, 4, \dots, 2s\}\} \in P(2s, 0)$, the partitions with an even number of points and each block having the same number of odd and even points modulo $s$ for all $s \geq 3$.
      \end{itemize}
  \item The partition categories $\mathcal{C}$ with $\sph \in \mathcal{C}$ are called \emph{group-theoretic} and are indexed by certain subgroups of $\Z^{*\infty}_2$.
  \item The partition categories $\mathcal{C}$ with $\spc \in \mathcal{C}$, $\spb \otimes \spb~ \not\in \mathcal{C}$ and $\sph \not\in \mathcal{C}$ are exactly the categories $\Pi_k = \lrangle{\pi_k}$, generated by the element
    \begin{center}
      \vspace{.4cm}
      \begin{tikzpicture}[scale=.6]
        \foreach \x in {0, 1} {
          \pgfmathsetmacro{\p}{ifthenelse(\x>0,1,2)}
          \pgfmathsetmacro{\s}{pow(-1, \x)}

          \begin{scope}[xscale=\s] 
            \coordinate (l1\p) at (.25, 0);
            \coordinate (la1\p) at (.25, 2);
            \coordinate (l2\p) at (.75, 0);
            \coordinate (la2\p) at (.75, 1.5);

            \coordinate (l3\p) at (2.75, 0);
            \coordinate (la3\p) at (2.75, 1);

            \coordinate (l4\p) at (3.25, 0);
            \coordinate (la4\p) at (3.25, .5);
            \coordinate (l5\p) at (3.75, 0);
            \coordinate (la5\p) at (3.75, .5);

            \coordinate (l6\p) at (4.25, 0);
            \coordinate (la6\p) at (4.25, 1);

            \coordinate (l7\p) at (6.25, 0);
            \coordinate (la7\p) at (6.25, 1.5);
            \coordinate (l8\p) at (6.75, 0);
            \coordinate (la8\p) at (6.75, 2);

            \node at (1.75, 0) {\dots};
            \node at (5.25, 0) {\dots};

            \coordinate (h1\p) at (1, 1);
            \coordinate (h2\p) at (1, .5);

            \coordinate (hh1\p) at (2.45, .5);
            \coordinate (hh2\p) at (3.05, .5);
          \end{scope}

          \draw 
            (l1\p) -- (la1\p)
            (l8\p) -- (la8\p)
            (l2\p) -- (la2\p)
            (l7\p) -- (la7\p)
            (la1\p) -- (la8\p)
            (la2\p) -- (la7\p)
            (l3\p) -- (la3\p)
            (l6\p) -- (la6\p)
            (la3\p) -- (la6\p)
            (l4\p) -- (la4\p)
            (l5\p) -- (la5\p)
            (la4\p) -- (la5\p)
            (h1\p) -- (la3\p)
            (la4\p) -- (hh2\p)
            (h2\p) -- (hh1\p)
            ;
        };

      \draw 
        (la11) -- (la12)
        (la21) to[bend left] (la22)
        (h11) to[bend left] (h12)
        (h21) to[bend left] (h22)
        (hh11) to[bend right] (hh21)
        (hh12) to[bend left] (hh22);

      \node at (8.6, .9) {$\in \P(0, 4k)$};
      \node at (-8, .9) {$\pi_k =$};
    \end{tikzpicture}
    \vspace{.2cm}
  \end{center}
    and $\Pi_\infty = \lrangle{\pi_k \mid k \in \N}$.
  \end{enumerate}
  The cases in 1, 2, 3 are pairwise distinct.
\end{theorem}

As noted above, these partition categories form tensor categories with duals and thus uniquely correspond to compact matrix quantum groups. Moreover, the compact matrix quantum groups with intertwiner \spa{} have commutative \Cstar-algebras and thus correspond to classical groups. 

\begin{proposition}
  The easy groups, i.e.\ the easy quantum groups that correspond to partition categories $\mathcal{C}$ with $\spa \in \mathcal{C}$ are
  \begin{enumerate}
    \item $G_n(P) = S_n$, the symmetric group,
    \item $G_n(P_2) = O_n$, the orthogonal group,
    \item $G_n(P_{even}) = H_n$, the hyperoctahedral group, i.e.\ the group of monomial orthogonal matrices, 
    \item $G_n(P') = S'_n$, the modified symmetric group, i.e.\ the group of signed permutation matrices,
    \item $G_n(P_b) = B_n$, the bistochastic group, i.e.\ the group of orthogonal matrices having row and column sum $1$,
    \item $G_n(P_b') = B_n'$, the modified bistochastic group, i.e.\ the group of signed bistochastic matrices.
  \end{enumerate}
\end{proposition}

Here a matrix is called \emph{monomial} if it has the same zero pattern as a permutation matrix, and a signed permutation (bistochastic) matrix is a permutation (bistochastic) matrix multiplied by $\pm 1$. 
Note that  the term ``bistochastic'' is sometimes used to refer to a matrix all whose row and column sums are $1$. In this article, \emph{bistochastic} means additionally that the matrix is orthogonal.

We obtain ``liberated'' quantum versions of these groups by dropping the commutativity constraint. This corresponds to dropping the $\spa$ partition from the generating sets.

\begin{proposition}
  The easy quantum groups that correspond to categories of non-crossing partitions are
  \begin{enumerate}
    \item $G_n(NC) = S_n^+$, the quantum symmetric group,
    \item $G_n(NC_2) = O_n^+$, the quantum orthogonal group,
    \item $G_n(NC_{even}) = H_n^+$, the quantum hyperoctahedral group,  
    \item $G_n(NC') = S_n'^+$, the quantum modified symmetric group,
    \item $G_n(NC_b) = B_n^+$, the quantum bistochastic group,
    \item $G_n(NC_b') = B_n'^+$, the quantum complexly modified bistochastic group,
    \item $G_n(NC_b^{\#}) = B_n^{\#+}$, the quantum modified bistochastic group.
  \end{enumerate}
\end{proposition}

An interesting special case occurs with the liberation of the modified bistochastic group. In the classical case, the group has two natural generating sets in $\langle \spa, \spb \otimes \spb \rangle$ and $\langle \spa, \spj \rangle$. Dropping the commutativity constraint, however, yields the distinct easy quantum groups $B_n'^+$ and $B^{\#+}_n$ respectively.

\begin{remark}
  In the literature on easy quantum groups, $B_n'^+$ and $B_n^{\#+}$ are usually called ``quantum modified bistochastic group'' and ``quantum freely modified bistochastic group'' respectively. However, as we will see later, the intertwiner $\spb \otimes \spb$ captures the notion of signed bistochastic matrices in a much more intuitive way than \spj{}. It thus makes sense to instead call $B_n^{\#+}$ the quantum modified bistochastic group. The name ``quantum complexly modified bistochastic group'' is justified by the fact that $B_n'^+$ is the so-called \emph{tensor 2-complexification} of $B_n^+$, see \cite[Remark 4.14]{tarrago2017unitary}.
\end{remark}

\subsection{Graph Homomorphisms and the Quantum Symmetric Group}
\label{ssec:graphhomsandqsym}
Man\v{c}inska and Roberson \cite{manvcinska2020quantum} discovered a surprising connection between these quantum groups and graph homomorphisms. Concretely, the intertwiners of easy quantum groups are equivalently characterised by \emph{homomorphism tensors} which count homomorphisms from bilabelled graphs.

As the name suggests, bilabelled graphs are graphs whose vertices may be annotated by two types of labels, the \emph{in-labels} and the \emph{out-labels}. A vertex can have both in-labels and out-labels and not every vertex has to be labelled. 

\begin{definition}
  A $(k, \ell)$-\emph{bilabelled graph} $\mathbf{K}$ is a tuple $(K, \tup{a}, \tup{b})$ where $K$ is a graph, $\tup{a} = (a_1, \dots, a_k) \in V(K)^k$ is the \emph{output vector}, and $\tup{b} = (b_1, \dots, b_\ell) \in V(K)^\ell$ is the \emph{input vector}. We let $V(\mathbf{K}) = V(K)$ and $E(\mathbf{K}) = E(K)$. We denote the set of all $(k, \ell)$-bilabelled graphs by $\gbG(k, \ell)$.
\end{definition}

  One may think of the labels as elements in $[k]$ and $[\ell]$ respectively, where a vertex $v \in V(K)$ has out-label $i \in [k]$ if $a_i = v$, and in-label $j \in [\ell]$ if $b_j = v$. In this sense, we will also use the following definition.

  \begin{definition}
    
    Let $\mathbf{K} \in \gbG(k, \ell)$ be a bilabelled graph. We denote by $l^{\mathbf{K}}_{in}(v) \subseteq [\ell]$, $l^\mathbf{K}_{out}(v) \subseteq [k]$, and $l^\mathbf{K}(v) \subseteq [\ell] \sqcup [k]$ the sets of in-labels, out-labels, and all labels of $v$ respectively. We omit the superscript when the graph is clear from the context.
  \end{definition}

\begin{figure}[h]
  \centering
  \begin{tikzpicture}
    \node[vertex,label={1}] (v) at (0, 0) {};
    \node[vertex,label={2}] (v2) [above right=of v] {};
    \node[vertex,label={3}] (v3) [above left=of v] {};
    \node[vertex,label={4}] (v4) [below left=of v] {};
    \node[vertex,label={5}] (v5) [below right=of v] {};

    \draw[edge]
      (v) -- (v2)
      (v) -- (v3)
      (v) -- (v4)
      (v) -- (v5);

    \node (l1) [right=of v2] {};
    \node (l3) [right=of v5] {};
    \node (l2) at ($(l1)!.5!(l3)$) {};

    \node (lo1) [left=of v3] {};
    \node (lo3) [left=of v4] {};
    \node (lo2) at ($(lo1)!.5!(lo3)$) {};

    \coordinate (h) at ($(v3)!.5!(v4)$);

    \draw[wire]
      (v) to[out=0,in=180] (l1)
      (v5) to[out=0,in=180] (l2)
      (v5) -- (l3);

    \draw[wire]
      (lo1) -- (v3)
      (v) -- (h)
      (lo3) to[out=0,in=180] (h)
      (lo2) to[out=0,in=180] (v4);

    \node at (0, -2.5) {$\mathbf{K} = (K_{1, 4}, (3, 4, 1), (1, 5, 5))$};
  \end{tikzpicture}
  \hspace{1cm}
  \begin{tikzpicture}
    \begin{scope}[yshift=-4] 
    \node[vertex,label={1}] (v1) at (-1.35, 0) {};
    \node[vertex,label={2}] (v2) at (1.35, 0) {};
    \node[vertex,label={3}] (v3) at ($(v1)!1!60:(v2)$) {};

    \node[vertex,label={-90:4}] (v4) at (0,.8) {};

    \draw[edge]
      (v1) -- (v2)
      (v1) -- (v3)
      (v1) -- (v4)
      (v2) -- (v3)
      (v2) -- (v4)
      (v3) -- (v4);

    \coordinate (lo1) at (-2.25, 2.33);
    \coordinate (lo2) at (-2.25, .8);
    \coordinate (lo3) at (-2.25, 0);
    \coordinate (l1) at (2.25, 2.33);
    \coordinate (l2) at (2.25, .8);

    \draw[wire]
      (v3) -- (lo1)
      (v4) -- (lo2)
      (v1) -- (lo3);

    \draw[wire]
      (v3) -- (l1)
      (v4) -- (l2);

    \end{scope}

    \node at (0, -1.5) {$\mathbf{K'} = (K_{4}, (3, 4, 1), (3, 4))$};
  \end{tikzpicture}
  \caption{Drawing bilabelled graphs}
  \label{fig:drawingbilabgraphs}    
\end{figure}
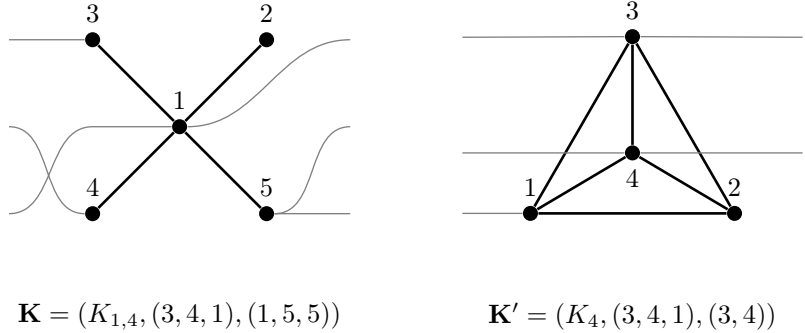

When drawing a bilabelled graph, the labels are represented as grey ``wires'' that are ordered from top to bottom, where input wires extend to the right and output wires to the left, as seen in \cref{fig:drawingbilabgraphs}.  

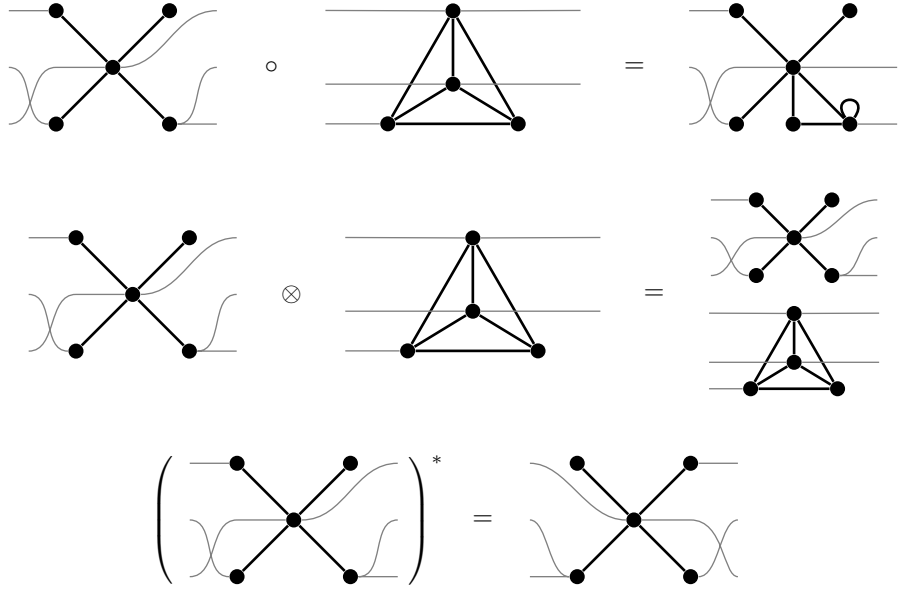
\begin{figure}[h]
  \centering
  \begin{tikzpicture}
    \node at (0, 0) {
      \begin{tikzpicture}[scale=.75]
          \node[vertex] (v) at (0, 0) {};
          \node[vertex] (v2) at (1, 1) {};
          \node[vertex] (v3) at (-1, 1) {};
          \node[vertex] (v4) at (-1, -1) {};
          \node[vertex] (v5) at (1, -1) {};

          \draw[edge]
            (v) -- (v2)
            (v) -- (v3)
            (v) -- (v4)
            (v) -- (v5);

          \node (l1) at (2, 1) {};
          \node (l3) at (2, -1) {};
          \node (l2) at ($(l1)!.5!(l3)$) {};

          \node (lo1) at (-2, 1) {};
          \node (lo3) at (-2, -1) {};
          \node (lo2) at ($(lo1)!.5!(lo3)$) {};

          \coordinate (h) at ($(v3)!.5!(v4)$);

          \draw[wire]
            (v) to[out=0,in=180] (l1)
            (v5) to[out=0,in=180] (l2)
            (v5) -- (l3);

          \draw[wire]
            (lo1) -- (v3)
            (v) -- (h)
            (lo3) to[out=0,in=180] (h)
            (lo2) to[out=0,in=180] (v4);
        \end{tikzpicture}
      };
  \node at (2.1, 0) {$\circ$};
  \node at (4.5, 0) {
    \begin{tikzpicture}[scale=.75]
    \node[vertex] (v1) at (-1.15, 0) {};
    \node[vertex] (v2) at (1.15, 0) {};
    \node[vertex] (v3) at ($(v1)!1!60:(v2)$) {};

    \node[vertex] (v4) at (0,.7) {};

    \draw[edge]
      (v1) -- (v2)
      (v1) -- (v3)
      (v1) -- (v4)
      (v2) -- (v3)
      (v2) -- (v4)
      (v3) -- (v4);

    \coordinate (lo1) at (-2.25, 2);
    \coordinate (lo2) at (-2.25, .7);
    \coordinate (lo3) at (-2.25, 0);
    \coordinate (l1) at (2.25, 2);
    \coordinate (l2) at (2.25, .7);

    \draw[wire]
      (v3) -- (lo1)
      (v4) -- (lo2)
      (v1) -- (lo3);

    \draw[wire]
      (v3) -- (l1)
      (v4) -- (l2);
    \end{tikzpicture}
  };
  \node at (6.9, 0) {$=$};
  \node at (9, 0) {
    \begin{tikzpicture}[scale=.75]
      \node[vertex] (v) at (0, 0) {};
      \node[vertex] (v2) at (1, 1) {};
      \node[vertex] (v3) at (-1, 1) {};
      \node[vertex] (v4) at (-1, -1) {};
      \node[vertex] (v5) at (1, -1) {};

      \node[vertex] (w) at (0, -1) {};

      \draw[edge]
        (w) -- (v5)
        (w) -- (v)
        (v) -- (v2)
        (v) -- (v3)
        (v) -- (v4)
        (v) -- (v5);

      \draw[edge]
        (v5) to[out=125,in=55,looseness=8] (v5);

      \node (l1) at (2, 1) {};
      \node (l3) at (2, -1) {};
      \node (l2) at ($(l1)!.5!(l3)$) {};

      \node (lo1) at (-2, 1) {};
      \node (lo3) at (-2, -1) {};
      \node (lo2) at ($(lo1)!.5!(lo3)$) {};

      \coordinate (h) at ($(v3)!.5!(v4)$);

      \draw[wire]
        (v) -- (l2)
        (v5) -- (l3);

      \draw[wire]
        (lo1) -- (v3)
        (v) -- (h)
        (lo3) to[out=0,in=180] (h)
        (lo2) to[out=0,in=180] (v4);
    \end{tikzpicture}
  };
  \end{tikzpicture} 
  \\
  \vspace{.5cm}
  \begin{tikzpicture}
    \node at (0, 0) {
      \begin{tikzpicture}[scale=.75]
          \node[vertex] (v) at (0, 0) {};
          \node[vertex] (v2) at (1, 1) {};
          \node[vertex] (v3) at (-1, 1) {};
          \node[vertex] (v4) at (-1, -1) {};
          \node[vertex] (v5) at (1, -1) {};

          \draw[edge]
            (v) -- (v2)
            (v) -- (v3)
            (v) -- (v4)
            (v) -- (v5);

          \node (l1) at (2, 1) {};
          \node (l3) at (2, -1) {};
          \node (l2) at ($(l1)!.5!(l3)$) {};

          \node (lo1) at (-2, 1) {};
          \node (lo3) at (-2, -1) {};
          \node (lo2) at ($(lo1)!.5!(lo3)$) {};

          \coordinate (h) at ($(v3)!.5!(v4)$);

          \draw[wire]
            (v) to[out=0,in=180] (l1)
            (v5) to[out=0,in=180] (l2)
            (v5) -- (l3);

          \draw[wire]
            (lo1) -- (v3)
            (v) -- (h)
            (lo3) to[out=0,in=180] (h)
            (lo2) to[out=0,in=180] (v4);
        \end{tikzpicture}
      };
  \node at (2.1, 0) {$\otimes$};
  \node at (4.5, 0) {
    \begin{tikzpicture}[scale=.75]
      \node[vertex] (v1) at (-1.15, 0) {};
      \node[vertex] (v2) at (1.15, 0) {};
      \node[vertex] (v3) at ($(v1)!1!60:(v2)$) {};

      \node[vertex] (v4) at (0,.7) {};

      \draw[edge]
        (v1) -- (v2)
        (v1) -- (v3)
        (v1) -- (v4)
        (v2) -- (v3)
        (v2) -- (v4)
        (v3) -- (v4);

      \coordinate (lo1) at (-2.25, 2);
      \coordinate (lo2) at (-2.25, .7);
      \coordinate (lo3) at (-2.25, 0);
      \coordinate (l1) at (2.25, 2);
      \coordinate (l2) at (2.25, .7);

      \draw[wire]
        (v3) -- (lo1)
        (v4) -- (lo2)
        (v1) -- (lo3);

      \draw[wire]
        (v3) -- (l1)
        (v4) -- (l2);
    \end{tikzpicture}
  };
  \node at (6.9, 0) {$=$};
  \node at (8.75, -.75) {
    \begin{tikzpicture}[scale=.5]
      \node[vertex] (v1) at (-1.15, 0) {};
      \node[vertex] (v2) at (1.15, 0) {};
      \node[vertex] (v3) at ($(v1)!1!60:(v2)$) {};

      \node[vertex] (v4) at (0,.7) {};

      \draw[edge]
        (v1) -- (v2)
        (v1) -- (v3)
        (v1) -- (v4)
        (v2) -- (v3)
        (v2) -- (v4)
        (v3) -- (v4);

      \coordinate (lo1) at (-2.25, 2);
      \coordinate (lo2) at (-2.25, .7);
      \coordinate (lo3) at (-2.25, 0);
      \coordinate (l1) at (2.25, 2);
      \coordinate (l2) at (2.25, .7);

      \draw[wire]
        (v3) -- (lo1)
        (v4) -- (lo2)
        (v1) -- (lo3);

      \draw[wire]
        (v3) -- (l1)
        (v4) -- (l2);
    \end{tikzpicture}
  };
  \node at (8.75, .75) {
    \begin{tikzpicture}[scale=.5]
      \node[vertex] (v) at (0, 0) {};
      \node[vertex] (v2) at (1, 1) {};
      \node[vertex] (v3) at (-1, 1) {};
      \node[vertex] (v4) at (-1, -1) {};
      \node[vertex] (v5) at (1, -1) {};

      \draw[edge]
        (v) -- (v2)
        (v) -- (v3)
        (v) -- (v4)
        (v) -- (v5);

      \node (l1) at (2.45, 1) {};
      \node (l3) at (2.45, -1) {};
      \node (l2) at ($(l1)!.5!(l3)$) {};

      \node (lo1) at (-2.45, 1) {};
      \node (lo3) at (-2.45, -1) {};
      \node (lo2) at ($(lo1)!.5!(lo3)$) {};

      \coordinate (h) at ($(v3)!.5!(v4)$);

      \draw[wire]
        (v) to[out=0,in=180] (l1)
        (v5) to[out=0,in=180] (l2)
        (v5) -- (l3);

      \draw[wire]
        (lo1) -- (v3)
        (v) -- (h)
        (lo3) to[out=0,in=180] (h)
        (lo2) to[out=0,in=180] (v4);
    \end{tikzpicture}
  };
  \end{tikzpicture} 
  \\
  \vspace{.5cm}
  \begin{tikzpicture}
    \node at (-1.75, 0) {$\Vast($};
    \node at (0, 0) {
      \begin{tikzpicture}[scale=.75]
          \node[vertex] (v) at (0, 0) {};
          \node[vertex] (v2) at (1, 1) {};
          \node[vertex] (v3) at (-1, 1) {};
          \node[vertex] (v4) at (-1, -1) {};
          \node[vertex] (v5) at (1, -1) {};

          \draw[edge]
            (v) -- (v2)
            (v) -- (v3)
            (v) -- (v4)
            (v) -- (v5);

          \node (l1) at (2, 1) {};
          \node (l3) at (2, -1) {};
          \node (l2) at ($(l1)!.5!(l3)$) {};

          \node (lo1) at (-2, 1) {};
          \node (lo3) at (-2, -1) {};
          \node (lo2) at ($(lo1)!.5!(lo3)$) {};

          \coordinate (h) at ($(v3)!.5!(v4)$);

          \draw[wire]
            (v) to[out=0,in=180] (l1)
            (v5) to[out=0,in=180] (l2)
            (v5) -- (l3);

          \draw[wire]
            (lo1) -- (v3)
            (v) -- (h)
            (lo3) to[out=0,in=180] (h)
            (lo2) to[out=0,in=180] (v4);
        \end{tikzpicture}
      };
    \node at (1.75, 0) {$\Vast)^*$};
    \node at (2.5, 0) {$=$};
    \node at (4.5, 0) {
      \begin{tikzpicture}[scale=.75,xscale=-1]
          \node[vertex] (v) at (0, 0) {};
          \node[vertex] (v2) at (1, 1) {};
          \node[vertex] (v3) at (-1, 1) {};
          \node[vertex] (v4) at (-1, -1) {};
          \node[vertex] (v5) at (1, -1) {};

          \draw[edge]
            (v) -- (v2)
            (v) -- (v3)
            (v) -- (v4)
            (v) -- (v5);

          \node (l1) at (2, 1) {};
          \node (l3) at (2, -1) {};
          \node (l2) at ($(l1)!.5!(l3)$) {};

          \node (lo1) at (-2, 1) {};
          \node (lo3) at (-2, -1) {};
          \node (lo2) at ($(lo1)!.5!(lo3)$) {};

          \coordinate (h) at ($(v3)!.5!(v4)$);

          \draw[wire]
            (v) to[out=0,in=180] (l1)
            (v5) to[out=0,in=180] (l2)
            (v5) -- (l3);

          \draw[wire]
            (lo1) -- (v3)
            (v) -- (h)
            (lo3) to[out=0,in=180] (h)
            (lo2) to[out=0,in=180] (v4);
        \end{tikzpicture}
      };
  \end{tikzpicture}
  \caption{Operations on bilabelled graphs}
  \label{fig:operationsongraphs}
\end{figure}

Based on the labels, it is possible to define tensor products, composition, and adjoints of bilabelled graphs. Let $\mathbf{K} = (K, \tup{a}, \tup{b}) \in \gbG(\ell, k)$ and $\mathbf{K'} = (K', \tup{a}', \tup{b}') \in \gbG(\ell', k')$. Then the tensor product $\mathbf{K} \otimes \mathbf{K'} \in \gbG(\ell + \ell', k + k')$ is defined as the bilabelled graph $(K \cup K', \tup{a}\tup{a'}, \tup{b}\tup{b}')$. If $\ell' = k$, the composition $\mathbf{K}' \circ \mathbf{K} \in \gbG(k', \ell)$ is formed by constructing the disjoint union $K \cup K'$, identifying the vertices $a_i'$ and $b_i$ for all $i \in [k]$, and removing any duplicate edges but keeping the loops. Finally, the adjoint $\mathbf{K}^* \in \gbG(k, \ell)$ is given by $(K, \tup{b}, \tup{a})$.
We denote the closure of a set of bilabelled graphs $\mathbf{K}_1, \dots, \mathbf{K}_n$ under these operations by $\intertwclosure{\mathbf{K}_1, \dots, \mathbf{K}_n}$.
As for partitions, we can view the operations graphically, as shown in \cref{fig:operationsongraphs}. Composition, in particular, corresponds to drawing $\mathbf{K}'$ on the left of $\mathbf{K}$ so that the output and input wires align, connecting the wires, and then contracting the resulting edges.

Homomorphisms between bilabelled graphs are required to preserve labels.

\begin{definition} 
  A \emph{bilabelled graph homomorphism} from $\mathbf{K} = (K, \tup{a}, \tup{b}) \in \gbG(k, \ell)$ to $\mathbf{K}' = (K', \tup{a}', \tup{b}') \in \gbG(k, \ell)$ is a homomorphism $h\colon K \to K'$ such that $h(\tup{a}) = \tup{a}'$ and $h(\tup{b}) = \tup{b}'$.
\end{definition}

We count homomorphisms from a bilabelled graph to an unlabelled graph by considering all possible label positions in the target graph, and arranging the resulting numbers of bilabelled graph homomorphisms in a tensor.

\begin{definition}
  Given a bilabelled graph $\mathbf{K} \in \gbG(\ell, k)$ and a graph $G$, the \emph{$G$\nobreakdash-homomorphism tensor} of $\mathbf{K}$ is the $V(G)^\ell \times V(G)^k$ matrix $\mathbf{K}_G$ with entries
  \begin{equation*}
    (\mathbf{K}_G)_{\tup{u}\tup{v}} = \abs{\{h\colon \mathbf{K} \to (G, \tup{u}, \tup{v}) \mid h \text{ bilabelled graph homomorphism}\}}.
  \end{equation*}
\end{definition}

\begin{example}
  Let $\mathbf{K} = (K, \tup{a}, \tup{b}) \in \gbG(\ell, k)$ be a bilabelled graph, and $G$ be an unlabelled graph. Then we have $\hom(K, G) = \soe(\mathbf{K}_G)$, where
  \begin{equation*}
    \soe(\mathbf{K}_G) = \sum_{\tup{u} \in V(G)^\ell} \sum_{\tup{v} \in V(G)^k} (\mathbf{K}_G)_{\tup{u}\tup{v}}
  \end{equation*}
  is the sum-of-entries function.
\end{example}

This procedure can be seen as assigning a linear map $(\C^n)^{\otimes k} \to (\C^n)^{\otimes \ell}$ to a bilabelled graph $\mathbf{K} \in \gbG(\ell, k)$ and an unlabelled graph $G$. Moreover, one may note the following. Consider an $(\ell, k)$-partition $P = \{p_1, \dots, p_n\}$. Construct a bilabelled graph $\mathbf{P} \eqqcolon \Gamma(P)$ by setting $V(\mathbf{P}) = \{v_i \mid i \in [n] \}$ and $E(\mathbf{P}) = \varnothing$, as well as $l_{in}(v_i) = \{i \mid i \in [\ell], i_U \in p_i\}$ and $l_{out}(v_i) = \{i \mid i \in [k], i_L \in p_i \}$. 
Then it is a routine calculation to verify that $\Gamma(P \otimes P') = \Gamma(P) \otimes \Gamma(P')$, $\Gamma(P \circ P') = \Gamma(P) \circ \Gamma(P')$, and $\Gamma(P^*) = \Gamma(P)^*$. Furthermore, $\Gamma$ has an inverse that assigns to an edgeless bilabelled graph $\mathbf{K} = (K, \tup{a}, \tup{b})$ the partition $P_\mathbf{K} = \Gamma\inverse(\mathbf{K}) = \{\{i_U \mid i \in l_{in}(v)\} \cup \{i_L \mid i \in l_{out}(v)\} \mid v \in V(K)\}$.

\begin{example}
  \label{ex:genmultgraph}
  Consider the edgeless bilabelled graphs $\bM^{\ell, k}$ associated to the single-block partitions $\{\{1_L, \dots, \ell_L, 1_U, \dots, k_U\}\}$. Then $\bM^{\ell,k}$ is a single-vertex graph with $k$ in-labels and $\ell$ out-labels. Since $\bM^{\ell, k}$ has no edges, $\bM^{\ell, k}_G$ depends only on $\abs{V(G)} \eqqcolon n$ and we have 
\begin{equation*}
  (\bM^{\ell, k}_G)_{\tup{u}\tup{v}} = \begin{cases}
    1 & \text{if } u_1 = \dots = u_\ell = v_1 = \dots = v_k,\\
    0 & \text{otherwise.}
  \end{cases}
\end{equation*}
In particular, we have $\bM^{\ell, k}_G = M^{(n), \ell, k}$ for all $\ell, k \in \N$. An important special case is $\bM^{1,1}_G$, which is just the identity matrix $I_n$. We thus also write $\mathbf{I}$ for $\bM^{1,1}$.
\end{example}

\begin{figure}[h]
  \centering
  \begin{tikzpicture}
    \begin{scope}[yshift=20]
      \node[vertex] (v) at (0, 0) {};
      \node[vertex] (w) at (1.75, 0) {};

      \coordinate (lo) at (-.75, 0);
      \coordinate (li) at (2.5, 0);

      \draw[edge] (v) -- (w);

      \draw[wire] 
        (v) -- (lo)
        (w) -- (li);
    \end{scope}

    \node (cap) at (.875, -1) {$\mathbf{A}$};
  \end{tikzpicture}
  \hspace{1cm}
  \begin{tikzpicture}
    \begin{scope}[yshift=20]
    \node[vertex] (v) at (0, 0) {};

    \coordinate (lo1) at (-.75, 0);
    \coordinate (li1) at (.75, 0);

    \draw[wire] 
      (v) -- (lo1)
      (v) -- (li1);

    \draw[wire]
      (v) -- (lo1)
      (v) -- (li1);
    \end{scope}

    \node (cap) at (0, -1) {$\mathbf{I}$};
  \end{tikzpicture}
  \hspace{1cm}
  \begin{tikzpicture}
    \begin{scope}[yshift=20]
    \node[vertex] (v) at (0, 0) {};

    \coordinate (lo1) at (-.75, -.5);
    \coordinate (lo2) at (-.75, .5);

    \draw[wire] 
      (v) to[bend left] (lo1)
      (v) to[bend right] (lo2);
    \end{scope}

    \node (cap) at (0, -1) {$\bM^{2, 0}$};
  \end{tikzpicture}
  \hspace{1cm}
  \begin{tikzpicture}
    \node[vertex] (v) at (0, 0) {};
    \node[vertex] (w) at (0, 1.5) {};

    \coordinate (lo1) at (-.75, 0);
    \coordinate (lo2) at (-.75, 1.5);

    \coordinate (li1) at (1.25, 0);
    \coordinate (li2) at (1.25, 1.5);

    \draw[wire] 
      (v) -- (lo1)
      (w) -- (lo2);

    \draw[wire]
      (v) to[out=0,in=180] (li2)
      (w) to[out=0,in=180] (li1);

    \node (cap) at (.5, -1) {$\mathbf{S}$};
  \end{tikzpicture}
  \caption{Some important bilabelled graphs.}
  \label{fig:importantblgraphs}
\end{figure}
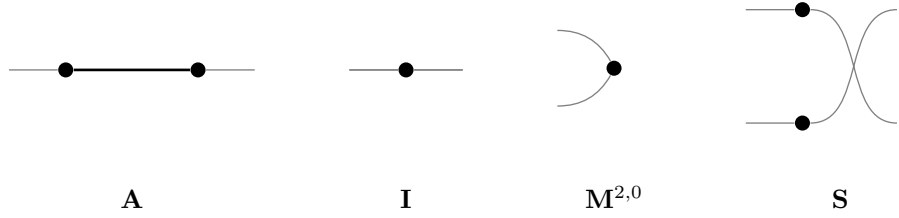

\begin{example}
  \label{ex:swapgraph}
  Consider the \emph{swap graph} $\mathbf{S}$ associated to the swap partition $\spa$, illustrated in \cref{fig:importantblgraphs}. Again, since $\mathbf{S}$ does not have any edges, $\mathbf{S}_G$ only depends on $n$, and we have
  \begin{equation*}
    (\mathbf{S}_G)_{u_1u_2,v_1v_2} = \delta_{u_1v_2}\delta_{u_2v_1}.
  \end{equation*}
  As a linear map, this correponds exactly to $S^{(n)}$.
\end{example}

The reader may convince themself that every partition can be constructed from single block partitions and the swap map using composition and tensor products. Examples \ref{ex:genmultgraph} and \ref{ex:swapgraph} thus imply that $\mathbf{K}_G = T_{\Gamma\inverse(\mathbf{K})}$ for all edgeless bilabelled graphs $\mathbf{K}$.

Generally, of course, bilabelled graphs can have edges. An important example is the \emph{bilabelled edge}.

\begin{example}
  \label{ex:bilabedge}
  The bilabelled edge is given by $\mathbf{A} = (K_2, (v_1), (v_2))$. For any graph $G$, we have
  \begin{equation*}
    (\mathbf{A}_G)_{vw} = \begin{cases}
      1 & \text{if } vw \in E(G),\\
      0 & \text{otherwise}.
    \end{cases}
  \end{equation*}
  In other words, $\mathbf{A}_G$ is just the adjacency matrix $A_G$ of $G$.
\end{example}

For arbitrary bilabelled graphs, Man\v{c}inska and Roberson \cite{manvcinska2020quantum} proved an important generalisation of \cref{prop:partmapcorrespondence}.

\begin{lemma}\label{lem:opcorrespondence}
  Let $\mathbf{K}, \mathbf{K}' \in \gbG$. Then we have $(\mathbf{K} \otimes \mathbf{K'})_G = \mathbf{K}_G \otimes \mathbf{K}'_G$, $(\mathbf{K}^*)_G = \mathbf{K}_G^*$, and if $\mathbf{K}$ and $\mathbf{K}'$ are compatible, $(\mathbf{K} \circ \mathbf{K}')_G = \mathbf{K}_G \circ \mathbf{K}'_G$.
\end{lemma}

Based on this result, they introduced \emph{graph categories}, which give rise to concrete tensor categories with duals.

\begin{definition}
  A \emph{graph category} $\gF$ is a collection of bilabelled graphs, such that 
  \begin{enumerate}
    \item if $\mathbf{K} \in \gF(k, \ell)$ and $\mathbf{K}' \in \gF(r, s)$, then $\mathbf{K} \otimes \mathbf{K}' \in \gF(k + r, \ell + s)$,
    \item if $\mathbf{K} \in \gF(k, \ell)$ and $\mathbf{K}' \in \gF(\ell, r)$, then $\mathbf{K} \circ \mathbf{K}' \in \gF(k, r)$,
    \item if $\mathbf{K} \in \gF(k, \ell)$ then $\mathbf{K}^* \in \gF(\ell, k)$,
    \item $\mathbf{I} \in \gF(1, 1)$,
    \item $\bM^{2, 0} \in \gF(2, 0)$,
  \end{enumerate}
  where $\gF(k, \ell) = \gF \cap \gbG(k, \ell)$.
\end{definition}

Just as for partition categories, we write $\intertwclosure{\bM_1, \dots, \bM_n}$ to denote the graph category generated by the bilabelled graphs $\bM_1, \dots \bM_n$, that is, the closure of $\mathbf{I}, \bM^{2,0}, \bM_1, \dots, \bM_n$ under tensor products, composition, and taking adjoints. \cref{lem:opcorrespondence} now implies that for every graph category $\gF$ and graph $G$, 
\begin{equation*}
  C(\ell, k) = \lspan\{\mathbf{K}_G \mid G \in \gF(\ell, k)\}
\end{equation*}
is a concrete tensor category with duals. In particular, every pair of a graph category $\gF$ and a graph $G$ corresponds to a unique compact matrix quantum group through Woronowicz' Tannaka-Krein duality (\cref{thm:tannakakrein}). Man\v{c}inska and Roberson call these quantum groups \emph{graph-theoretical}. 
An important example is the \emph{quantum automorphism group} of a graph. 

\begin{definition}
  Let $G$ be a graph. The quantum automorphism group $\Qut(G)$ of $G$ is the unique compact matrix quantum group with intertwiners
  \begin{equation*}
    C^{\Qut(G)} = \lspan\{T_K \mid K \in \lrangle{M^{1, 0}, M^{1, 2}, A_G}\}.
  \end{equation*}
\end{definition}

In other words, the quantum automorphism group $\Qut(G)$ is obtained by starting with the quantum symmetric group $S_n^+$, and adding the adjacency matrix of $G$ as an intertwiner. In particular, by Examples \ref{ex:genmultgraph} and \ref{ex:bilabedge}, it can be phrased entirely in terms of homomorphism tensors: It is the graph-theoretical quantum group obtained by starting with the graph category underlying $S_n^+$, and adding the bilabelled edge $\mathbf{A}$.
Based on these observations, Man\v{c}inska and Roberson eventually prove the following result.

\begin{theorem}[Theorem 7.16 in \cite{manvcinska2020quantum}]
  \label{thm:mrqiso}
  Let $G$ and $H$ be graphs. Then the following are equivalent.
  \begin{enumerate}
    \item There exists a quantum permutation matrix $\U$ such that $\U A_G = A_H \U$, that is, $G$ and $H$ are quantum isomorphic.
    \item There exists a weak isomorphism between $\Qut(G)$ and $\Qut(H)$.
    \item $G$ and $H$ are homomorphism indistinguishable over the $(0, 0)$-bilabelled graphs in $\intertwclosure{\bM^{1, 0}, \bM^{1, 2}, \mathbf{A}}$, which are precisely the planar graphs.
  \end{enumerate}
\end{theorem}

A \emph{quantum permutation matrix} is a matrix in $\mathcal{Z}^{n \times n}$ for some \Cstar-algebra $\mathcal{Z}$, satisfying the generating relations of the quantum symmetric group in \cref{eq:symrep}. Generally, we will call matrices with entries in a \Cstar-algebra \emph{quantum matrices}. A \emph{weak isomorphism} is a notion introduced to capture that the intertwiner spaces of two quantum groups are essentially the same, cf. \cite[Definition 7.2]{manvcinska2020quantum} and later \cref{def:weakiso}.

  \section{Counting Homomorphisms from Arbitrary Intertwiner Graphs}
  The idea of adding the adjacency matrix of a graph to the intertwiners of an easy quantum group $Q$ is of course not unique to the quantum symmetric group. This procedure  gives rise to a class of ``graph-dependent'' quantum subgroups of $Q$, generalising the concept of quantum automorphism groups. The question thus arises whether \cref{thm:mrqiso} can also be generalised in this way, replacing the $\Qut(G)$ and $\Qut(H)$ with the corresponding subgroups of $Q$, the matrix $\U$ with a matrix satisfying the same intertwiner relations as the fundamental representation of $Q$, and the graph class with the bilabelled graphs corresponding to the intertwiners of $Q$.

\subsection{The Graph Instatiation}
Let us begin by formalising these ideas. The first is the idea of ``adding the bilabelled edge''. For brevity, we simplify some definitions. Let $Q$ be an orthogonal easy quantum group corresponding to some partition category $\lrangle{P_1, \dots, P_k}$. Then we call the linear maps $\identity, M^{2, 0}, M_1, \dots, M_k$ corresponding to $\spd, \spe, P_1, \dots, P_k$ the \emph{generators} of $Q$. Moreover, we denote the corresponding edgeless bilabelled graphs by bold letters, in this case $\bI, \bM^{2, 0}, \bM_1, \dots, \bM_k$. 

\begin{definition}
  Let $Q$ be an easy quantum group with generators $M_1, \dots, M_k$ and let $G$ be a graph. Then the \emph{$G$-instantiation} $Q(G)$ of $Q$ is the unique compact matrix quantum group with intertwiners $C^Q_G(k, \ell) \coloneqq \lspan \{ \mathbf{K}_G \mid \mathbf{K} \in \intertwclosure{\bM_1, \dots, \bM_k, \mathbf{A}} \cap \gbG(k, \ell)\}$.
\end{definition}

In particular, the $G$-instantiation of $S^+_n$ is simply the quantum automorphism group $\Qut(G)$, and we have $C^{S^+_n}_G = C^{\Qut(G)}$.  

\begin{definition}
  Let $Q$ be an easy quantum group with generators $M_1, \dots, M_k$. We call the set of graphs $\gF = \intertwclosure{\bM_1, \dots, \bM_k, \mathbf{A}}$ the \emph{intertwiner graphs} of $Q$. Moreover, for all $k, \ell \in \N$ we denote by $\gF(\ell, k) \subseteq \gF$ the set of $(\ell, k)$-bilabelled graphs in $\gF$.
\end{definition}

For example, Man\v{c}inska and Roberson showed that the $(0, 0)$-bilabelled graphs among the intertwiner graphs of $S_n^+$ are precisely the planar graphs.
 Next, we define a shorthand notation for homomorphism indistinguishability over the $(0, 0)$-bilabelled graphs in a graph category.

\begin{definition}
  Let $\gF$ be a collection of bilabelled graphs. We say that two graphs $G$, $H$ are \emph{$\gF$-isomorphic}, in symbols $G \cong_\gF H$, if $G$ and $H$ are homomorphism indistinguishable over $\gF(0, 0)$, that is, 
  \begin{equation*}
  \hom(F, G) = \hom(F, H)
  \end{equation*}
  for all $F \in \gF(0, 0)$.
\end{definition}

Central to Man\v{c}inska and Roberson's proof is the concept of a weak isomorphism between quantum automorphism groups. Their definition can be generalised to graph instantiations of arbitrary easy quantum groups in a straightforward way.

\begin{definition}
  \label{def:weakiso}
  Let $Q = (C(Q), \U, n)$ be an easy quantum group with generators $M_1, \dots, M_k$, and let $G$, $H$ be two graphs of order $n$. 
  A \emph{weak isomorphism} between $Q(G)$ and $Q(H)$ is a bijection $\Phi \colon C^Q_G \to C^Q_H$ such that for all $T, T' \in C^Q_G$, 
  \begin{enumerate}
    \item $\Phi(A_G) = A_H$ and $\Phi(M_i) = M_i$ for all $i \in [k]$,
    \item $\Phi(TT') = \Phi(T)\Phi(T')$ if $T, T'$ are compatible,
    \item $\Phi(T \otimes T') = \Phi(T) \otimes \Phi(T')$,
    \item $\Phi(T^*) = \Phi(T)^*$,
    \item For all $\ell, k \in \N$, the restriction of $\Phi$ to $C^Q_G(\ell, k)$ is a bijective linear map to $C^Q_H(\ell, k)$.
  \end{enumerate}
\end{definition}

These functions essentially capture that the intertwiner spaces of the two graph instantiations have the same structure. In particular, every construction of an intertwiner from the generators of $Q(G)$ can be mirrored in $Q(H)$. 

\begin{remark}
  Technically, we should write $\Phi_{M_1, \dots, M_k}$ since the definition depends on the choice of generators. However, it will later become clear that for any two choices of generators $M_1, \dots, M_k$ and $M'_1, \dots, M'_\ell$ we have that $\Phi(M_i) = M_i$ for all $i \in [k]$ implies $\Phi(M'_j) = M'_j$  for all $j \in [\ell]$.
\end{remark}

In the next two sections, we will prove the following theorem, which is exactly our desired generalisation of \cref{thm:mrqiso}.

\begin{restatable}{thm}{mainresult}\label{thm:mainresult}
  Let $Q$ be an orthogonal easy quantum group with generators $M_1, \dots, M_m$ and let $\gF$ be the set of its intertwiner graphs. Let $G$ and $H$ be graphs. Then the following are equivalent.
  \begin{enumerate}
    \item $G \cong_\gF H$.
    \item There exists a weak isomorphism between $Q(G)$ and $Q(H)$.
    \item There exists a quantum orthogonal matrix $\U$ with $\Utp{k} M = M \Utp{\ell}$ for all $M \in C^Q(k, \ell)$, such that $\U A_G = A_H \U$.
    \item There exists a quantum orthogonal matrix $\U$ with $\Utp{k} M_i = M_i \Utp{\ell}$ for all $i \in [m]$ and $M_i \in C^Q(\ell, k)$, such that $\U A_G = A_H \U$.
  \end{enumerate}
  Moreover, if these statements hold, then 
  \begin{enumerate}
    \setcounter{enumi}{4}
    \item $Q(G)$ and $Q(H)$ are unitarily monoidally equivalent.
  \end{enumerate}
\end{restatable}

A \emph{quantum orthogonal matrix} is a quantum matrix satisfying the definining relations of the quantum orthogonal group in \cref{eq:orthrep}.
The proof of the implications 4 $\implies$ 2 $\implies$ 1 is more or less an obvious generalisation of Man\v{c}inska and Roberson's original proof using our just established definitions. We nevertheless include the proofs for the sake of completeness. The converse direction is much more involved. 
The original proof uses results on quantum group orbits and orbitals that do not generalise beyond the quantum symmetric group. Here we take a different approach, using previous results on monoidal categories, or, more precisely, \Cstar-tensor categories. This approach also yields statement 5, which was already known to hold for the quantum symmetric group \cite{brannan2020bigalois}.

We split the proof into two sections. In the remainder of this section, we examine the connection between homomorphism indistinguishability and the structure of intertwiner spaces through weak isomorphisms. We show that graphs $G$ and $H$ are homomorphism indistinguishable over the unlabelled fragments of the intertwiner graphs of an easy quantum group $Q$ if and only if there exists a weak isomorphism between $Q(G)$ and $Q(H)$, hence proving 1 $\iff$ 2.

In the next section, we want to show that the existence of a weak isomorphism is equivalent to a statement that more closely resembles the formulation of quantum isomorphism as a \Cstar-valued linear system of equations $\U A_G = A_H \U$.
To that end, we first show 4 $\implies$ 2, proving that any quantum matrix satisfying the intertwiner relations of the underlying easy quantum group induces a weak isomorphism between $Q(G)$ and $Q(H)$. Since the implication 3 $\implies$ 4 is trivial, we conclude by proving that 2 $\implies$ 3 using category-theoretic machinery in the form of a theorem by Neshveyev and Tuset \cite{neshveyev_compact_2013}. On the way, we obtain 2 $\implies$ 5.

\subsection{\texorpdfstring{$\gF$}{F}-isomorphism implies weak isomorphism of graph instantiations}
Suppose then that $G \cong_\gF H$. We want to show that there exists a weak isomorphism between $Q(G)$ and $Q(H)$. The first requirement is that $G$ and $H$ must be of the same order, which follows without much difficulty.

\begin{observation}\label{obs:graphsameorder}
  Let $Q$ be a compact matrix quantum group and $\gF$ the class of its intertwiner graphs. Then for all graphs $G$, $H$ with $G \cong_{\gF} H$ it holds that $\abs{G} = \abs{H}$.
\begin{proof}
  Note that $\gF(0, 0)$ contains the single vertex graph $K_1$. Indeed, we have $K_1 = \mathbf{M}^{0,2}\mathbf{M}^{2,0}$. The result now follows from the fact that $\hom(K_1, G) = \abs{G}$ for all graphs $G$.
\end{proof}
\end{observation}

The central observation is now that for any graphs $G$ and $H$, $C^Q_G$ and $C^Q_H$ already look quite similar on the surface, since the intertwiner spaces are generated by the homomorphism tensors of the same bilabelled graphs. In particular, the individual vector spaces $C^Q_G(k, \ell)$ and $C^Q_H(k, \ell)$ are essentially spanned by the ``same'' vectors. Now if the inner product between any two vectors in $C^Q_G(k, \ell)$ always equals that of the corresponding vectors in $C^Q_H(k, \ell)$, then we can conclude by a Gram-Schmidt argument that there exists a unitary linear map between the two vector spaces that maps a vector in $C^Q_G(k, \ell)$ to its analogue in $C^Q_H(k, \ell)$. This is formalised in the following lemma.

\begin{lemma}[Lemma 2.4.3 in \cite{seppelt_homomorphism_2024}]\label{lem:gramschmidt}
	Let $V$ and $W$ be finite-dimensional complex vector spaces.
	Let $I$ be a possibly infinite set.
	Let $(v_i)_{i \in I}$ and $(w_i)_{i \in I}$ be two sequences of vectors such that $v_i \in V$ and $w_i \in W$ for all $i \in I$.
	Suppose that 
	\begin{enumerate}
		\item the $v_i$ for $i \in I$ span $V$, the $w_i$ for $i \in I$ span $W$, and
		\item $\langle v_i, v_j \rangle = \langle w_i, w_j \rangle$ for all $i, j \in I$.
	\end{enumerate}
	Then there exists a unitary linear map $\Phi \colon V \to W$ such that $\Phi(v_i) = w_i$ for all $i \in I$.
\end{lemma}

The next lemma further implies that the inner product of vectors in $C^Q_G$ and $C^Q_H$ is precisely given by homomorphism counts of unlabelled intertwiner graphs of $Q$.

\begin{lemma}\label{lem:scalarprod}
  Let $Q$ be a compact matrix quantum group. For all $k, \ell$, there exists an inner product on $C^Q(k, \ell)$ defined in terms of $\identity$ and $M^{2, 0}$.
  \begin{proof}
    Let $T_1, T_2 \in C^Q(k, \ell)$. First note that if $k = 0$, then letting $\lrangle{T_1, T_2} \coloneqq T_1^*T_2$ is sufficient.
    Otherwise we show that we can construct the Hilbert-Schmidt inner product on the space of linear maps $(\C^n)^{\otimes k} \to (\C^n)^{\otimes \ell}$. For $k \geq 1$, we let $\tilde{B}_k = \identity^{\otimes k-1} \otimes M^{2, 0} \otimes \identity^{\otimes k-1}$ and $B_k = \tilde{B}_k \tilde{B}_{k-1} \cdots \tilde{B}_1$. We then define the inner product of $T_1, T_2$ by
    \begin{equation*}
      \langle T_1, T_2 \rangle \coloneqq B_k^* (\identity^{\otimes k} \otimes T_1^*T_2) B_k.
    \end{equation*}
     A routine computation then shows that $\lrangle{T_1, T_2} = \operatorname{Tr} T_1^* T_2$, as desired.
  \end{proof}
\end{lemma}

Since the construction only involves the intertwiners $M^{2,0}$ and $\identity$, which correspond to bilabelled graphs, we can extend $\lrangle{\argp, \argp}$ to $(k, \ell)$-bilabelled graphs $\mathbf{K}, \mathbf{K'}$ for all $k, \ell \in \N$. Then $\mathbf{\lrangle{K, K'}}$ is an unlabelled graph, and we have $\mathbf{\lrangle{K, K'}}_G = (\lrangle{\mathbf{K}_G, \mathbf{K'}_G}) \in C^Q_G(0, 0)$. We can now prove the following lemma.

\begin{lemma}\label{lem:homindtoweakiso}
  Let $Q$ be an easy quantum group, $\gF$ the set of its intertwiner graphs, and let $G$ and $H$ be graphs such that $G \cong_{\gF} H$. Then there exists a weak isomorphism $\Phi \colon C^Q_G \to C^Q_H$.
  \begin{proof}
    First note that by \cref{obs:graphsameorder}, we have $C^Q_G(k, \ell), C^Q_H(k, \ell) \subseteq \C^{n^k \times n^\ell}$ for all $k, \ell \in \N$. 
    We now define sequences $(v_\mathbf{K})_{\mathbf{K} \in \gF} \subseteq C^Q_G$, $(w_\mathbf{K})_{\mathbf{K} \in \gF} \subseteq C^Q_H$ by setting $v_\mathbf{K} \coloneqq \mathbf{K}_G$ and $w_\mathbf{K} \coloneqq \mathbf{K}_H$. Then $\{v_\mathbf{K} \mid \mathbf{K} \in \gF(\ell, k)\}$ spans $C^Q_G(\ell, k)$ and $\{w_\mathbf{K} \mid \mathbf{K} \in \gF(\ell, k) \}$ spans $C^Q_H(\ell, k)$. Moreover, we have for all $k, \ell \in \N$ and $\mathbf{K}, \mathbf{K'} \in \gF(\ell, k)$
    \begin{align*}
      \lrangle{v_\mathbf{K}, v_{\mathbf{K'}}} &= \soe(\mathbf{\lrangle{K, K'}}_G)\\
                          &= \hom(\mathbf{\lrangle{K, K'}}, G)\\
                          &= \hom(\mathbf{\lrangle{K, K'}}, H)\\
                          &= \soe(\mathbf{\lrangle{K, K'}}_H)\\
                          &= \lrangle{w_\mathbf{K}, w_{\mathbf{K'}}},
    \end{align*}
    where $\lrangle{\argp, \argp}$ is the inner product defined in \cref{lem:scalarprod}. Consequently, by \cref{lem:gramschmidt} there exists, for all $\ell, k \in \N$, a unitary linear map $\Phi_{\ell, k}\colon C^Q_G(\ell, k) \to C^Q_H(\ell, k)$, mapping $v_\mathbf{K}$ to $w_\mathbf{K}$ for all $\mathbf{K} \in \gF$.

    \begin{claim}
      $\Phi \coloneqq \bigcup_{k, \ell \in \N} \Phi_{\ell, k}$ is a weak isomorphism $C^Q_G \to C^Q_H$.
      \begin{subproof}
        By linearity, it suffices to show that for all $\mathbf{K}_G, \mathbf{K'}_G \in C^Q_G$ with $\mathbf{K}, \mathbf{K'} \in \gF$ it holds that
        \begin{enumerate}[(a)]
          \item $\Phi(A_G) = A_H$ and $\Phi(M_i) = M_i$ for all $i \in [k]$,
          \item $\Phi(\mathbf{K}_G\mathbf{K'}_G) = \Phi(\mathbf{K}_G)\Phi(\mathbf{K'}_G)$ if $\mathbf{K}_G, \mathbf{K'}_G$ are compatible,
          \item $\Phi(\mathbf{K}_G \otimes \mathbf{K'}_G) = \Phi(\mathbf{K}_G) \otimes \Phi(\mathbf{K'}_G)$,
          \item $\Phi(\mathbf{K}^*_G) = \Phi(\mathbf{K}_G)^*$,
          \item For all $\ell, k \in \N$, the restriction of $\Phi$ to $C^Q_G(\ell, k)$ is a bijective linear map to $C^Q_H(\ell, k)$.
        \end{enumerate}

        First note that the restriction of $\Phi$ to $C^Q_G(\ell, k)$ is just $\Phi_{\ell, k}$, which is a unitary linear map and thus bijective. Moreover, we have $A_G = \mathbf{A}_G = v_A$ and $A_H = \mathbf{A}_H = w_A$ and thus $\Phi(A_G) = A_H$; as well as $M_i = v_{\mathbf{M}_i} = w_{\mathbf{M}_i}$ (since $\mathbf{M}_i$ is edgeless and $G$, $H$ have the same order) and hence $\Phi(M_i) = M_i$. To see that (b) holds, observe that for compatible $\mathbf{K}_G, \mathbf{K'}_G \in C^Q_G$,
        \begin{align*}
        \Phi(\mathbf{K}_G\mathbf{K'}_G) &= \Phi((\mathbf{KK'})_G)
        = \Phi(v_{\mathbf{KK'}})
        = w_{\mathbf{KK'}}
                                        = (\mathbf{KK'})_H
                                        = \mathbf{K}_H\mathbf{K'}_H
                                        = w_\mathbf{K}w_{\mathbf{K'}}\\
                                        &= \Phi(\mathbf{K}_G)\Phi(\mathbf{K'}_G).
        \end{align*}
        The proof of (c) is analogous. Finally, we have 
        \begin{align*}
          \Phi(\mathbf{K}_G^*) = \Phi((\mathbf{K}^*)_G) = \Phi(v_{\mathbf{K}^*}) = w_{\mathbf{K}^*} = (\mathbf{K}^*)_H = \mathbf{K}_H^* = \Phi(\mathbf{K}_G)^*,
        \end{align*}
        so (d) is also satisfied, and $\Phi$ is a weak isomorphism.
      \end{subproof}
    \end{claim} 
    This concludes the proof.
  \end{proof}
\end{lemma}

\subsection{Weak isomorphism of graph instantiations implies \texorpdfstring{$\gF$}{F}-isomorphism}

The converse direction is a direct generalisation of Man\v{c}inska and Roberson's proof. The main observation is again that a weak isomorphism mirrors the construction of intertwiners using composition, tensor product, and adjoints. In particular, the following holds.

\begin{lemma}\label{lem:constructionmirror}
  Let $Q$ be an easy quantum group, $\gF$ the set of its intertwiner graphs, and $G$, $H$ be graphs such that there exists a weak isomorphism $\Phi\colon C^Q_G \to C^Q_H$. Then for all $\mathbf{K} \in \gF$ it holds that $\Phi(\mathbf{K}_G) = \mathbf{K}_H$.
  \begin{proof}
    Since $\mathbf{K} \in \gF$, there exists a construction of $\mathbf{K}$ from generators $\mathbf{M}_1, \dots, \allowbreak \mathbf{M}_k \in \gF$ and $\mathbf{A}$ using the operations $\circ, \otimes, *$. We fix such a construction and proceed inductively.
  Per definition of weak isomorphisms, we have $\Phi((\mathbf{M}_i)_G) = \Phi(M_i) =M_i = (\mathbf{M}_i)_H$ for all $i \in [k]$ and $\Phi(\mathbf{A}_G) = \Phi(A_G) = A_H = \mathbf{A}_H$.
  Now suppose $\mathbf{K} = \mathbf{K}_1 \circ \mathbf{K}_2$. Then
    \begin{align*}
      \Phi(\mathbf{K}_G) &= \Phi((\mathbf{K}_1)_G(\mathbf{K}_2)_G) = \Phi((\mathbf{K}_1)_G)\Phi((\mathbf{K}_2)_G) = (\mathbf{K}_1)_H(\mathbf{K}_2)_H = (\mathbf{K}_1\mathbf{K}_2)_H\\ 
                         &= \mathbf{K}_H.
    \end{align*}
    The cases $\mathbf{K} = \mathbf{K}_1 \otimes \mathbf{K}_2$ and $\mathbf{K} = \mathbf{K}_1^*$ are analogous.
  \end{proof}
\end{lemma}

For easy quantum groups, we can always assume that $M^{2,0}$ is among the generators. This allows us to determine the $(0,0)$-slice of any weak isomorphism.

\begin{lemma}\label{lem:weakisoisidentity}
  Let $(C(Q), \U, n)$ be an easy quantum group, and $G$, $H$ be graphs that admit a weak isomorphism $\Phi \colon C^Q_G \to C^Q_H$. Then $\Phi(\identity^{\otimes 0}) = \identity^{\otimes 0}$. In particular, the restriction of $\Phi$ to $C^Q_G(0, 0)$ is the identity map.
  \begin{proof}
    Per definition of weak isomorphisms, we have $\abs{G} = \abs{H} \eqqcolon n$. We use that the map $M \coloneqq M^{2,0} \in C^Q_G$ satifies $M^*M = n\identity^{\otimes 0}$. Then we get
    \begin{equation*}
      n\Phi(\identity^{\otimes 0}) = \Phi(n\identity^{\otimes 0}) = \Phi(M^*M) = \Phi(M^*)\Phi(M) = M^*M = n\identity^{\otimes 0}.
    \end{equation*}
    Dividing by $n$ on both sides gives us the desired result. The linearity of $\Phi$ then further implies that its restriction to $C^Q_G(0,0)$ must be the identity map.
  \end{proof}
\end{lemma}

This allow us to prove implication 2 $\implies$ 1 of \cref{thm:mainresult}.

\begin{lemma}\label{lem:weakisotohomind}
  Let $Q$ be an easy quantum group, $\gF$ the set of its intertwiner graphs, and $G$ and $H$ be graphs that admit a weak isomorphism $\Phi$ between $Q(G)$ and $Q(H)$. Then $G \cong_\gF H$.
  \begin{proof}
    By \cref{lem:weakisoisidentity}, the weak isomorphism $\Phi\colon C^Q_G \to C^Q_H$ is the identity on $C^Q_G(0,0)$. Now let $\mathbf{F} = (F, (), ()) \in \gF(0, 0)$ be arbitrary. Then we have $\mathbf{F}_G \in C^Q_G(0,0)$ and thus
    \begin{equation*}
      \hom(F, G) = \soe(\mathbf{F}_G) = \soe(\Phi(\mathbf{F}_G)) = \soe(\mathbf{F}_H) = \hom(F, H). \qedhere
    \end{equation*}
  \end{proof}
\end{lemma}

  \section{Monoidal Equivalence and Intertwining Quantum Matrices}
  \label{sec:monoidequiv}
  So far we have seen that the characterisation of intertwiners as homomorphism tensors implies that homomorphism indistinguishability of graphs $G$ and $H$ over the unlabelled intertwiner graphs of an easy quantum group $Q$ means that the intertwiner spaces of the graph instantiations $Q(G)$ and $Q(H)$ ``look the same''. While this is already an interesting result, we would like to obtain a characterisation of $G$ and $H$ in terms of a \Cstar-algebra-valued linear system, analogous to quantum isomorphism. 
In this section we derive such a characterisation, proving the equivalences 2 $\iff$ 3 $\iff$ 4, as well as $2 \implies 5$ of \cref{thm:mainresult}.
 
\subsection{Quantum Matrices induce Weak Isomorphisms}

The first direction is again a straightforward generalisation of Man\v{c}inska and Roberson's proof that we include for the sake of completeness.

\begin{lemma}\label{lem:qmatrixtoweakiso}
  Let $Q$ be an easy quantum group with generators $M_1, \dots, M_m$ and let $G$ and $H$ be graphs. If there exists a quantum orthogonal matrix $\U$ with $\U A_G = A_H \U$ such that $\Utp{k} M_i = M_i \Utp{\ell}$ for all $i \in [m]$ and $M_i \in C^Q(k, \ell)$, then there exists a weak isomorphism $\Phi\colon C^Q_G \to C^Q_H$ between $Q(G)$ and $Q(H)$.
  \begin{proof}
    We show that the map $\Phi \colon C^Q_G \to C^Q_H$ given by $\Utp{\ell}T = \Phi(T)\Utp{k}$ for $T \in C^Q_G(\ell, k)$ for all $k, \ell \in \N$ is a weak isomorphism.
    To see that $\Phi$ is well-defined, note that $T\Utp{k} = T'\Utp{k}$ implies that $T = T'$, since $\Utp{k}$ is orthogonal and thus invertible.

    Now consider the generators $M_1, \dots, M_k$ of $Q$. Per assumption $\Utp{\ell}M_i = M_i\Utp{k}$ for all $i \in [k]$ with $M_i \in C^Q(\ell, k)$ and $\U A_G = A_H\U$, so we have $\Phi(M_i) = M_i$ and $\Phi(A_G) = A_H$. Since the $M_i$ and $A_G$ generate $C^Q_G$, to prove that $\Phi$ is defined everywhere on $C^Q_G$, it suffices to show that $\Phi$ is linear and commutes with tensor product, composition and hermitian adjoint.

    Suppose that $\Utp{\ell}T_i = T_i'\Utp{k}$ and $\lambda_i \in \C$ for $i \in [m]$. Then it is straightforward to see that
    \begin{equation*}
      \Utp{\ell}\left(\sum_{i=1}^m \lambda_i T_i\right) = 
      \left(\sum_{i=1}^m \lambda_i T'_i\right)\Utp{k}.
    \end{equation*}
    Now let $T_1, T_1' \in C^Q_G(\ell, k), T_2, T_2' \in C^Q_G(k, m)$ with $\Utp{\ell}T_1 = T_1'\Utp{k}$ and $\Utp{k}T_2 = T_2'\Utp{m}$. Then
    \begin{equation*}
      \Utp{\ell}(T_1T_2) = T_1'\Utp{k}T_2 = T_1'T_2'\Utp{m}.
    \end{equation*}
    Next let $T_1, T_1' \in C^Q_G(\ell, k), T_2, T_2' \in C^Q_G(r, s)$ with $\Utp{\ell}T_1 = T_1'\Utp{k}$ and $\Utp{r}T_2 = T_2'\Utp{s}$. Then 
    \begin{align*}
      \Utp{(\ell + r)} (T_1 \otimes T_2) &= (\Utp{\ell} \otimes \Utp{r})(T_1\otimes T_2) = \Utp{\ell}T_1 \otimes \Utp{r}T_2\\
                                         &= T_1'\Utp{k} \otimes T_2'\Utp{s} = (T_1' \otimes T_2')\Utp{(k + s)}
    \end{align*}
    Finally, let $T, T' \in C^Q_G(\ell, k)$ with $\Utp{\ell}T = T'\Utp{k}$. Then
    \begin{align*}
      \Utp{k}T^* &= \Utp{k}[T^*(\Utp{\ell})^*]\Utp{\ell} = \Utp{k}[\Utp{\ell}T]^*\Utp{\ell} = \Utp{k}[T'\Utp{k}]^*\Utp{\ell}\\ &= \Utp{k}(\Utp{k})^*T'^*\Utp{\ell} = T'^*\Utp{\ell}.
    \end{align*}

    We thus have shown that $\Phi$ is linear and commutes with the necessary operations. Since $M_1, \dots, M_k, A_H$ generate $C^Q_H$ and these intertwiners are in the image of $\Phi$, it follows that $\Phi$ is surjective. To see that $\Phi$ is injective, suppose that $\Phi(T) = \Phi(T')$. Then $\Utp{k}T = \Phi(T)\Utp{\ell} = \Phi(T')\Utp{\ell} = \Utp{k}T'$, thus the orthogonality of $\Utp{k}$ implies that $T = T'$.
    It follows that $\Phi$ is a weak isomorphism, which completes the proof.
  \end{proof}
\end{lemma}

\subsection{A brief excursion into Category Theory}
In order to prove the converse direction, we need to introduce a few more concepts from category theory. In the following let $Q$ be a compact matrix quantum group and $G$, $H$ be graphs. Our first goal is to show that a weak isomorphism between $Q(G)$ and $Q(H)$ is really just a \emph{monoidal equivalence} of $Q(G)$ and $Q(H)$ in disguise.

Recall the definition of the representation category $\Rep(Q)$ from \cref{ssec:easyqgroups}. It is known that $\Rep(Q)$ forms a so-called \emph{\Cstar-tensor category}. For a full definition of \Cstar-tensor categories see e.g. Definition 2.3.13 in \cite{maassen2021representation}\footnote{For simplicity and ease of notation, we tacitly assume that every \Cstar-tensor category is \emph{strict}, i.e.\ that the associator $\alpha$ as well as the left and right unitors $\lambda$ and $\rho$ are the identity. This does not change the validity of the results, since $\Rep(Q)$ is strict for all compact quantum groups $Q$.}---for our purposes it suffices to know that they are monoidal categories that allow to take linear combinations and adjoints of morphisms.

\begin{definition}
  Let $\mathcal{C}, \mathcal{C}'$ be \Cstar-tensor categories. A \emph{tensor functor} $\mathcal{C} \to \mathcal{C}'$ is a functor $F\colon \mathcal{C} \to \mathcal{C}'$ that is linear on morphisms, together with natural isomorphisms
  \begin{equation*}
     F_2\colon F(U) \otimes F(V) \to F(U \otimes V)
  \end{equation*}
  such that the diagram
  \begin{center}
    \begin{tikzcd}
      (F(U)\otimes F(V)) \otimes F(W) \arrow[r, "F_2 \otimes \id"] \arrow[d, "\id"]  & F(U \otimes V) \otimes F(W) \arrow[r, "F_2"] & F((U \otimes V) \otimes W) \arrow[d, "\id"] \\
      F(U) \otimes (F(V) \otimes F(W)) \arrow[r, "\id \otimes F_2"]                     & F(U) \otimes F(V \otimes W) \arrow[r, "F_2"] & F(U \otimes (V \otimes W))                       
    \end{tikzcd}
  \end{center}
commutes, and $F(\mathbf{1}) \cong \mathbf{1}'$. We call $F$ \emph{unitary} if additionally $F(g^*) = F(g)^*$ for all morphisms $g$ and $F_2$ is unitary.
\end{definition}

\begin{definition}
We call two compact quantum groups $Q_1$ and $Q_2$ \emph{(unitarily) monoidally equivalent} if there exists a (unitary) tensor functor $\Rep(Q_1) \to \Rep(Q_2)$ that is fully faithful and essentially surjective.
\end{definition}

Now suppose we have a weak isomorphism $\Phi\colon C^Q_G \to C^Q_H$. Recall that the sets $C^Q_G$ and $C^Q_H$ comprise the morphisms of the reduced representation categories $\Rep_0(Q(G))$ and $\Rep_0(Q(H))$ respectively. The objects of these categories are the tensor powers of the fundamental representations $\U_G$ of $Q(G)$ and $\U_H$ of $Q(H)$, so given $\Phi$ we can construct a functor $F\colon \Rep_0(Q(G)) \to \Rep_0(Q(H))$ with $F(\U_G^{\otimes k}) = \U_H^{\otimes k}$ and $F(T) = \Phi(T)$ for all  $T \in \Mor(\U_G^{\otimes k}, \U_G^{\otimes \ell})$ and $k, \ell \in \N$.

We will show that we can extend this functor from $\Rep_0$ to the category $\Rep$. The main observation is that for any compact matrix quantum group $Q$, the category $\Rep(Q)$ is the pseudoabelian completion of $\Rep_0(Q)$. Roughly speaking, in a pseudoabelian category every idempotent endomorphism decomposes into the direct sum of kernel and image. Now for compact matrix quantum groups, every finite-dimensional representation is a direct summand of a tensor power of the fundamental representation.
Moreover, the projectors onto the corresponding subspaces are exactly the idempotent endomorphisms. Consequently, moving to the pseudoabelian completion of the reduced representation category recovers the other finite-dimensional representations as the images of the projectors.

\begin{definition}
  The \emph{pseudoabelian completion} of a category $\mathcal{C}$ is defined as the  category given by
  \begin{quote}
    \begin{tabular}{ l l }
      Objects: & $(A, e)$, $A \in \operatorname{Obj}(\mathcal{C})$, $e \in \Mor(A, A)$ idempotent.\\
      Morphisms: & $\operatorname{Mor}((A, e), (B, f)) = f\Mor(A, B)e$.\\
      Composition: & induced by the composition in $\mathcal{C}$.
    \end{tabular}
  \end{quote}
\end{definition}

Intuitively, the object $(A, e)$ corresponds to the ``subspace'' of $A$ that $e$ projects onto.
Concretely, this means that for $Q = (C(Q), \U, n)$, we can assume $\Rep(Q)$ to be of the following form.
\begin{quote}
\begin{tabular}{ l l }
  Objects: & $(\U^{\otimes k}, T)$, $T\colon (\C^n)^{\otimes k} \to (\C^n)^{\otimes k}$ linear and idempotent,\\ &$\Utp{k} T = T \Utp{k}$, $k \in \N$. \\
  Morphisms: & $\Mor((\U^{\otimes k}, T), (\U^{\otimes \ell}, T')) = T'ST$,\\ &$S \in \Mor(\U^{\otimes k}, \U^{\otimes \ell}) = C(k, \ell)$.\\
  Composition: & natural composition.
\end{tabular}
\end{quote}

This allows us to prove the following lemma.

\begin{lemma}\label{lem:weakisotfunctor}
  Let $Q$ be a compact matrix quantum group with generators $M_i \in \Mor((\Utp{k_i}_G, \identity^{\otimes k_i}), (\Utp{\ell_i}, \identity^{\otimes \ell_i}))$, $m, k_i, \ell_i \in \N, i \in [m]$, and let $G$, $H$ be graphs that admit a weak isomorphism $\Phi$ between $Q(G)$ and $Q(H)$. Then there exists a unitary tensor functor $F\colon \Rep(Q(G)) \to \Rep(Q(H))$ that is fully faithful and essentially surjective, such that 
  \begin{enumerate}
    \item $F(A_G) = A_H$,
    \item $F(M_i) = M_i$ for all $i \in [m]$,
    \item $F(\Utp{k}_G, \identity^{\otimes k}) = (\Utp{k}_H, \identity^{\otimes k})$,
    \item $F_2$ is the identity.
  \end{enumerate}
  In particular, $Q(G)$ and $Q(H)$ are unitarily monoidally equivalent.
  \begin{proof}
    We construct a functor $F\colon \Rep(Q(G)) \to \Rep(Q(H))$ as follows.  
    \begin{quote}
      \begin{tabular}{ l l }
        On objects: & $(\U_G^{\otimes k}, T) \mapsto (\U_H^{\otimes k}, \Phi(T))$.\\
        On morphisms: & $T'ST \mapsto \Phi(T')\Phi(S)\Phi(T)$,\\ &for $T'ST \in \Mor((\U_G^{\otimes k}, T), (\U_G^{\otimes \ell}, T'))$.
      \end{tabular}
    \end{quote}
    To prove that this is well defined, we need to show that $T \in \Mor_{\Rep_0(Q(G))}(\U^{\otimes k}, \U^{\otimes k})$ is idempotent if and only if $\Phi(T)$ is idempotent. If $T$ is idempotent, we have $\Phi(T)\Phi(T) = \Phi(TT) = \Phi(T)$, so $\Phi(T)$ is idempotent.
    For the converse suppose that $\Phi(T)$ is idempotent. Then $\Phi(TT) = \Phi(T)\Phi(T) = \Phi(T)$ and hence $T = TT$ by injectivity of $\Phi$.

    To see that $F$ is a valid functor $\Rep(Q(G)) \to \Rep(Q(H))$, we need to show that $F(\id_U) = \id_{F(U)}$ for all objects $U$ and that $F(g \circ f) = F(g) \circ F(f)$ for all morphisms $G$, $H$. The identity of $U = (\U_G^{\otimes k}, T)$ is $T$, so we have $F(T) = \Phi(T) = \id_{F(U)}$. Moreover, for morphisms
      $T_2ST_1 \in \Mor((\Utp{k_1}, T_1), (\Utp{k_2}, T_2))$,
      $T_3S'T_2 \in \Mor((\Utp{k_2}, T_2), (\Utp{k_3}, T_3))$
     we have 
     \begin{align*}
       F(T_2ST_1 \circ T_3S'T_2) &= \Phi(T_2ST_1 \circ T_3S'T_2)\\ 
                                 &= \Phi(T_2ST_1) \circ \Phi(T_3S'T_2)\\
                                 &= \Phi(T_2)\Phi(S)\Phi(T_1) \circ \Phi(T_3)\Phi(S')\Phi(T_2)\\
                                 &= F(T_2ST_1) \circ F(T_3S'T_2)
     \end{align*}

    \begin{claim}
      $F$ is a unitary tensor functor.
      \begin{subproof}
        It is not hard to see that $F$ is linear on morphisms. This follows almost immediately from the linearity of $\Phi$ and the fact that it commutes with composition. Moreover, the identities at each object indeed form a natural isomorphism $F_2\colon F(U) \otimes F(V) \to F(U \otimes V)$: For objects $(\Utp{k}_G, T)$ and $(\Utp{\ell}_G, T')$, we have 
        \begin{align*}
          F(\Utp{k}_G, T) \otimes F(\Utp{\ell}_G, T') &= (\Utp{k}_H, \Phi(T)) \otimes (\Utp{\ell}_H, \Phi(T'))\\
                                                      &= (\Utp{k}_H \otimes \Utp{\ell}_H, \Phi(T) \otimes \Phi(T'))\\
                                                      &= (\Utp{(k + \ell)}_H, \Phi(T \otimes T'))\\
                                                      &= F(\Utp{(k + \ell)}_G, T \otimes T')\\
                                                      &= F(\Utp{k}_G \otimes \Utp{\ell}_G, T \otimes T').
        \end{align*}
        where the second equality uses the natural tensor structure on $\Rep(Q(H))$. For morphisms $
      T_2ST_1 \in \Mor((\Utp{k_1}, T_1), (\Utp{k_2}, T_2)),
      T_4S'T_3 \in \Mor((\Utp{k_3}, T_2),$ $(\Utp{k_4}, T_3))$, we have
      \begin{align*}
        F(T_2ST_1 \otimes T_4S'T_3) &= \Phi(T_2ST_1 \otimes T_4S'T_3) = \Phi(T_2ST_1)\otimes \Phi(T_4S'T_3)\\ &= F(T_2ST_1) \otimes F(T_4S'T_3).
      \end{align*}
      The unit $\mathbf{1}_G$ is the object $(\Utp{0}_G, \identity^{\otimes 0})$ which, by \cref{lem:weakisoisidentity}, gets mapped to $F(\Utp{0}_G, \identity^{\otimes 0}) = (\Utp{0}_H, \Phi(\identity^{\otimes 0})) = (\Utp{0}_H, \identity^{\otimes 0})$, which is the unit $\mathbf{1}_H$ in $\Rep(Q(H))$. In particular, $F(\mathbf{1}_G) \cong \mathbf{1}_H$.
      Finally, the unitarity of $F$ follows immediately from the fact that $\Phi$ commutes with the hermitian adjoint.
      \end{subproof}
    \end{claim}

    \begin{claim}
      $F$ is fully faithful and essentially surjective.
      \begin{subproof}
        We have already seen that $\Phi$ is bijective on the idempotent intertwiners. This implies that $F$ is essentially surjective, i.e.\ surjective on objects. Indeed, suppose $(\Utp{k}_H, T) \in \Rep(Q(H))$. Since $T \in C^Q_G(k, k)$ is idempotent, there must exist an idempotent $T' \in C^Q_H(k, k)$ such that $\Phi(T') = T$, and thus $F(\Utp{k}_G, T') = (\Utp{k}_H, \Phi(T')) = (\Utp{k}_H, T)$.

        That $F$ is fully faithful, i.e.\ bijective on morphisms, follows immediately from the fact that the restrictions of $\Phi$ to $C^Q_G(k, \ell)$ are bijective for all $k, \ell \in \N$.
      \end{subproof}
    \end{claim}

      We have thus shown that there exists a unitary tensor functor $F\colon \Rep(Q(G)) \to \Rep(Q(H))$ that is fully faithful and essentially surjective. Per definition, it follows that $Q(G)$ and $Q(H)$ are unitarily monoidally equivalent. Conditions 1, 2, 3 follow immediately from the construction of $F$.
    \end{proof}
\end{lemma}

In this sense, our weak isomorphisms serve as a combinatorial view on the category-theoretic structure of representation categories. 

\subsection{Weak Isomorphisms induce Quantum Matrices}
We have seen that a weak isomorphism between two graph instatiations $Q(G)$ and $Q(H)$ witnesses that $Q(G)$ and $Q(H)$ are monoidally equivalent. 
The central result that will allow us to prove the converse of \cref{lem:qmatrixtoweakiso} is that any two monoidally equivalent \Cstar-tensor categories admit a \emph{linking algebra} that is compatible with the morphisms in a specific sense. This linking algebra will provide the entries for our desired quantum matrix.
 We first need to clarify the dual structure of the representation categories. Recall the following definition of dual objects.

\begin{definition}
  Let $\mathcal{C}$ be a \Cstar-tensor category. An object $\bar{U}$ is called the \emph{dual} of an object $U \in \mathcal{C}$ if there exists morphisms $R\colon \mathbf{1} \to \bar{U} \otimes U$ and $\bar{R}\colon \mathbf{1} \to U \otimes \bar{U}$ such that 
  \begin{center}
    \begin{tikzcd}
      U \arrow[r, "\id \otimes R"] & U \otimes \bar{U} \otimes U \arrow[r, "\bar{R}^*\otimes \id"] & U
    \end{tikzcd}
    and
    \begin{tikzcd}
      \bar{U} \arrow[r, "\id \otimes \bar{R}"] & \bar{U} \otimes U \otimes \bar{U} \arrow[r, "R^*\otimes \id"] & \bar{U}
    \end{tikzcd}
  \end{center}
  are the identity morphisms. We call these conditions the \emph{conjugate equations}.
\end{definition}

\begin{definition}
  A \emph{rigid \Cstar-tensor category} is a \Cstar-tensor category where every element has a dual.
\end{definition}

An important result is that for every compact quantum group $Q$, its representation category $\Rep(Q)$ is a rigid \Cstar-tensor category---see for instance \cite[Example 2.2.3]{neshveyev_compact_2013}.

Finally, we need to introduce the concept of a \emph{fiber functor}. Essentially, this is a special type of forgetful functor, mapping representations to their underlying spaces. We give a definition for the case of rigid \Cstar-tensor categories.

\begin{definition}
  Let $\mathcal{C}$ be a rigid \Cstar-tensor category. A \emph{fiber functor} is a tensor functor $F \colon \mathcal{C} \to \operatorname{fdHilb}$, where $\operatorname{fdHilb}$ denotes the category of finite-dimensional Hilbert spaces.
\end{definition}

The representation category $\Rep(Q)$ of any compact quantum group $Q$ admits a canonical fiber functor $F_Q \colon \Rep(Q) \to \operatorname{fdHilb}$ defined as follows.
    \begin{quote}
      \begin{tabular}{ l l }
        On objects: & $\V \mapsto \C^m$ for all representations $\V \in C(Q)^{m \times m}$.\\
        On morphisms: & $T \mapsto T$  for all intertwiners $T$ of $Q$.
      \end{tabular}
    \end{quote}

    Now the following is straightforward to verify.

\begin{lemma}
  Let $\mathcal{C}, \mathcal{C}', \mathcal{C}''$ be \Cstar-tensor categories and let $E\colon \mathcal{C} \to \mathcal{C}'$ and $F\colon \mathcal{C}' \to \mathcal{C}''$ be (unitary) tensor functors. Then $F \circ E\colon \mathcal{C} \to \mathcal{C}''$ is also a (unitary) tensor functor.
\end{lemma}

\begin{corollary}\label{cor:twofiberfunctors}
  Let $Q$ be an easy quantum group and $G$, $H$ be graphs. Suppose $Q(G)$ and $Q(H)$ are monoidally equivalent, witnessed by a tensor functor $F \colon \Rep(Q(G)) \to \Rep(Q(H))$. Then $F_{Q(H)} \circ F$ is a fiber functor $\Rep(Q(G)) \to \operatorname{fdHilb}$.
\end{corollary}

Consequently, the monoidal equivalence of $Q(G)$ and $Q(H)$ yields another fiber functor $\Rep(Q(G)) \to \operatorname{fdHilb}$ in addition to the canonical fiber functor $F_{Q(G)}$. This allows us to use the following theorem by Neshveyev and Tuset, which is stated in terms of fiber functors.

\begin{theorem}[Theorem 2.3.11 in \cite{neshveyev_compact_2013}]\label{thm:neshveyevlinking}
  Let $\mathcal{C}$ be a \Cstar-tensor category with duals and $E, F \colon \mathcal{C} \to \operatorname{fdHilb}$ be two fiber functors. Then there exists a \Cstar-algebra $\mathcal{Z}$, and for every object $U \in \mathcal{C}$ a unitary element $X^U \in \mathcal{Z} \otimes B(E(U), F(U))$ such that 
  \begin{enumerate}
    \item If $T \in \Mor(U, V)$, then $(\mathbf{1} \otimes F(T)) X^U = X^V (\mathbf{1} \otimes E(T))$.
    \item $(\mathbf{1} \otimes F_2)X^U_{12}X^V_{13} = X^{U \otimes V}(\mathbf{1} \otimes E_2)$.
    \item If $(R, \bar{R})$ is a solution of the conjugate equations for $U$ and $\bar{U}$, then 
      \begin{equation*}
        (X^U_{12})^*(\mathbf{1} \otimes F_2^*F(\bar{R})) = X^{\bar{U}}_{13}(\mathbf{1} \otimes E_2^*E(\bar{R})).
      \end{equation*}
  \end{enumerate}
\end{theorem}

Here $B(E(U), F(U))$ denotes the set of bounded linear operators from $E(U)$ to $F(U)$. The expression $X_{12}$ uses the so-called leg numbering notation, which is used to ``pad'' operators such that they only act on certain subspaces: If $X = \sum_i z_i \otimes T_i \in \mathcal{Z} \otimes B(E(U), F(U))$, then $X_{12} = \sum_i z_i \otimes T_i \otimes \identity$ and $X_{13} = \sum_i z_i \otimes \identity \otimes T_i$.

Now if there exists a weak isomorphism between $Q(G)$ and $Q(H)$, \cref{lem:weakisotfunctor} guarantees the existence of a tensor functor $\Rep(Q(G)) \to \Rep(Q(H))$, which in turn, as we have already noted, yields two fiber functors $\Rep(Q(G)) \to \operatorname{fdHilb}$. 
\cref{thm:neshveyevlinking} thus already looks quite similar to what we want to prove. To bring it into the form that we need, we first note that we already know the solutions to the conjugate equations for $(\U, I) \in \Rep(Q(G))$.


\begin{lemma}\label{lem:solvingconjeqs}
  Let $(C(Q), \U, n)$ be a compact matrix quantum group. Then $(\U, \identity)$ is self-dual in $\Rep(Q)$ and $R = \bar{R} = M^{2, 0}$ solves the conjugate equations.
  \begin{proof}
    First note that $M^{2, 0}$ is a $(2, 0)$-intertwiner, thus in particular a morphism $\mathbf{1} \to (\U, \identity) \otimes (\U, \identity)$ in $\Rep(Q)$. The identity at $(\U, \identity)$ is $\identity$, so we have to show that $((M^{2, 0})^* \otimes \identity)(\identity \otimes M^{2, 0}) = \identity$. Indeed we have
    \begin{align*}
      ((M^{2, 0})^* \otimes \identity)(\identity \otimes M^{2, 0}) &= \big(\sum_i e_i\transpose \otimes e_i\transpose \otimes \identity\big)\big(\sum_i \identity \otimes e_i \otimes e_i\big)\\
                                                                   &= \sum_{ij} e_i\transpose \otimes e_i\transpose e_j \otimes e_j\\
                                                                  &= \sum_{ij} e_i\transpose \otimes \delta_{ij} \otimes e_j\\
                                                                  &= \sum_i e_i\transpose \otimes e_i 
                                                                = \identity. \qedhere
    \end{align*}
  \end{proof}
\end{lemma}

This allows us to prove the following theorem. For simplicity, we will identify $(\Utp{k}_G, \identity^{\otimes k}) \in \Rep(Q(G))$ with $\Utp{k}_G$. We can do this without loss of generality, since the subcategory of objects $(\Utp{k}_G, \identity^{\otimes k})$ is just (equal to) the reduced representation category $\Rep_0(Q(G))$.

\begin{theorem}\label{thm:simplifiedlinking}
  Let $(C(Q), \U, n)$ be an easy quantum group and $G$ and $H$ be graphs of size $n$ that admit a weak isomorphism $\Phi$ between $Q(G)$ and $Q(H)$. Then there exists a \Cstar-algebra $\mathcal{Z}$ and for all $\Utp{k}_G \in \Rep(Q(G))$ a unitary matrix $Z^{(k)} \in \mathcal{Z}^{n^k \times n^k}$, such that for all $k, \ell \in \N$
  \begin{enumerate}
    \item If $T \in \Mor(\Utp{k}_G, \Utp{\ell}_G)$, then $Z^{(\ell)}T = \Phi(T)Z^{(k)}$.
    \item $Z^{(k)} \otimes Z^{(\ell)} = Z^{(k + \ell)}$.
  \end{enumerate}
  Moreover, the entries $z_{ij}$ of $Z^{(1)}$ satisfy $z_{ij}^* = z_{ij}$.
  \begin{proof}
    \newcommand{\FGH}{F^{H, G}}
    We start with some observations. The existence of a weak isomorphism $\Phi$ between $Q(G)$ and $Q(H)$ implies by \cref{lem:weakisotfunctor} that there exists a unitary tensor functor $\FGH \colon \Rep(Q(G)) \to \Rep(Q(H))$ such that 
    \begin{enumerate}[(a)]
      \item $\FGH(A_G) = A_H$,
      \item $\FGH(M_i) = M_i = \Phi(M_i)$ for all generators $M_i$ of $Q$,
      \item $\FGH_2$ is the identity,
      \item $\FGH(\Utp{k}_G, \identity) = (\Utp{k}_H, \identity)$.
    \end{enumerate}
    By \cref{cor:twofiberfunctors}, there exist two fiber functors $E, F \colon \Rep(Q(G)) \to \operatorname{fdHilb}$ given by $E = F_{Q(G)}$ and $F = F_{Q(H)} \circ \FGH$. By property (d), we have $E(\Utp{k}_G, \identity) = F(\Utp{k}_G, \identity) = (\C^n)^{\otimes k} \cong \C^{n^k}$. This means that we can identify $\mathcal{Z} \otimes B(E(\Utp{k}_G), F(\Utp{k}_G)) = \mathcal{Z} \otimes B(\C^{n^k}, \C^{n^k})$ with the $n^k \times n^k$ matrices over $\mathcal{Z}$. 
    We use the symbol $\triangleq$ to denote this identification.
    In particular, we have for $Z \in \mathcal{Z} \otimes B(\C^{n^k}, \C^{n^k})$,
    \begin{align*}
        Z_{12} &= \sum_{ij} z_{ij} \otimes e_{ij} \otimes \identity = \left(\sum_{ij} z_{ij} \otimes e_{ij} \right) \otimes \identity \triangleq Z \otimes \identity, \numberthis\label{eqs:legtotensora}\\
        Z_{13} &= \sum_{ij} z_{ij} \otimes \identity \otimes e_{ij} \triangleq \sum_{ij} z_{ij}\cdot (\delta_{i_2i}\delta_{j_2j}\delta_{i_1j_2})_{i_1i_2, j_1j_2} = \identity \otimes Z.\numberthis\label{eqs:legtotensorb}
    \end{align*}

    Since the canonical fiber functors $F_{Q(G)}, F_{Q(H)}$ are the identity on morphisms, we get that for all $T \in \Mor(\Utp{k}_G, \Utp{\ell}_G)$ we have $E(T) = T$,  and for all generators $M_i$ of $Q$ we have $F(M_i) = \Phi(M_i)$. We also get $F(A_G) = \Phi(A_G) = A_H$. An argument similar to \cref{lem:constructionmirror} then shows that $F(T) = \Phi(T)$ for all $T \in \Mor(\Utp{k}_G, \Utp{\ell}_G)$.

    We can now apply \cref{thm:neshveyevlinking} to the functors $E$ and $F$. We let $Z^{(k)} \coloneqq X^{\Utp{k}_G}$.
    Then property 1 of \cref{thm:neshveyevlinking} simplifies under the above observations and the identification of $\mathcal{Z} \otimes B(E(\Utp{k}_G), F(\Utp{k}_G))$ with $\mathcal{Z}^{n^k \times n^k}$ to
    \begin{equation*}
      \Phi(T)Z^{(k)} = Z^{(\ell)}T,
    \end{equation*}
    as desired. By property (c) of $\FGH$, we get that $E_2$ and $F_2$ are the identity. Thus, using \cref{eqs:legtotensora,eqs:legtotensorb}, property 2 simplifies to
    \begin{equation*}
      Z^{(k)} \otimes Z^{(\ell)} =(Z^{(k)} \otimes \identity)(\identity \otimes Z^{(\ell)}) \triangleq Z^{(k)}_{12}Z^{(\ell)}_{13} \triangleq Z^{\Utp{k}_G \otimes \Utp{\ell}_G} = Z^{\Utp{(k + \ell)}_G} = Z^{(k + \ell)}.
    \end{equation*}

    It remains to show that the entries $z_{ij}$ of $Z^{(1)}$ satisfy $z_{ij}^* = z_{ij}$. By \cref{lem:solvingconjeqs}, $R = \bar{R} = M^{2,0}$ solves the conjugate equations for $\U_G \in \Rep(Q(G))$. Thus by property 3 of \cref{thm:neshveyevlinking}, we get
    \begin{equation*}
      (Z^{(1)} \otimes \identity)^* M^{2, 0} = (\identity \otimes Z^{(1)})M^{2, 0},
    \end{equation*}
    and we have
    \begin{align*} 
      (Z^{(1)} \otimes \identity)^* M^{2, 0} &= \left(\sum_{ij} z_{ij}^* e_{ji} \otimes \identity \right) M^{2, 0} = \sum_{ijk} z_{ij}^* (e_{ji}e_k \otimes e_k)\\ 
                                             &= \sum_{ijk} z_{ij}^* (\delta_{ik}e_j \otimes e_k) = \sum_{ij} z_{ij}^* (e_j \otimes e_i)
    \end{align*}
    and
    \begin{align*}
      (\identity \otimes Z^{(1)})M^{2, 0} &= \left(\sum_{ij} \identity \otimes z_{ij} e_{ij} \right)M^{2, 0} = \sum_{ijk} z_{ij} (e_k \otimes e_{ij}e_k)\\ 
                                          &= \sum_{ijk} z_{ij} (e_k \otimes \delta_{jk} e_i) = \sum_{ij} z_{ij} (e_j \otimes e_i).
    \end{align*}
    Hence $\sum_{ij} z_{ij} (e_j \otimes e_i) = \sum_{ij} z_{ij}^* (e_j \otimes e_i)$, which implies $z_{ij}^* = z_{ij}$ for all $i, j \in [n]$.
  \end{proof}
\end{theorem}

Finally, we are able to prove the converse of \cref{lem:qmatrixtoweakiso}.

\begin{corollary}\label{cor:weakisotoqmatrix}
  Let $Q$ be an easy quantum group and let $G$ and $H$ be graphs. If there exists a weak isomorphism $\Phi$ between $Q(G)$ and $Q(H)$, then there exists a quantum orthogonal matrix $\U$ with $\U A_G = A_H \U$, such that $\Utp{k} M = M \Utp{\ell}$ for all $M \in C^Q(k, \ell)$.
  \begin{proof}
    By \cref{thm:simplifiedlinking}, there exist a \Cstar-algebra $\mathcal{Z}$ and matrices $Z^{(k)} \in \mathcal{Z}^{n^k \times n^k}$ for all $k \in \N$, such that 
    \begin{enumerate}[(a)]
      \item $Z^{(\ell)}T = \Phi(T)Z^{(k)}$ for all $T \in C^Q_G(\ell, k)$,
      \item $Z^{(k)} \otimes Z^{(\ell)} = Z^{(k + \ell)}$,
      \item the entries $z_{ij}$ of $Z^{(1)}$ satisfy $z_{ij}^* = z_{ij}$.
    \end{enumerate}
    Set $\U = Z^{(1)}$. Since $Z^{(1)}$ is unitary, (c) implies that it is orthogonal.
    \begin{claim}
      For all $k \in \N$, $\Utp{k} = Z^{(k)}$.
      \begin{subproof}
        First suppose $k = 0$. We know that $Z^{(0)}$ is a $1 \times 1$ matrix and thus of the form $(\lambda)$ where $\lambda \in \mathcal{Z}$. Moreover, by (b) we have $\lambda Z^{(1)} = Z^{(1)} \otimes Z^{(0)} = Z^{(1)}$. Since $Z^{(1)}$ is unitary, multiplying both sides by $(Z^{(1)})^*$ yields $\lambda \identity = \identity$, which implies $\lambda = \mathbf{1}$. Consequently, $Z^{0} = (\mathbf{1}) = \Utp{0}$. 
        By induction we then have
        \begin{equation*}
          Z^{(k+1)} = Z^{(k)} \otimes Z^{(1)} = \Utp{k} \otimes \U = \Utp{(k + 1)}. \qedhere
        \end{equation*} 
      \end{subproof}
    \end{claim}
    Finally suppose $M \in C^Q(\ell, k) \subseteq C^Q_G(\ell, k)$. Then by (a) we have 
    \begin{equation*}
      \Utp{\ell} M = Z^{(\ell)} M = \Phi(M) Z^{(k)} = M Z^{(k)} = M\Utp{k}.
    \end{equation*}
    Since $A_G \in C^Q_G(1, 1)$, we analogously get $\U A_G = A_H \U$ as desired. 
  \end{proof}
\end{corollary}

  \section{Cycles and More: Constructing Intertwiner Graphs}
  \label{sec:constructing}
  In the previous two sections we have established the following homomorphism indistinguishability characterisation.

\mainresult*
\begin{proof}
  We proved 1 $\implies$ 2 in \cref{lem:homindtoweakiso}, 2 $\implies$ 1 in \cref{lem:weakisotohomind}, and 2 $\implies$ 3 is in \cref{cor:weakisotoqmatrix}. The implication 3 $\implies$ 4 is trivial, and 4 $\implies$ 2 is proved in \cref{lem:qmatrixtoweakiso}. Finally, 2 $\implies$ 5 is proved in \cref{lem:weakisotfunctor}.
\end{proof}

The natural question to ask is what the intertwiner graph classes of the orthogonal quantum groups actually look like. Man\v{c}inska and Roberson showed that for the quantum symmetric group one obtains the class of planar bilabelled graphs, whose $(0,0)$-fragment are simply the planar graphs. In this section, we answer this question for the remaining orthogonal easy quantum groups. While Man\v{c}inska and Roberson identified the entire class of intertwiner graphs, we will here only focus on the subclass $\gF(0, 0)$ of unlabelled graphs in $\gF$. This simplifies our presentation somewhat, and is sufficient to state the resulting homomorphism indistinguishability characterisations. We find that the resulting graph classes are either all graphs, all planar graphs, all cycles, or all paths and cycles. To structure this section, we follow the classification of \cref{thm:classifyingpartitions}.

\begin{figure}[t]
  \centering
  \begin{tikzpicture}
    \foreach \x in {1, 2} {
      \pgfmathsetmacro{\sx}{pow(-1, \x)}
      \begin{scope}[yscale=\sx] 
      \node[vertex] (v1) at (0, 1.875) {};
      \coordinate (v1f) at (1, 1.875);
      \coordinate (v1b) at (-1, 1.875);
      \draw[wire] 
        (v1) -- (v1f)
        (v1) -- (v1b)
        ;

      \node[vertex] (v2) at (0, 1.125) {};
      \coordinate (v2f) at (1, 1.125);
      \coordinate (v2b) at (-1, 1.125);
      \draw[wire] 
        (v2) -- (v2f)
        (v2) -- (v2b)
        ;

      \node[vertex] (v31) at (.5, .375) {};
      \node[vertex] (v32) at (-.5, .375) {};
      \coordinate (v31f) at (1, .375);
      \coordinate (v32b) at (-1, .375);
      \draw[wire] 
        (v31) -- (v31f)
        (v32) -- (v32b)
        ;
      \draw[edge] (v31) -- (v32);
      \end{scope}
    }

    \foreach \x in {1, 2} {
      \pgfmathsetmacro{\sx}{pow(-1, \x)}
      \begin{scope}[xscale=\sx, yscale=\sx]
      \node[vertex] (v7) at (2.25, .75) {};
      \coordinate (v7b) at (2, .75);
      \coordinate (v7bu) at (2, 1.875);
      \coordinate (v7buh) at (2, 1.125);
      \coordinate (v7buhb) at (1.5, 1.125);
      \coordinate (v7bub) at (1.5, 1.875);
      \coordinate (v7bl) at (2, -.375);
      \coordinate (v7blh) at (2, .375);
      \coordinate (v7blhb) at (1.5, .375);
      \coordinate (v7blb) at (1.5, -.375);
      \coordinate (v7f) at (2.5, .75);
      \coordinate (v7fu) at (2.5, 1.125);
      \coordinate (v7fl) at (2.5, .375);
      \coordinate (v7fuf) at (3, 1.125);
      \coordinate (v7flf) at (3, .375);

      \draw[wire]
        (v7) -- (v7b)
        (v7b) -- (v7bu)
        (v7b) -- (v7bl)
        (v7bu) -- (v7bub)
        (v7bl) -- (v7blb)
        (v7buh) -- (v7buhb)
        (v7blh) -- (v7blhb)
        (v7) -- (v7f)
        (v7f) -- (v7fu)
        (v7f) -- (v7fl)
        (v7fu) -- (v7fuf)
        (v7fl) -- (v7flf)
        ;

      \node[vertex] (v8) at (2.25, -1.125) {};
      \coordinate (v8b) at (1.5, -1.125);
      \coordinate (v8f) at (3, -1.125);

      \draw[wire]
        (v8) -- (v8b)
        (v8) -- (v8f)
        ;

      \node[vertex] (v9) at (2.25, -1.875) {};
      \coordinate (v9b) at (1.5, -1.875);
      \coordinate (v9f) at (3, -1.875);

      \draw[wire]
        (v9) -- (v9b)
        (v9) -- (v9f)
        ;
      \end{scope}
    }

    \node at (4.15, 0) {$=$};

    \node at (6, 0) {
        \begin{tikzpicture}
          \foreach \y in {1, 2} {
            \pgfmathsetmacro{\sy}{pow(-1, \y)}
            \node[vertex] (w\y) at (0, \sy) {};
            \foreach \x in {1, 2} {
              \pgfmathsetmacro{\sx}{pow(-1, \x)}
              \begin{scope}[xscale=\sx, yscale=\sy]
                \coordinate (wf) at (.25, 1);
                \coordinate (wfu) at (.25, 1.375);
                \coordinate (wfl) at (.25, .625);
                \coordinate (wfuf) at (.75, 1.375);
                \coordinate (wflf) at (.75, .625);

                \draw[wire] 
                  (w\y) -- (wf)
                  (wf) -- (wfu)
                  (wf) -- (wfl)
                  (wfu) -- (wfuf)
                  (wfl) -- (wflf)
                  ;
              \end{scope}
            }
          }

          \draw[edge] (w1) -- (w2);
        \end{tikzpicture}
      };

  \end{tikzpicture}
  \caption{Constructing the edge gadget $\mathbf{\hat{A}}$.}
  \label{fig:edgegadget}
\end{figure}
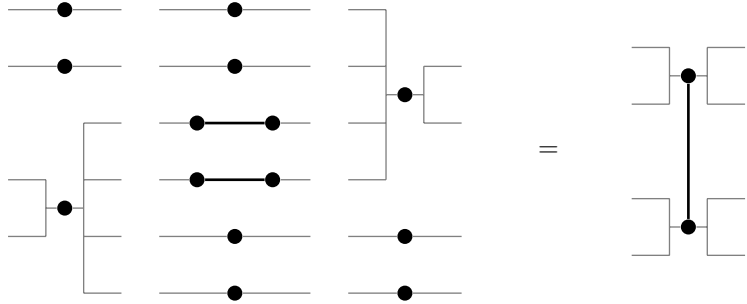

In the following we understand all mentioned classes to be closed under disjoint unions. This is because for any two constructed graphs $\mathbf{K}, \mathbf{K}'$, their disjoint union is $\mathbf{K} \otimes \mathbf{K}'$, and thus always constructible. In the context of homomorphism indistinguishability this makes no difference, since $\hom(F_1 \cup F_2, G) = \hom(F_1, G) \cdot \hom(F_2, G)$ for all graphs $F_1, F_2, G$.
For brevity, we refrain from drawing all the involved intermediate and gadget graphs, and only highlight a few central ones in \cref{fig:edgegadget,fig:fatswap}. However, most of the constructions should be clear from a quick sketch on a piece of paper.

\subsection{Classical, Liberated, and Half-Liberated Quantum Groups}

Let us begin by investigating the classical, liberated, and the half-liberated quantum groups---the latter being the easy quantum groups $G_n(\mathcal{C})$ with $\spg \in \mathcal{C}$ and $\spa \not\in \mathcal{C}$. 

The first observation is that the bilabelled graphs $\bI$, $\bM^{2, 0}$, and $\mathbf{A}$ already allow us to construct arbitrary cycles. Since every easy quantum group has $\identity$ and $M^{2, 0}$ as intertwiners, their corresponding intertwiner graphs always include all cycles. In particular, the class of unlabelled intertwiner graphs of the quantum orthogonal group---which is generated precisely by $\identity$ and $M^{2, 0}$---is exactly the class of all cycles. Moreover, the swap graph $\bS$ does not allow us to construct any additional graphs, so the same holds for the classical orthogonal group, generated by $\identity, M^{2,0}, S$.

\begin{lemma}\label{lem:constructingcycles}
  Let $\gF$ be the class of intertwiner graphs of the orthogonal group and $\gF^+$ be the class of intertwiner graphs of the quantum orthogonal group. Then $\gF(0, 0) = \gF^+(0, 0) = \mathcal{C}$, the class of all cycles.
  \begin{proof}
    We have $\gF = \intertwclosure{\mathbf{I}, \bM^{2, 0}, \mathbf{S}, \mathbf{A}}$ and $\gF^+ =\intertwclosure{\mathbf{I}, \mathbf{M}^{2, 0}, \mathbf{A}}$. We first show that we can construct arbitrary cycles. We have $C_1 = \mathbf{M}^{0, 2}\mathbf{M}^{2, 0}$ and for $n \geq 2$,
    \begin{equation*}
      C_n = \bM^{0, 2}(\mathbf{A} \otimes \mathbf{A}^{n-1})\mathbf{M}^{2, 0}.
    \end{equation*}

    To see that $\gF(0, 0) \supseteq \gF^+(0, 0)$ only contains cycles, we show that every vertex has degree $2$, or is part of either $C_1$ or $C_2$. For finite graphs, this is equivalent to being a disjoint union of cycles.
    Let $\mathbf{K} \in \gF$ and denote for $\mathbf{K}^v$ the connected component containing $v \in V(\mathbf{K})$. By induction, we show the slightly stronger statement that all vertices $v \in V(\mathbf{K})$ satisfy either $\deg(v) + \abs{l(v)} = 2$, $\mathbf{K}^v = C_1$, or $\mathbf{K}^v = C_2$. 

    Obviously this holds for the generators $\mathbf{I}, \bM^{2, 0}, \mathbf{S}, \mathbf{A}$, so assume it holds for $\mathbf{K}_1, \mathbf{K}_2 \in \gF$. The inductive step is trivial for $\mathbf{K} = \mathbf{K}_1 \otimes \mathbf{K}_2$ and $\mathbf{K} = \mathbf{K}_1^*$, so assume $\mathbf{K} = \mathbf{K}_1 \mathbf{K}_2$. The composition leaves unlabelled vertices unchanged, and by assumption identifies at most $3$ vertices at a time. Let $v, v_1, v_2 \in \mathbf{K}_1$ and $w \in \mathbf{K}_2$. We make the following case distinction.

    $\abs{l_{out}(v)} = \abs{l_{in}(w)} = 2$: In this case $v$ and $w$ must be isolated vertices, and thus their identification $vw$ is an isolated vertex without labels. Consequently, $\mathbf{K}^{vw} = C_1$.

    $\abs{l_{out}(v)} = \abs{l_{in}(w)} = 1$: In this case $v$ and $w$ have exactly one adjacent vertex in $\mathbf{K}_1$ and $\mathbf{K}_2$ respectively, so $vw$ is a labelless vertex of degree $2$.

    $\abs{l_{out}(v_1)} = \abs{l_{out}(v_2)} = 1$, $\abs{l_{in}(w)} = 2$: In this case $w$ is an isolated vertex, and $v_1, v_2$ both have exactly one adjacent vertex in $\mathbf{K}_1$. Denote the adjacent vertices by $v_1'$ and $v_2'$ respectively. If $v_1' \neq v_2'$, then $v_1v_2w$ is a labelless vertex of degree 2 as desired. If $v_1' = v_2' \coloneqq v'$, then $\deg(v') = 2$. By assumption this implies that $v'$ is labelless and has no other neighbours. Consequently, $\mathbf{K}^{v'} = \mathbf{K}^{v_1v_2w} = C_2$.

    The remaining case is symmetric.
  \end{proof}
\end{lemma}

Obviously, this also holds for quantum groups $Q$ with $O_n \subseteq Q \subseteq O^+_n$.

\begin{corollary}
  Let $\gF$ be the class of intertwiner graphs of $G_n(E_o)$. Then $\gF(0, 0) = \mathcal{C}$, the class of all cycles.
  \begin{proof}
    We have $E_o = \lrangle{\spg}$. Let $\mathbf{\bar{S}}$ denote the bilabelled graph corresponding to $\spg$. Then $\gF = \intertwclosure{\bI, \bM^{2, 0}, \mathbf{\bar{S}}}$, and we have
    \begin{equation*}
      \mathbf{\bar{S}} = (\mathbf{S} \otimes \bI) \circ (\bI \otimes \mathbf{S}) \circ (\mathbf{S} \otimes \bI).
    \end{equation*}
    In particular, we get
    $\mathcal{C} = \intertwclosure{\bI, \bM^{2,0}, \mathbf{A}} \subseteq \intertwclosure{\bI, \bM^{2,0}, \mathbf{\bar{S}}, \mathbf{A}} \subseteq \intertwclosure{\bI, \bM^{2,0}, \mathbf{S}, \mathbf{A}} = \mathcal{C}$. 
  \end{proof}
\end{corollary}

Along the same lines as \cref{lem:constructingcycles}, we can prove the following result for the bistochastic and quantum bistochastic groups. 

\begin{lemma}\label{lem:constructingpathsandcycles}
  Let $\gF$ be the class of intertwiner graphs of the bistochastic group and $\gF^+$ be the class of intertwiner graphs of the quantum bistochastic group. Then $\gF(0, 0) = \gF^+(0,0)$ is the class of all paths and cycles. 
  \begin{proof}
    We have $\gF = \intertwclosure{\bI, \bM^{2, 0},\bM^{1, 0}, \mathbf{S}, \mathbf{A}}$ and $\gF^+ = \intertwclosure{\bI, \bM^{2, 0}, \bM^{1, 0}, \mathbf{A}}$. We proceed similarly to the previous lemma. We already know how to construct arbitary cycles. $P_1$ is again equal to $\bM^{0, 2}\mathbf{M}^{2, 0}$, and for $n \geq 2$, we have
    \begin{equation*}
      P_n = \bM^{0, 1} \circ \mathbf{A}^{n-1} \circ \bM^{1, 0}.
    \end{equation*}
    To see that $\gF(0, 0) \supseteq \gF^+(0, 0)$ only contains paths and cycles, we show that every vertex has degree at most $2$. For finite graphs, this is equivalent to being a disjoint union of paths and cycles. We show the slightly stronger statement that for all $\mathbf{K} \in \gF$ all vertices $v \in V(K)$ satisfy $\deg(v) + \abs{l(v)} \leq 2$.

    Obviously, this holds for $\bI, \bM^{2, 0}, \bM^{1, 0}, \mathbf{S}$, so assume it holds for $\mathbf{K}_1, \mathbf{K}_2 \in \gF$. The inductive step is again trivial for $\mathbf{K} = \mathbf{K}_1 \otimes \mathbf{K}_2$ and $\mathbf{K} = \mathbf{K}_1^*$, so assume $\mathbf{K} = \mathbf{K}_1\mathbf{K}_2$. The composition leaves unlabelled vertices unchanged and by assumption identifies at most $3$ vertices at a time. Let $v, v_1, v_2 \in \mathbf{K}_1$ and $w \in \mathbf{K}_2$. We make the following case distinction.

    $\abs{l_{out}(v)} = \abs{l_{in}(w)} = 2$: In this case $v$ and $w$ must be isolated vertices, and thus their identification $vw$ is an isolated vertex without labels. In particular, $\deg(vw) + \abs{l(vw)} = 0$.

    $\abs{l_{out}(v)} = \abs{l_{in}(w)} = 1$: In this case $v$ and $w$ have at most one adjacent vertex in $V(K_1)$ and $V(K_2)$ respectively. It follows that $vw$ is an unlabelled vertex with degree $\deg(v) + \deg(w) \leq 2$.

    $\abs{l_{out}(v_1)} = \abs{l_{out}(v_2)} = 1$, $\abs{l_{in}(w)} = 2$: In this case $w$ is an isolated vertex, and $v_1, v_2$ both have at most one adjacent vertex in $\mathbf{K}_1$. Thus $v_1v_2w$ is an unlabelled vertex with $\deg(v_1v_2w) \leq \deg(v_1) + \deg(v_2) \leq 2$.

    The remaining case is symmetric.
  \end{proof}
\end{lemma}

Since paths are fairly easy to construct, we obtain the following result for supergroups of the (quantum) bistochastic group.

\begin{lemma}
  Let $Q$ be an easy quantum group and $\gF$ the class of its intertwiner graphs. If $Q$ is either 
  \begin{enumerate}
    \item the bistochastic group $B_n$,
    \item the quantum bistochastic group $B_n^+$,
    \item the modified bistochastic group $B_n'$,
    \item the quantum modified bistochastic group $B_n^{\#+}$,
    \item the quantum complexly modified bistochastic group $B_n'^+$, or
    \item $G_n(E'_b)$,
  \end{enumerate}
  then $\gF(0, 0)$ is the class of all paths and cycles.
  \begin{proof}
    The cases 1 and 2 are proved in \cref{lem:constructingpathsandcycles}. For the remaining groups, their definining partition categories are either subsets of those defining the bistochastic group or those defining the quantum bistochastic group (depending on whether they contain crossing partitions). This means that their intertwiner graphs must be disjoint unions of paths or cycles. Moreover, the partition categories are obviously supersets of those defining the orthogonal group, so we can construct arbitrary cycles. Consequently, it suffices to show that we can also construct arbitrary paths.

    For $B_n'$, $B_n^{\#+}$, and $G_n(E'_b)$ we have $\bM^{1, 0} \otimes \bM^{1, 0} \in \gF$. This allows us to construct $P_n$ as
\begin{equation*}
  P_n = (\bM^{0, 1} \otimes \bM^{0, 1}) \circ (\mathbf{A}^n \otimes \bI) \circ \bM^{2, 0}.
\end{equation*}
For $B_n'^+$, we instead have access to $\bM^{1, 0} \otimes \bI \otimes \bM^{0, 1}$, which allows us to construct $P_n$ as
\begin{equation*}
  \bM^{0, 2} \circ (\bM^{1, 0} \otimes \bI \otimes \bM^{0, 1}) \circ (\mathbf{A}^n \otimes \bI) \circ \bM^{2, 0}. \qedhere
\end{equation*}
  \end{proof}
\end{lemma}

The remaining easy quantum groups enjoy much larger classes of intertwiner graphs. This is in part due to a general construction that allows us to compose any unlabelled graph when we have access to $\bM^{2,2}$ and some ``swapping'' partition. 

\begin{lemma}\label{lem:constructingallgraphs}
  Let $\mathcal{G}$ denote the class of all (unlabelled) graphs. Then $\mathcal{G} \subseteq \intertwclosure{\bI, \bM^{2, 0}, \bM^{2, 2}, \mathbf{S}}$ and $\mathcal{G} \subseteq \intertwclosure{\bI, \bM^{2, 0}, \bM^{2, 2}, \mathbf{\bar{S}}}$.
  \begin{proof}
    We first show that $\mathcal{G} \subseteq \intertwclosure{\bI, \bM^{2, 0}, \bM^{2, 2}, \mathbf{\bar{S}}}$. Let $G \in \mathcal{G}$ with $\abs{G} \eqqcolon n$. We construct $G$ as follows. Fix some order $v_1, \dots, v_n$ on $V(G)$ and start with $n$ copies of $\bM^{2, 0}$ that we identify with the vertices of $G$. We now construct two gadgets. The first, $\hat{\mathbf{A}}$, allows us to add an edge between vertices $v_i$ and $v_{i + 1}$ for all $i \in [n - 1]$. First note that $\mathbf{M}^{4, 2} = (\bM^{2, 2} \otimes \bM^{2, 2}) \circ (\bI \otimes \mathbf{M}^{2, 0} \otimes \bI)$. We let
    \begin{equation*}
      \hat{\mathbf{A}} = (\bI^{\otimes 2} \otimes \bM^{2, 4}) \circ (\bI^{\otimes 2} \otimes \mathbf{A}^{\otimes 2} \otimes \bI^{\otimes 2}) \circ (\bM^{4, 2} \otimes \bI^{\otimes 2}).
    \end{equation*}
    Composition with $\bI^{\otimes 2(i-1)} \otimes \hat{\mathbf{A}} \otimes \bI^{\otimes 2(n - i - 1)}$ then adds an edge between $v_i$ and $v_{i + 1}$ while preserving the out-labels. The second gadget allows us to swap any two vertices adjacent in the ordering. We let 
    \begin{equation*}
      \hat{\mathbf{S}} = (\bI \otimes \mathbf{\bar{S}}) \circ (\mathbf{\bar{S}} \otimes \bI).
    \end{equation*}
    Composition with $\hat{\mathbf{S}}_i \coloneqq \bI^{\otimes 2(i-1)} \otimes \hat{\mathbf{S}} \otimes \bI^{\otimes 2(n - i - 1)}$ swaps $v_i$ and $v_{i + 1}$. More precisely, let $\mathbf{K} = (K, (v_1, v_1, \dots, v_n, v_n), ())$. Then it is straightforward to verify that
    \begin{equation*}
      \hat{\mathbf{S}}_i \circ \mathbf{K} = (K, (v_1, v_1, \dots, v_{i + 1}, v_{i + 1}, v_i, v_i, \dots, v_n, v_n), ()).
    \end{equation*}
    Together, these two gadgets allow us to add edges between arbitrary vertices. After adding the edges, we can remove the $2n$ out-labels by composition with $(\bM^{0, 2})^{\otimes n}$.
    Finally, since
    \begin{equation*}
      \mathbf{\bar{S}} = (\mathbf{S} \otimes \bI) \circ (\bI \otimes \mathbf{S}) \circ (\mathbf{S} \otimes \bI),
    \end{equation*}
    we get that $\mathcal{G} \subseteq \intertwclosure{\bI, \bM^{2, 0}, \bM^{2,2}, \mathbf{\bar{S}}} \subseteq \intertwclosure{\bI, \bM^{2, 0}, \bM^{2,2}, \mathbf{S}}$.
  \end{proof}
\end{lemma}

In the context of the classification results of \cref{ssec:fullclassification}, we can thus state the following corollary.

\begin{corollary}
  Let $Q$ be an easy quantum group and $\gF$ the class of its intertwiner graphs. If $Q$ is either
  \begin{enumerate}
    \item the symmetric group $S_n$,
    \item the hyperoctahedral group $H_n$,
    \item the modified symmetric group $S'_n$,
    \item $G_n(E_h)$, or
    \item $G_n(E_h^s)$,
  \end{enumerate}
  then $\gF(0, 0) = \mathcal{G}$, the class of all graphs.
\end{corollary}

\subsection{The Case of Planar Graphs}

In the case of the quantum modified symmetric- and the quantum hyperoctahedral group, the $(0,0)$-intertwiner graphs turn out to be all planar graphs. In fact, an approach analogous to Man\v{c}inska and Roberson's proof for the quantum symmetric group goes through if we replace single labels by double labels. Their proof can be found in Sections 5 and 6 of \cite{manvcinska2020quantum}, we keep our proof slightly more concise. 

We first need a notion of planarity for bilabelled graphs. The idea is to not only consider the underlying graph, but also require the wires representing the labels not to cross in any embedding. The way to formalise this is to turn wires into full-edges by adding the missing incident vertices, enforce their ordering by connecting them in a cycle, and then reason about the planarity of the resulting (classical) graph.

\begin{definition}
  For a bilabelled graph $\mathbf{K} = (K, \tup{a}, \tup{b}) \in \gbG(k, \ell)$ we define $K^\circ \coloneqq K^\circ(\tup{a}, \tup{b})$ as the graph obtained from $K$ by adding the cycle $C = \alpha_1,\dots, \alpha_k, \beta_\ell, \dots, \beta_1$ and adding the edges $a_i\alpha_i$ and $b_j\beta_j$ for all $i \in [k]$ and $j \in [\ell]$. We refer to $C$ as the \emph{enveloping cycle}\footnote{In the case $k + \ell = 2$, Man\v{c}inska and Roberson take the enveloping cycle to consist of two vertices connected by two edges, hence $K^\circ$ is a multigraph. This is done to avoid special cases in their proofs and does not change the definition of $\gbP$.} of $K^\circ$. 
  Moreover, we define $K^\odot \coloneqq K^\odot(\tup{a}, \tup{b})$ by adding an additional vertex adjacent to all vertices on the enveloping cycle.
\end{definition}

Although $K^\circ$ and $K^\odot$ depend on the input and output vectors of $\mathbf{K}$, they will usually be clear from the context, so we will omit them in our notation. Using these graphs, we can define planar bilabelled graphs.

\begin{definition}
  For any $k, \ell \in \N$, the planar $(k, \ell)$-bilabelled graphs are given by
  \begin{equation*}
    \gbP(k, \ell) = \{\mathbf{K} \in \gbG(k, \ell) \mid K^\odot \text{ is planar}\},
  \end{equation*}
  and we let $\gbP = \bigcup_{k, \ell \in \N} \gbP(k, \ell)$. In particular, $\gbP(0, 0)$ are just the planar graphs.
\end{definition}

The additional vertex in $K^\odot$ ensures that there exists an embedding where the enveloping cycle is the boundary of the outer face. In such an embedding, the new edges really correspond to the original wires. Cycles with this property are called \emph{facial}.
Hence, an alternative definition of planar bilabelled graphs is the following.

\begin{lemma}
  For any $k, \ell \in \N$, the planar $(k, \ell)$-bilabelled graphs are given by
  \begin{equation*}
    \gbP(k, \ell) = \{\mathbf{K} \in \gbG(k, \ell) \mid K^\circ \text{ is planar and its enveloping cycle is facial}\}.
  \end{equation*}
\end{lemma}

The proof is now by induction on the size of the graph. Concretely, one can show that any planar bilabelled graph contains a vertex, which, when removed, can be added back through a collection of planar bilabelled gadget graphs and the usual operations of composition, tensor products, and taking adjoints. 

\newcommand{\dbS}{{^2\mathbf{S}}}
\begin{definition}
  \label{def:gadgetgraphs}
  Let $S^d$ be the star graph on $d+1$ vertices with vertex set $\{v, v_1, \dots, v_d\}$ and edges $\{vv_i \mid i \in [d]\}$. For all $m, d \in \N$, we define $\dbS^{m, d} \coloneqq (S^{2d}, (v^{2m}), (v_1, \dots, v_{2d}))$, $\dbS^{m, d}_L \coloneqq (S^{2d}, (v^{2m}), (v, v, v_1, \dots, v_{2d}))$, and $\dbS^{m,d}_R \coloneqq (S^{2d}, (v^{2m}), (v_1, \dots, v_{2d}, v, v))$.
\end{definition}

\newcommand{\dbP}{{^2\boldsymbol{\mathcal{P}}}}

Our gadget graphs differ from those introduced in \cite{manvcinska2020quantum} in the duplication of all labels. We let $\dbP(\ell, k)$ denote the class of planar bilabelled graphs $\mathbf{K}$ of the form $(K, (a_1, a_1, \dots, a_\ell, a_\ell),$ $(b_1, b_1, \dots, b_k, b_k))$. 

For the inductive step, we need certain basic operations on bilabelled graphs that preserve planarity. The proofs are straightforward, so we omit them here.

\begin{lemma}\label{c4:lem:basicplanarityops}
  Let $G$ be a planar graph, $u, v \in V(G)$, and $uv \in E(G)$. If $G'$ is a graph obtained from $G$ by either removing $u$, removing $uv$, subdividing $uv$, or contracting $uv$, then $G'$ is planar.
\end{lemma}

\begin{lemma}\label{c4:lem:addparvertexplanar}
  Let $G$ be a planar graph, $u, v, w \in V(G)$, and $uv, vw \in E(G)$. Let $G'$ be the graph obtained by subdividing the edge $vw$ with vertex $w_1$, and adding the edge $uw_1$. Then $G'$ is planar.
\end{lemma}

\begin{lemma}[Lemma 4.11 in \cite{manvcinska2020quantum}]\label{c4:lem:addcycleplanar}
  Let $G$ be a planar graph and $v \in V(G)$, such that there exists a planar embedding in which the edges incident to $v$ occur in cyclic order $e_1, \dots, e_k$. Let $G'$ be the graph obtained by subdividing each $e_i$ with vertices $v_i$ and adding the edges $v_kv_1$ and $v_iv_{i + 1}$ for $i < k$. Then $G'$ is planar.
\end{lemma}

\begin{lemma}[Lemma 4.12 in \cite{manvcinska2020quantum}]\label{c4:lem:connectingcyclesplanar}
  Let $G$ and $H$ be planar graphs with facial cycles $C = u_1, \dots, u_k$ and $D = v_1, \dots, v_\ell$ respectively. Let $X$ be the graph obtained from $G \cup H$ by adding the edges $u_1v_1, \dots u_mv_m$ for some $m \leq \min(k, \ell)$. Then $X$ is planar and $u_m, u_{m+1}, \dots, u_\ell, u_1, v_1, v_k, v_{k-1}, \dots, v_m$ is a facial cycle.
\end{lemma}

With slightly more work, one can show that composition, tensor products, and adjoint can, in fact, be reduced to these operations.

\begin{lemma}[cf. Section 5.1 in \cite{manvcinska2020quantum}]
  $\gbP$ is closed under composition, tensor products, and taking adjoints.
\end{lemma}

We now note a necessary condition regarding the labels of a planar bilabelled graph.
Recall the bijection $\Gamma$ between partitions and edgeless bilabelled graphs introduced in \cref{ssec:graphhomsandqsym}. We can extend the domain of $\Gamma$ to arbitrary bilabelled graphs by simply ignoring their edges.
 A bit of thought then shows that any partition associated to a planar bilabelled graph in this way must be non-crossing.

\begin{lemma}[Lemma 6.3 in \cite{manvcinska2020quantum}]
  For any bilabelled graph $\mathbf{K} \in \gbP$, the partition $\mathbb{P}_\mathbf{K}$ is non-crossing.
\end{lemma}

Central to our construction of planar bilabelled graphs will be the following technical statements about non-crossing partitions. To state them concisely we consider indices modulo $n$.

\begin{lemma}[Lemma 6.4 in \cite{manvcinska2020quantum}]\label{c4:lem:consecutivevertices}
  Let $V$ be a set and let $\tup{c} = (c_1, \dots, c_n) \in V^n$ such that $\mathbb{P}_\tup{c}$ is non-crossing. If there exists a $v \in V$ and $i, r \in [n]$ such that $c_i = v = c_{i + r}$ and $c_{i + s} \neq v$ for all $s \in [r - 1]$, then there exists a $u \in V$ that occurs consecutively in $\tup{c}$ and only occurs among $c_{i + 1}, \dots, c_{i + r - 1}$.
\end{lemma}

\begin{corollary}[Corollary 6.5 in \cite{manvcinska2020quantum}]\label{c4:cor:consecutivevertices}
  Let $V$ be a set and $\tup{c} \in V^n$ such that $\mathbb{P}_\tup{c}$ is non-crossing. Then there exists a $v \in V$ that occurs consecutively in $\tup{c}$. In particular, if $\mathbb{P}_\tup{c} = \mathbb{P}_\mathbf{K}$ for some $\mathbf{K} = (K, \tup{a}, \tup{b}) \in \mathcal{P}(\ell, k)$, then there exists a vertex $v \in V(K)$ that occurs consecutively in $(a_1, \dots, a_\ell, b_k, \dots, b_1)$.
\end{corollary}

In particular, these results imply that for any planar bilabelled graph, there must exist a vertex whose labels are not ``interleaved'' with the remaining labels. The central observation is that we can add such a vertex to a planar bilabelled graph using the gadget graphs introduced in Definition \ref{def:gadgetgraphs}. This allows us to prove the following lemma.

\begin{lemma}\label{lem:planargraphblueprint}
  Let $\mathbf{K} \in \dbP$  with $V(\mathbf{K}) \geq 2$. Then there exists a bilabelled graph $\mathbf{K}' \in \dbP$ with $\abs{V(\mathbf{K}')} = \abs{V(\mathbf{K})} - 1$ and $m, d, r, t \in \N$ such that one of the following holds:
  \begin{align}
    \mathbf{K} &= (\bI^{\otimes 2r} \otimes \dbS^{m, d} \otimes \bI^{\otimes 2t}) \circ \mathbf{K}' \label{eq:planar1} \\
    \mathbf{K} &= \mathbf{K}' \circ (\bI^{\otimes 2r} \otimes \dbS^{m, d} \otimes \bI^{\otimes 2t})^* \label{eq:planar2} \\
    \mathbf{K} &= (\dbS^{m, d}_L \otimes \bI^{\otimes 2t}) \circ (\bM^{2, 2r} \otimes \mathbf{K}') \label{eq:planar3} \\
    \mathbf{K} &= (\bI^{\otimes 2t} \otimes \dbS^{m, d}_R) \circ (\mathbf{K}' \otimes \bM^{2, 2r}) \label{eq:planar4}
  \end{align}
  \begin{proof}
    Suppose first that $\mathbf{K} = (K, \tup{a}, \tup{b}) \in \dbP(\ell, k)$ with $\ell+ k \geq 1$. By \cref{c4:cor:consecutivevertices}, there exists a $v \in V(\mathbf{K})$ that occurs consecutively in $(a_1, \dots, a_\ell, b_k, \dots, b_1)$. We will construct $\mathbf{K}'$ by removing $v$. We make the following case distinction.

    Case 1: \emph{$v$ occurs only in $\tup{a}$}. In this case, $v$ might occur consecutively by appearing in $a_1, \dots, a_p$ and $a_q, \dots, a_\ell$ for some $p, q \in [\ell]$. But if this is the case, we can find, by \cref{c4:cor:consecutivevertices}, a $v' \in V(\mathbf{K})$ that occurs consecutively and only among $a_{p + 1}, \dots a_{q - 1}$. We may thus assume that the occurrences of $v$ in $\tup{a}$ are precisely $a_p, \dots, a_q$ for some $p, q \in [\ell]$.
    Now let $v_1, \dots, v_d$ be the neighbours of $v$, and set $\mathbf{K}' = (K', \tup{a'}, \tup{b})$, where $K' = K - v$ and $\tup{a'} = (a_1, a_1, \dots, a_{p-1}, v_1, v_1, \dots, v_d, v_d, a_{q+1}, \dots, a_\ell)$. Then
  \begin{equation*}
    \mathbf{K} = (\bI^{\otimes 2r} \otimes \dbS^{m, d} \otimes \bI^{\otimes 2t}) \circ \mathbf{K}',
  \end{equation*}
  where $r = p - 1$, $m = q - p + 1$, and $t = \ell - q$.

  It remains to show that $\mathbf{K}'$ is planar. This is proven by constructing $K'^\circ$ from $K^\circ$ using the operations listed in \cref{c4:lem:basicplanarityops,c4:lem:addcycleplanar}. For brevity, we show this procedure graphically in \cref{c4:fig:kprimeplanarity}. Compared to the proof in \cite{manvcinska2020quantum}, we need an additional step to duplicate the out-labels of $v_1, \dots, v_d$.


  Case 2: \emph{$v$ occurs only in $\tup{b}$}. In this case $\mathbf{K}^*$ falls into Case 1. Consequently, there exists a $\mathbf{K}'$ such that
  \begin{equation*}
    \mathbf{K}^* = (\bI^{\otimes 2r} \otimes \dbS^{m, d} \otimes \bI^{\otimes 2t}) \circ \mathbf{K}'.
  \end{equation*}
  Taking the adjoint of both sides yields
  \begin{equation*}
    \mathbf{K} = {\mathbf{K}'}^* \circ (\bI^{\otimes 2r} \otimes \dbS^{m, d} \otimes \bI^{\otimes 2t})^*,
  \end{equation*}
  and it is not hard to see that ${\mathbf{K}'}^* \in \dbP$.

  Case 3: \emph{$a_1 = b_1 = v$}. In this case $v$ occurs in both $\tup{a}$ and $\tup{b}$, so we need to do slightly more work to ensure the correct labelling. First we consider the special subcase where $a_\ell = b_k = v$, but not all entries of $\tup{a}$ and $\tup{b}$ are equal to $v$. In that case there exists by \cref{c4:cor:consecutivevertices} a $v' \in V(\mathbf{K})$ that occurs consecutively and only among $\tup{a}$ or $\tup{b}$. Thus either Case 1 or Case 2 apply.

  Otherwise, we can assume that the occurrences of $v$ are exactly $a_1, \dots, a_m$ and $b_1, \dots, b_r$ respectively. Again let $v_1, \dots, v_d$ be the neighbours of $v$ in $\mathbf{K}$ and set $\mathbf{K}' = (K', \tup{a}', \tup{b}')$, where $K' = K - v$, $\tup{a}' = (v_1, v_1, \dots, v_d, v_d, a_{m + 1}, a_{m + 1}, \dots, a_\ell)$, and $\tup{b}' = (b_{r + 1}, b_{r + 1}, \dots, b_k)$. Let $t = \ell - m$. Then we have
  \begin{equation*}
    \mathbf{K} = (\dbS^{m, d}_L \otimes \bI^{\otimes 2t}) \circ (\bM^{2, 2r} \otimes \mathbf{K'}).
  \end{equation*}
  To see that $\mathbf{K}'$ is planar, we reduce to Case 1. Consider $\mathbf{\hat{K}} = \mathbf{K} \circ (\bM^{2r, 0} \otimes \bI^{\otimes 2(k - r)})$. This removes the occurrences of $v$ from $\tup{b}$, so it falls into Case 1. Moreover, the $\mathbf{\hat{K}}'$ constructed in Case 1 is exactly the $\mathbf{K}'$ constructed here. In particular, $\mathbf{K}' \in \dbP$.

  Case 4: \emph{$a_\ell = b_k = v$}. This case is symmetric to Case 3.

  Case 5: \emph{$k + \ell = 0$}. In this case there is no enveloping cycle. We thus choose an arbitrary vertex $v \in V(\mathbf{K})$ with neighbours $v_1, \dots v_d$ and let $\mathbf{K}' = (K', (v_1, v_1, \dots, v_d, v_d), ())$ with $K' = K - v$. Then 
  $\mathbf{K} = \dbS^{0, 2d} \circ \mathbf{K}'$. To see that $\mathbf{K}'$ is planar, fix some planar embedding of $K$ where $v_1, \dots, v_d$ appear in cyclic order. Subdivide the edges $vv_i$ with vertices $\alpha_i$ and add edges to create a cycle $C = \alpha_1, \dots, \alpha_{d}$. By \cref{c4:lem:addcycleplanar}, the resulting graph is planar and is in fact isomorphic to $K'^\odot$. It follows that $\mathbf{K}' \in \dbP$.
  \end{proof}
\end{lemma}

\begin{figure}
  \noindent\makebox[\linewidth][c]{%
  \begin{tikzpicture}[baseline={(v)},scale=.9]
    \draw[fill=gray!20] plot [smooth cycle,tension=.75] coordinates {(-.5, 0) (.1, .9) (1.6, 1) (2, 0) (1.6, -1) (.1, -.9) };

    \node[vertex,label={90:$v$}] (v) at (-.25, 0) {};
    \node[vertex,label={0:$v_1$}] (v1) at (.25, .75) {};
    \node[vertex,label={0:$v_2$}] (v2) at (.25, 0) {};
    \node[vertex,label={0:$v_3$}] (v3) at (.25, -.75) {};

    \draw 
      (v) -- (v1)
      (v) -- (v2)
      (v) -- (v3);

    \node at (1.45, .25) {$K$};

    \draw[name path=circle] (0, 0) circle (2.25);
    \coordinate (vr) at (-2.25, 0);

    \node[vertex,label={170:$\alpha_{p - 1}$}] (ap1) at ($(0, 0)!1!-40:(vr)$) {};
    \node[vertex,label={170:$\alpha_p$}] (ap) at ($(0, 0)!1!-20:(vr)$) {};
    \node[vertex] (a1) at ($(0, 0)!1!-10:(vr)$) {};
    \node[vertex] (a2) at ($(0, 0)!1!10:(vr)$) {};
    \node[vertex,label={190:$\alpha_q$}] (aq) at ($(0, 0)!1!20:(vr)$) {};
    \node[vertex,label={190:$\alpha_{q + 1}$}] (aq1) at ($(0, 0)!1!40:(vr)$) {};

    \draw 
      (v) -- (ap)
      (v) -- (a1)
      (v) -- (a2)
      (v) -- (aq);

    \coordinate (c1) at ($(0, 0)!.9!-10:(vr)$) {};
    \coordinate (c2) at ($(0, 0)!.9!10:(vr)$) {};

    \draw[dotted] (c1) to[bend right,looseness=.55] (c2);

    \node[text width=.42\textwidth,anchor=north] at (0, -2.75) {The vertex $v$ and its neighbours in $K^\circ$.};
  \end{tikzpicture}

  \hspace{.6cm}

  \begin{tikzpicture}[baseline={(v)}, scale=.9]
    \draw[fill=gray!20] plot [smooth cycle,tension=.75] coordinates {(-.5, 0) (.1, .9) (1.6, 1) (2, 0) (1.6, -1) (.1, -.9) };

    \node[vertex,label={0:$v_1$}] (v1) at (.25, .75) {};
    \node[vertex,label={0:$v_2$}] (v2) at (.25, 0) {};
    \node[vertex,label={0:$v_3$}] (v3) at (.25, -.75) {};

    \node at (1.45, .25) {$K'$};

    \draw[name path=circle] (0, 0) circle (2.25);
    \coordinate (vr) at (-2.25, 0);

    \node[vertex,label={180:$v$}] (v) at (vr) {};
    \node[vertex,label={170:$\alpha_{p - 1}$}] (ap1) at ($(0, 0)!1!-40:(vr)$) {};
    \node[vertex,label={190:$\alpha_{q + 1}$}] (aq1) at ($(0, 0)!1!40:(vr)$) {};

    \draw 
      (v) -- (v1)
      (v) -- (v2)
      (v) -- (v3);

    \node[text width=.47\textwidth,anchor=north] at (0, -2.75) {Remove edges $v\alpha_i$ for $i \in \{p+1, \dots, q\}$, contract $v\alpha_p$, and unsubdivide $v, \dots, \alpha_{q + 1}$. };
  \end{tikzpicture}
}

\vspace{.5cm}

  \noindent\makebox[\linewidth][c]{%
  \begin{tikzpicture}[baseline={(v)}, scale=.9]
    \draw[fill=gray!20] plot [smooth cycle,tension=.75] coordinates {(-.5, 0) (.1, .9) (1.6, 1) (2, 0) (1.6, -1) (.1, -.9) };

    \node[vertex,label={0:$v_1$}] (v1) at (.25, .75) {};
    \node[vertex,label={0:$v_2$}] (v2) at (.25, 0) {};
    \node[vertex,label={0:$v_3$}] (v3) at (.25, -.75) {};

    \node at (1.45, .25) {$K'$};

    \draw[name path=circle] (0, 0) circle (2.25);
    \coordinate (vr) at (-2.25, 0);

    \node[vertex,label={180:$v$}] (v) at (vr) {};
    \node[vertex,label={170:$\alpha_{p - 1}$}] (ap1) at ($(0, 0)!1!-40:(vr)$) {};
    \node[vertex,label={190:$\alpha_{q + 1}$}] (aq1) at ($(0, 0)!1!40:(vr)$) {};

    \draw[name path=scircle] (v) circle (1);

    \path [name intersections={of=circle and scircle, by={A, B}}];

    \node[vertex,label={160:$w_1$}] (w1) at (A) {};
    \node[vertex,label={200:$w_2$}] (w2) at (B) {};

    \coordinate (svr) at (-1.25, 0);
    \node[vertex,label={[label distance=0]45:$u_1$}] (u1) at ($(v)!1!33:(svr)$) {};
    \node[vertex,label={[label distance=0]30:$u_2$}] (u2) at (svr) {};
    \node[vertex,label={[label distance=0]-10:$u_3$}] (u3) at ($(v)!1!-33:(svr)$) {};

    \draw 
      (v) -- (u1)
      (v) -- (u2)
      (v) -- (u3)
      (v1) -- (u1)
      (v2) -- (u2)
      (v3) -- (u3);

    \node[text width=.47\textwidth,anchor=north] at (0, -2.75) {Subdivide edges incident to $v$ and add cycle through new vertices.};
  \end{tikzpicture}

  \begin{tikzpicture}[baseline={(v)}, scale=.9]
    \draw[fill=gray!20] plot [smooth cycle,tension=.75] coordinates {(-.5, 0) (.1, .9) (1.6, 1) (2, 0) (1.6, -1) (.1, -.9) };

    \node[vertex,label={0:$v_1$}] (v1) at (.25, .75) {};
    \node[vertex,label={0:$v_2$}] (v2) at (.25, 0) {};
    \node[vertex,label={0:$v_3$}] (v3) at (.25, -.75) {};

    \node at (1.45, .25) {$K'$};

    \coordinate (vr) at (-2.25, 0);
    \coordinate (v) at (vr);

    \begin{scope}
      \clip (0,0) circle (2.27);
      \draw[name path=circle] (0, 0) circle (2.25);
      \draw[name path=scircle] (v) circle (1);
      \path [name intersections={of=circle and scircle, by={A, B}}];
    \end{scope}
    \draw[draw=none,fill=white] (v) circle (.9);

    \node[vertex,label={170:$\alpha_{p - 1}$}] (ap1) at ($(0, 0)!1!-40:(vr)$) {};
    \node[vertex,label={190:$\alpha_{q + 1}$}] (aq1) at ($(0, 0)!1!40:(vr)$) {};
    \node[vertex,label={160:$w_1$}] (w1) at (A) {};
    \node[vertex,label={200:$w_2$}] (w2) at (B) {};

    \coordinate (svr) at (-1.25, 0);
    \node[vertex,label={[label distance=0]45:$u_1$}] (u1) at ($(v)!1!33:(svr)$) {};
    \node[vertex,label={[label distance=0]30:$u_2$}] (u2) at (svr) {};
    \node[vertex,label={[label distance=0]-10:$u_3$}] (u3) at ($(v)!1!-33:(svr)$) {};

    \draw 
      (v1) -- (u1)
      (v2) -- (u2)
      (v3) -- (u3);

    \node[align=center,text width=.47\textwidth,anchor=north] at (0, -2.75) {Remove vertex $v$ and edge $w_1w_2$.};
  \end{tikzpicture}
}

\vspace{.5cm}

  \noindent\makebox[\linewidth][c]{%

  \begin{tikzpicture}[baseline={(v)}, scale=.9]
    \draw[fill=gray!20] plot [smooth cycle,tension=.75] coordinates {(-.5, 0) (.1, .9) (1.6, 1) (2, 0) (1.6, -1) (.1, -.9) };

    \node[vertex,label={0:$v_1$}] (v1) at (.25, .75) {};
    \node[vertex,label={0:$v_2$}] (v2) at (.25, 0) {};
    \node[vertex,label={0:$v_3$}] (v3) at (.25, -.75) {};

    \node at (1.45, .25) {$K'$};

    \draw[name path=circle] (0, 0) circle (2.25);
    \coordinate (vr) at (-2.25, 0);

    \node[vertex,label={170:$\alpha_{p - 1}$}] (ap1) at ($(0, 0)!1!-40:(vr)$) {};
    \node[vertex,label={190:$\alpha_{q + 1}$}] (aq1) at ($(0, 0)!1!40:(vr)$) {};

    \node[vertex,label={[label distance=0]170:$u_1$}] (u1) at ($(0, 0)!1!-20:(vr)$) {};
    \node[vertex,label={[label distance=0]180:$u_2$}] (u2) at (vr) {};
    \node[vertex,label={[label distance=0]190:$u_3$}] (u3) at ($(0, 0)!1!20:(vr)$) {};

    \draw 
      (v1) -- (u1)
      (v2) -- (u2)
      (v3) -- (u3);

    \node[align=center,anchor=north] at (0, -2.75) {Unsubdivide $\alpha_{p-1}w_1u_1$ and $u_3w_2,\alpha_q$.};
  \end{tikzpicture}

  \begin{tikzpicture}[baseline={(v)}, scale=.9]
    \draw[fill=gray!20] plot [smooth cycle,tension=.75] coordinates {(-.5, 0) (.1, .9) (1.6, 1) (2, 0) (1.6, -1) (.1, -.9) };

    \node[vertex,label={0:$v_1$}] (v1) at (.25, .75) {};
    \node[vertex,label={0:$v_2$}] (v2) at (.25, 0) {};
    \node[vertex,label={0:$v_3$}] (v3) at (.25, -.75) {};

    \node at (1.45, .25) {$K'$};

    \draw[name path=circle] (0, 0) circle (2.25);
    \coordinate (vr) at (-2.25, 0);

    \node[vertex,label={170:$\alpha_{p - 1}$}] (ap1) at ($(0, 0)!1!-50:(vr)$) {};
    \node[vertex,label={190:$\alpha_{q + 1}$}] (aq1) at ($(0, 0)!1!50:(vr)$) {};

    \node[vertex,label={[label distance=0]170:$u_1$}] (u1) at ($(0, 0)!1!-20:(vr)$) {};
    \node[vertex,label={[label distance=0]170:$u'_1$}] (up1) at ($(0, 0)!1!-30:(vr)$) {};

    \node[vertex,label={[label distance=0]180:$u_2$}] (u2) at ($(0, 0)!1!5:(vr)$) {};
    \node[vertex,label={[label distance=0]180:$u'_2$}] (up2) at ($(0, 0)!1!-5:(vr)$) {};

    \node[vertex,label={[label distance=0]190:$u_3$}] (u3) at ($(0, 0)!1!20:(vr)$) {};
    \node[vertex,label={[label distance=0]190:$u'_3$}] (up3) at ($(0, 0)!1!30:(vr)$) {};

    \draw 
      (v1) -- (u1)
      (v2) -- (u2)
      (v3) -- (u3)
      (v1) -- (up1)
      (v2) -- (up2)
      (v3) -- (up3);

    \node[text width=.47\textwidth,anchor=north] at (0, -2.75) {Subdivide an edge incident to $u_i$ with vertex $u_i'$ and adding the edge $v_iu_i'$.}; 
  \end{tikzpicture}
}
  \caption{Proving that $\mathbf{K}' \in \gbP$.}
  \label{c4:fig:kprimeplanarity}
\end{figure}

The lemma now allows us to inductively construct our desired planar graphs. 

\begin{lemma}
  Let $\gF$ be the class of intertwiner graphs of the quantum hyperoctahedral group, and $\gF'$ be the class of intertwiner graphs of the quantum modified symmetric group. Then $\gF(0, 0) = \gF'(0, 0) = \mathcal{P}$, the class of all planar graphs.
  \begin{proof}
    We show the stronger result that $\dbP \subseteq \gF \cap \gF'$ by induction. The one-vertex graphs in $\dbP$ are precisely the $\bM^{2r, 2s}$ for $r, s \in \N$. The induction base thus follows from the next claim.
  \begin{claim}
    Let $r, s \in \N$. Then $\bM^{2r, 2s} \in \intertwclosure{\bI, \bM^{2, 0}, \bM^{2, 2}}$.
    \begin{subproof}
      Suppose we already have $\bM^{2r, 2s}$ for some $r, s \geq 1$. Then we can construct $\bM^{2r, 2s + 2}$ and $\bM^{2r + 2, 2s}$. Indeed, we have
      \begin{align*}
        \bM^{2r, 2s + 2} &= (\bI^{\otimes (2r-1)} \otimes \bM^{0, 2} \otimes \bI) \circ (\bM^{2r, 2s} \otimes \bM^{2, 2}), \\
        \bM^{2r + 2, 2s} &= (\bM^{2r, 2s} \otimes \bM^{2, 2}) \circ (\bI^{\otimes (2s-1)} \otimes \bM^{2, 0} \otimes \bI).
      \end{align*}
      It thus follows by induction that we can construct $\bM^{2r, 2s}$ for all $r, s \geq 1$. Finally, we have $\bM^{2r, 0} = \bM^{2r, 2} \circ \bM^{2, 0}$ and $\bM^{0, 2s} = (\bM^{2s, 0})^*$, which completes the proof.
    \end{subproof}
  \end{claim}

  Now consider some $\mathbf{K} \in \dbP$ with $V(\mathbf{K}) \geq 2$. By \cref{lem:planargraphblueprint}, there exists a bilabelled graph $\mathbf{K}' \in \dbP$ with $\abs{V(\mathbf{K}')} = \abs{V(\mathbf{K})} - 1$ such that one of \cref{eq:planar1,eq:planar2,eq:planar3,eq:planar4} holds. By induction hypothesis we can construct $\mathbf{K}'$, so it suffices to show that we can construct $\dbS^{m, d}$, $\dbS^{m, d}_L$, and $\dbS^{m, d}_R$.
  \begin{claim}
    Let $m, d \in \N$. Then $\dbS^{m, d}, \dbS^{m, d}_L, \dbS^{m, d}_R \in \intertwclosure{\bI, \bM^{2, 0}, \bM^{2, 2}, \mathbf{A}}$.
    \begin{subproof}
      We have seen that we can construct $\bM^{2m, 2d}$, which allows us to construct $\dbS^{m, d}$ as 
      \begin{equation*}
        \dbS^{m, d} = \bM^{2m, 2d} \circ \mathbf{A}^{\otimes 2d}.
      \end{equation*}
      Similarly, we can construct $\dbS^{m, d}_L$ and $\dbS^{m, d}_R$ as
      \begin{align*}
        \dbS^{m, d}_L &= \bM^{2m, 2d + 2} \circ (\bI^{\otimes 2} \otimes \mathbf{A}^{\otimes 2d}),\\
        \dbS^{m, d}_R &= \bM^{2m, 2d + 2} \circ (\mathbf{A}^{\otimes 2d} \otimes \bI^{\otimes 2}). \qedhere
      \end{align*}
    \end{subproof}
  \end{claim}

  We have thus shown that $\dbP \subseteq \gF \cap \gF'$. In particular, this means that $\mathcal{P} = \dbP(0, 0) \subseteq \gF(0, 0) \cap \gF'(0, 0) \subseteq \mathcal{P}$, where the second inclusion follows from the fact that the generators of $\gF$ and $\gF'$ are planar, and that $\boldsymbol{\mathcal{P}}$ is closed under composition, tensor products, and taking adjoints.
  \end{proof}
\end{lemma}

\subsection{Group-Theoretical- and \texorpdfstring{$\Pi$}{Π}-Quantum Groups}

It remains to investigate the (infinitely many) quantum groups induced by the partition categories in part (iv) and (v) of \cref{thm:classifyingpartitions}. We prove that the $(0, 0)$-fragment of their respective intertwiner graphs contain all graphs. 

The case for group-theoretical quantum groups follows from a result by Gromada \cite{gromada_grouptheoretical_2022}. 
Let $\mathbf{K}$ be a bilabelled graph and $\pi$ be a partition of its vertex set. Then we denote by $\mathbf{K}/\pi$ the graph obtained by identifying the blocks of $\pi$. Let $\mathbf{G}$ denote the bilabelled graph associated to the partition \sph{}. Then the following result holds.

\begin{lemma}[Lemma 2.31 in \cite{gromada_grouptheoretical_2022}]
  Let $\gF$ be a superset of $\intertwclosure{\mathbf{G}}$. Then $\gF$ is closed under computing quotients $\mathbf{K} \mapsto \mathbf{K}/\pi$, where all unlabelled vertices of $\mathbf{K}$ lie in singletons of $\pi$, i.e.\ are not identified in $\mathbf{K}/\pi$.
\end{lemma}

In particular, we can identify arbitrary labelled vertices. The general idea now is to start with $m$ copies of $\mathbf{A}$ and identify some of their endpoints. This way we can construct all graphs with $m$ edges. However, we need to be slightly careful with this construction to be able to remove any remaining labels afterwards. We achieve this by using double labels.

\begin{lemma}\label{lem:allgraphsfromgrouptgroups}
  Let $Q$ be a group-theoretical quantum group and $\gF$ the set of its intertwiner graphs. Then $\gF(0, 0)$ is the class of all graphs.
  \begin{proof}
    We first construct a modified version of $\mathbf{A}$ that has two in- and two out-labels. Note that $\bM^{1, 3} = (\bI \otimes \bM^{0, 2}) \circ \mathbf{G}$ and $\bM^{2, 4} = (\bM^{1, 3} \otimes \bM^{1, 3}) \circ (\bI^{\otimes 2} \otimes \bM^{2, 0} \otimes \bI^{\otimes 2})$. We then let 
    \begin{equation*}
      \mathbf{\tilde{A}} \coloneqq \bM^{2, 4} \circ \mathbf{A}^{\otimes 4} \circ \bM^{4, 2}.
    \end{equation*}
    We now show that we can construct any connected graph, the statement then follows from the fact that  $\gF(0, 0)$ is closed under under tensor products. 

    First note that $K_1 = \bM^{0, 2} \bM^{2, 0} \in \gF(0, 0)$.
    Suppose then that $K$ is a connected graph with $V(K) = \{v_1, \dots, v_n\}$, $n \geq 2$, and $\abs{E(K)} = m$. We let $\mathbf{K}_0 = \mathbf{\tilde{A}}^{\otimes m}$. Then $\mathbf{K}_0$ has $2m$ in-labels and out-labels respectively, and every vertex is labelled. We identify the $m$ copies of $\mathbf{\tilde{A}}$ with the edges $v_iv_j \in E(K)$ and fix a function $\lambda \colon V(\mathbf{K}_0) \to V(K)$ mapping the vertices of $v_iv_j$ to $\{v_i, v_j\}$. We then define the partition $\pi = \{\{\lambda(w) = v \mid w \in V(\mathbf{K}_0)\} \mid v \in V(K)\}$. Then $\mathbf{K}_0 / \pi = (K, \tup{l}, \tup{l}) \eqqcolon \mathbf{K}$, where $\tup{l}$ is of the form $(w^{v_1}, w^{v_1}, w^{v_1}, w^{v_1}, \dots, w^{v_n}, w^{v_n})$, with an even number of copies of $w^{v_i}$ for all $i \in [n]$ and $\abs{\tup{l}} = 2m$. Hence, 
    \begin{equation*}
      K = (\bM^{0, 2})^{\otimes m} \circ \mathbf{K} \circ (\bM^{2, 0})^{\otimes m}. 
    \end{equation*}
    yields the desired graph.
  \end{proof}
\end{lemma}

Finally, let us turn to the quantum groups $G_n(\Pi_\infty)$ and $G_n(\Pi_k)$. First, we note that $\Pi_1 = \lrangle{\spc}$, which corresponds to the quantum hyperoctahedral group and thus to the class of planar graphs.
For $G_n(\Pi_2)$, we follow a similar approach as in \cref{lem:constructingallgraphs}. We will again start with $n$ isolated vertices that each have two out-labels and then construct gadgets that allow us to add edges between arbitrary vertices. In fact, only the swap gadget will differ slightly.

\newcommand{\bP}{\boldsymbol{\Pi}}
\begin{lemma}
  Let $\gF$ be the class of intertwiner graphs of $G_n(\Pi_2)$. Then $\gF(0, 0)$ is the class of all graphs.
  \begin{proof}
    We first show that we have access to $\bM^{2, 2}$. Let $\bP_2$ denote the bilabelled graph associated to $\pi_2$. Then we can construct $\bM^{4, 0}$ as 
    \begin{equation*}
      \bM^{4, 0} = (\bI^{\otimes 4} \otimes \bM^{0, 2} \otimes \bM^{0, 2}) \circ \bP_2.
    \end{equation*}
    $\bM^{2, 2}$ is now obtained by ``rotating'' $\bM^{4, 0}$: Observe that for any bilabelled graph $(F, (a_1, \dots, a_k)$, $(b_1, \dots, b_\ell)) \in \gbG(k, \ell)$ we can obtain the graph $(F, (a_1, \dots, a_{k - 1}),$ $(b_1, \dots, b_\ell, a_k)) \in \gbG(k-1, \ell + 1)$ by parallel composition with $\bI$ and then composing with $\bI^{\otimes (k - 1)} \otimes \bM^{0, 2}$.
    Concretely, we have
    \begin{equation*}
      \bM^{2, 2} = (\bI^{\otimes 2} \otimes \bM^{0, 2}) \circ \left(\left((\bI^{\otimes 3} \otimes \bM^{0, 2}) \circ (\bM^{4, 0} \otimes \bI)\right) \otimes \bI\right).
    \end{equation*}
    We show this procedure in \cref{fig:rotating}, where we omit some of the identity graphs to improve legibility. 
    Recall then the construction of $\mathbf{\hat{A}}$ from \cref{lem:constructingallgraphs} using $\bM^{2, 2}$ and $\mathbf{A}$. Then $\mathbf{\hat{A}}$ is an edge whose vertices both have two in-labels and out-labels respectively. 
    It remains to show that we can construct a graph that acts as a swap map. It turns out that we can construct a \emph{fat swap} $\mathbf{\tilde{S}}$ that acts similarly to the $\mathbf{\hat{S}}$ constructed in \cref{lem:constructingallgraphs}. We again achieve this by rotating $\bP_2$, see \cref{fig:fatswap}.

    Now to construct an arbitrary graph $K \in \mathcal{G}$ with $\abs{G} \eqqcolon n$, we again start with $n$ copies of $\bM^{2, 0}$ that we identify with the vertices $v_1, \dots, v_n \in V(G)$. Then composition with $\bI^{\otimes 2(i - 1)} \otimes \mathbf{\hat{A}} \otimes \bI^{\otimes 2(n - i - 1)}$ adds an edge between $v_i$ and $v_{i + 1}$ while preserving the out-labels. 
    Composition with $\mathbf{\tilde{S}}_i \coloneqq \bI^{\otimes 2(i - 1)} \otimes \mathbf{\tilde{S}} \otimes \bI^{\otimes 2(n - i - 1)}$ swaps $v_i$ and $v_{i + 1}$ in the sense that for $\mathbf{K} = (K, (v_1, v_1, \dots, v_n, v_n), ())$, we have
    \begin{equation*}
      \mathbf{\tilde{S}}_i \circ \mathbf{K} = (K, (v_1, v_1, \dots, v_{i + 1}, v_{i + 1}, v_i, v_i, \dots, v_n, v_n), ()).
    \end{equation*}
    Together, the gadgets $\mathbf{\hat{A}}$ and $\mathbf{\tilde{S}}$ thus allow us to add edges between arbitrary vertices. Finally, composition with $(\bM^{0, 2})^{\otimes n}$ removes any remaining labels.
  \end{proof}
\end{lemma}

\begin{figure}[h]
  \centering
  \begin{tikzpicture}
    \node at (0, 0) {
        \begin{tikzpicture}
          \node[vertex] (v1) at (0, 0) {};
          \coordinate (vl1) at (0, 1);
          \coordinate (vl11) at (.5, 1);
          \coordinate (vl2) at (0, -1);
          \coordinate (vl21) at (.5, -1);
          \coordinate (vk1) at (0, .35);
          \coordinate (vk11) at (.5, .35);
          \coordinate (vk2) at (0, -.35);
          \coordinate (vk21) at (.5, -.35);
          \draw[wire] 
            (v1) -- (vl1)
            (v1) -- (vl2)
            (vl1) -- (vl11)
            (vl2) -- (vl21)
            (vk1) -- (vk11)
            (vk2) -- (vk21);

          \node[vertex] (v2) at (0, -1.65) {};
          \coordinate (v2l) at (-.5, -1.65);
          \coordinate (v2r) at (.5, -1.65);
          \draw[wire]
            (v2) -- (v2l)
            (v2) -- (v2r);

          \node[vertex] (v3) at (1.5, -1.325) {};
          \coordinate (v3u) at (1.5, -1);
          \coordinate (v3uf) at (1, -1);
          \coordinate (v3l) at (1.5, -1.65);
          \coordinate (v3lf) at (1, -1.65);
          \draw[wire]
            (v3) -- (v3l)
            (v3) -- (v3u)
            (v3l) -- (v3lf)
            (v3u) -- (v3uf);
        \end{tikzpicture}
      };

      \node at (4, 0) {
        \begin{tikzpicture}
          \node[vertex] (v1) at (0, 0) {};
          \coordinate (v1u) at (0, .75);
          \coordinate (v1uf) at (.5, .75);
          \coordinate (v1l) at (0, -.75);
          \coordinate (v1lf) at (.5, -.75);
          \coordinate (v1f) at (.5, 0);
          \coordinate (v1b) at (-.5, 0);
          \draw[wire] 
            (v1) -- (v1u)
            (v1) -- (v1l)
            (v1) -- (v1f)
            (v1) -- (v1b)
            (v1l) -- (v1lf)
            (v1u) -- (v1uf);

          \node[vertex] (v2) at (0, -1.4) {};
          \coordinate (v2l) at (-.5, -1.4);
          \coordinate (v2r) at (.5, -1.4);
          \draw[wire]
            (v2) -- (v2l)
            (v2) -- (v2r);

          \node[vertex] (v3) at (1.5, -1.08) {};
          \coordinate (v3u) at (1.5, -.75);
          \coordinate (v3uf) at (1, -.75);
          \coordinate (v3l) at (1.5, -1.4);
          \coordinate (v3lf) at (1, -1.4);
          \draw[wire]
            (v3) -- (v3l)
            (v3) -- (v3u)
            (v3l) -- (v3lf)
            (v3u) -- (v3uf);
        \end{tikzpicture}
      };

      \node at (8, 0) {
        \begin{tikzpicture}
          \node[vertex] (v1) at (0, 0) {};
          \coordinate (v1f) at (.35, 0);
          \coordinate (v1fu) at (.35, .5);
          \coordinate (v1fuf) at (.85, .5);
          \coordinate (v1fl) at (.35, -.5);
          \coordinate (v1flf) at (.85, -.5);
          \coordinate (v1b) at (-.35, 0);
          \coordinate (v1b) at (-.35, 0);
          \coordinate (v1bu) at (-.35, .5);
          \coordinate (v1buf) at (-.85, .5);
          \coordinate (v1bl) at (-.35, -.5);
          \coordinate (v1blf) at (-.85, -.5);

          \draw[wire]
            (v1) -- (v1f)
            (v1) -- (v1b)
            (v1f) -- (v1fu)
            (v1f) -- (v1fl)
            (v1fu) -- (v1fuf)
            (v1fl) -- (v1flf)
            (v1b) -- (v1bu)
            (v1b) -- (v1bl)
            (v1bu) -- (v1buf)
            (v1bl) -- (v1blf);
        \end{tikzpicture}
      };

      \draw[dotted, thick, gray]
         (.2, 1.45) -- (3.25, 1.2)
         (1, -1.45) -- (3.3, -.6)
         (4.2, 1.2) -- (7, .65)
         (5, -1.2) -- (7, -.65);
  \end{tikzpicture}
  \caption{Constructing $\bM^{2,2}$ from $\bM^{0, 4}$.}
  \label{fig:rotating} 
\end{figure}

\begin{figure}[h]
  \centering
  \begin{tikzpicture}[scale=.9]
    \node (g1) at (.25, 0) {
        \begin{tikzpicture}
          \node[vertex] (v1) at (0, 0) {};
          \coordinate (v1u) at (0, 1.5) {};
          \coordinate (v1uf) at (1, 1.5) {};
          \coordinate (v1uh) at (0, .35) {};
          \coordinate (v1uhf) at (1, .35) {};
          \coordinate (v1l) at (0, -1.5) {};
          \coordinate (v1lf) at (1, -1.5) {};
          \coordinate (v1lh) at (0, -.35) {};
          \coordinate (v1lhf) at (1, -.35) {};

          \draw[wire]
            (v1) -- (v1u)
            (v1) -- (v1l)
            (v1u) -- (v1uf)
            (v1uh) -- (v1uhf)
            (v1l) -- (v1lf)
            (v1lh) -- (v1lhf)
            ;

          \node[vertex] (v2) at (.5, 0) {};
          \coordinate (v2u) at (.5, 1.15) {};
          \coordinate (v2uf) at (1, 1.15) {};
          \coordinate (v2uh) at (.5, .7) {};
          \coordinate (v2uhf) at (1, .7) {};
          \coordinate (v2l) at (.5, -1.15) {};
          \coordinate (v2lf) at (1, -1.15) {};
          \coordinate (v2lh) at (.5, -.7) {};
          \coordinate (v2lhf) at (1, -.7) {};

          \draw[wire]
            (v2) -- (v2u)
            (v2) -- (v2l)
            (v2u) -- (v2uf)
            (v2uh) -- (v2uhf)
            (v2l) -- (v2lf)
            (v2lh) -- (v2lhf)
            ;
        \end{tikzpicture}
      };
      
      \draw[->] 
      (g1.north east) to[bend right] (g1.north west);
      \draw[->] 
      (g1.south east) to[bend left] (g1.south west);

      \node at (2, 0) {$\mapsto$};

      \node (g2) at (4, 0) {
        \begin{tikzpicture}
          \node[vertex] (v1) at (0, 0) {};
          \coordinate (v1fu) at (0, .35);
          \coordinate (v1fuf) at (1, .35);
          \coordinate (v1b) at (-.35, 0);
          \coordinate (v1bu) at (-.35, .35);
          \coordinate (v1bub) at (-.85, .35);
          \coordinate (v1fl) at (0, -.35);
          \coordinate (v1flf) at (1, -.35);
          \coordinate (v1b) at (-.35, 0);
          \coordinate (v1bl) at (-.35, -.35);
          \coordinate (v1blb) at (-.85, -.35);

          \draw[wire]
            (v1) -- (v1fu)
            (v1fu) -- (v1fuf)
            (v1) -- (v1b)
            (v1b) -- (v1bu)
            (v1bu) -- (v1bub)
            (v1) -- (v1fl)
            (v1fl) -- (v1flf)
            (v1) -- (v1b)
            (v1b) -- (v1bl)
            (v1bl) -- (v1blb)
            ;

          \node[vertex] (v2) at (.5, 0) {};
          \coordinate (v2u) at (.5, 1.15) {};
          \coordinate (v2uf) at (1, 1.15) {};
          \coordinate (v2uh) at (.5, .7) {};
          \coordinate (v2uhf) at (1, .7) {};
          \coordinate (v2l) at (.5, -1.15) {};
          \coordinate (v2lf) at (1, -1.15) {};
          \coordinate (v2lh) at (.5, -.7) {};
          \coordinate (v2lhf) at (1, -.7) {};

          \draw[wire]
            (v2) -- (v2u)
            (v2) -- (v2l)
            (v2u) -- (v2uf)
            (v2uh) -- (v2uhf)
            (v2l) -- (v2lf)
            (v2lh) -- (v2lhf)
            ;
        \end{tikzpicture}
      };

      \draw[->] 
         (g2.north east) to[bend right] (g2.north west);

      \node at (6, 0) {$\mapsto$};

      \node (g3) at (8, 0) {
        \begin{tikzpicture}
          \node[vertex] (v1) at (0, 0) {};
          \coordinate (v1fu) at (0, .35);
          \coordinate (v1fuf) at (1, .35);
          \coordinate (v1b) at (-.35, 0);
          \coordinate (v1bu) at (-.35, .35);
          \coordinate (v1bub) at (-.85, .35);
          \coordinate (v1fl) at (0, -.35);
          \coordinate (v1flf) at (1, -.35);
          \coordinate (v1b) at (-.35, 0);
          \coordinate (v1bl) at (-.35, -.35);
          \coordinate (v1blb) at (-.85, -.35);

          \draw[wire]
            (v1) -- (v1fu)
            (v1fu) -- (v1fuf)
            (v1) -- (v1b)
            (v1b) -- (v1bu)
            (v1bu) -- (v1bub)
            (v1) -- (v1fl)
            (v1fl) -- (v1flf)
            (v1) -- (v1b)
            (v1b) -- (v1bl)
            (v1bl) -- (v1blb)
            ;

          \node[vertex] (v2) at (.5, 0) {};
          \coordinate (v2u) at (.5, 1.15) {};
          \coordinate (v2uf) at (-.85, 1.15) {};
          \coordinate (v2uh) at (.5, .7) {};
          \coordinate (v2uhf) at (1, .7) {};
          \coordinate (v2l) at (.5, -1.15) {};
          \coordinate (v2lf) at (1, -1.15) {};
          \coordinate (v2lh) at (.5, -.7) {};
          \coordinate (v2lhf) at (1, -.7) {};

          \draw[wire]
            (v2) -- (v2u)
            (v2) -- (v2l)
            (v2u) -- (v2uf)
            (v2uh) -- (v2uhf)
            (v2l) -- (v2lf)
            (v2lh) -- (v2lhf)
            ;
        \end{tikzpicture}
      };

      \draw[->] 
         (g3.north east) to[bend right] (g3.north west);

      \node at (10, 0) {$\mapsto$};

      \node (g4) at (12, 0) {
        \begin{tikzpicture}
          \node[vertex] (v1) at (0, 0) {};
          \coordinate (v1fu) at (0, .35);
          \coordinate (v1fuf) at (1, .35);
          \coordinate (v1b) at (-.35, 0);
          \coordinate (v1bu) at (-.35, .35);
          \coordinate (v1bub) at (-.85, .35);
          \coordinate (v1fl) at (0, -.35);
          \coordinate (v1flf) at (1, -.35);
          \coordinate (v1b) at (-.35, 0);
          \coordinate (v1bl) at (-.35, -.35);
          \coordinate (v1blb) at (-.85, -.35);

          \draw[wire]
            (v1) -- (v1fu)
            (v1fu) -- (v1fuf)
            (v1) -- (v1b)
            (v1b) -- (v1bu)
            (v1bu) -- (v1bub)
            (v1) -- (v1fl)
            (v1fl) -- (v1flf)
            (v1) -- (v1b)
            (v1b) -- (v1bl)
            (v1bl) -- (v1blb)
            ;

          \node[vertex] (v2) at (.5, 0) {};
          \coordinate (v2u) at (.5, 1.15) {};
          \coordinate (v2uf) at (-.85, 1.15) {};
          \coordinate (v2uh) at (.5, .7) {};
          \coordinate (v2uhf) at (-.85, .7) {};
          \coordinate (v2l) at (.5, -1.15) {};
          \coordinate (v2lf) at (1, -1.15) {};
          \coordinate (v2lh) at (.5, -.7) {};
          \coordinate (v2lhf) at (1, -.7) {};

          \draw[wire]
            (v2) -- (v2u)
            (v2) -- (v2l)
            (v2u) -- (v2uf)
            (v2uh) -- (v2uhf)
            (v2l) -- (v2lf)
            (v2lh) -- (v2lhf)
            ;
        \end{tikzpicture}
      };

      \node at (4, -3.75) {$=$};

      \node (g5) at (6.25, -3.75) {
        \begin{tikzpicture}
          \foreach \x in {1, 2} {
            \foreach \y in {1, 2} {
              \pgfmathsetmacro{\sx}{pow(-1, \x)}
              \pgfmathsetmacro{\sy}{pow(-1, \y)}
              \begin{scope}[yscale=\sx, xscale=\sy] 
                \ifthenelse{\equal{\y}{2}}{\node[vertex]}{\coordinate} 
                  (v\x\y) at (.75, .75) {};
                \coordinate (v\x\y u) at (.75, 1.1);
                \coordinate (v\x\y uf) at (1.25, 1.1);
                \coordinate (v\x\y l) at (.75, .4);
                \coordinate (v\x\y lf) at (1.25, .4);

                \draw[wire]
                  (v\x\y) -- (v\x\y u)
                  (v\x\y) -- (v\x\y l)
                  (v\x\y u) -- (v\x\y uf)
                  (v\x\y l) -- (v\x\y lf)
                  ; 
              \end{scope}
            }
          }

          \draw[wire]
            (v11) -- (v22)
            (v12) -- (v21)
            ;
        \end{tikzpicture}
      };
    \end{tikzpicture}
  \caption{Constructing the Fat Swap $\mathbf{\tilde{S}}$ by rotating $\bP_2$.}
  \label{fig:fatswap}
\end{figure}
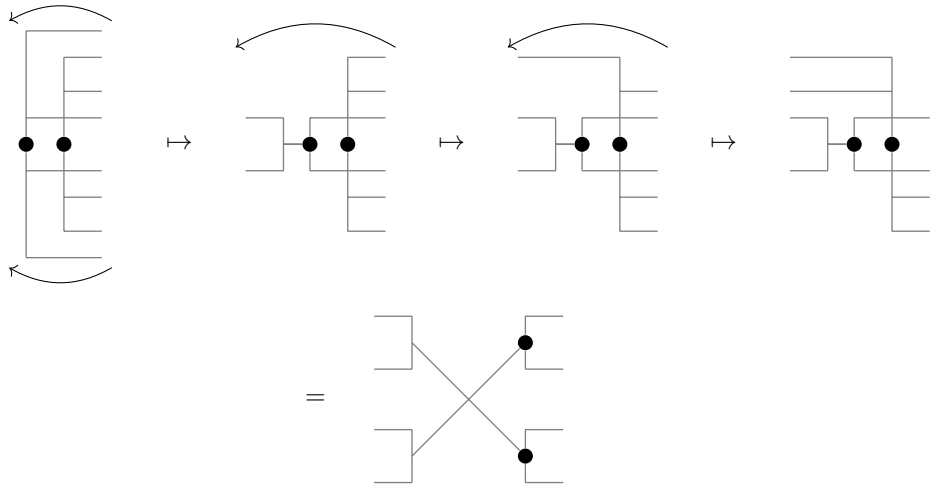

The case for arbitrary $k \geq 2$ now follows using a result on partition categories by Raum and Weber.

\begin{lemma}[Theorem 3.9 in \cite{raum_full_2016}]\label{lem:picategories}
  The partition categories $\Pi_k$ satisfy
  \begin{equation*}
    \Pi_2 \subsetneq \Pi_3 \subsetneq \dots \subsetneq \Pi_\infty \subsetneq \lrangle{\sph}.
  \end{equation*}
\end{lemma}

\begin{corollary}
  Let $\gF$ be the class of intertwiner graphs of $G_n(\Pi_k)$ for $k \geq 2$ and $\gF^\infty$ the class of intertwiner graphs of $G_n(\Pi_\infty)$. Then $\gF(0, 0) = \gF^\infty(0, 0) = \mathcal{G}$, the class of all graphs.
\end{corollary}

Raum and Weber's result moreover yields an alternative, less constructive, proof of \cref{lem:allgraphsfromgrouptgroups}.
We summarize the findings of this section in the following theorem.

\begin{theorem}\label{thm:graphclassclassification}
  Let $Q$ be an easy quantum group and $\gF$ the class of its intertwiner graphs. Then
  \begin{enumerate}
    \item $\gF(0, 0)$ is the class of all graphs if $Q$ is either 
      \begin{itemize}
        \item the symmetric group $S_n$,
        \item the hyperoctahedral group $H_n$,
        \item the modified symmetric group $S'_n$,
        \item $G_n(E_h)$,
        \item $G_n(E^s_h)$,
        \item a group-theoretical quantum group,
        \item $G_n(\Pi_k)$ for some $k \geq 2$, or
        \item $G_n(\Pi_\infty)$.
      \end{itemize}
    \item $\gF(0, 0)$ is the class of all planar graphs if $Q$ is either
      \begin{itemize}
        \item the quantum symmetric group $S^+_n$,
        \item the quantum modified symmetric group $S_n'^+$, or
        \item the quantum hyperoctahedral group $H_n^+$.
      \end{itemize}
    \item $\gF(0, 0)$ is the class of all paths and cycles if $Q$ is either
      \begin{itemize}
        \item the bistochastic group $B_n$,
        \item the quantum bistochastic group $B_n^+$,
        \item the modified bistochastic group $B'_n$,
        \item the quantum modified bistochastic group $B_n^{\#+}$, or
        \item the quantum complexly modified bistochastic group $B_n'^+$,
        \item $G_n(E'_b)$.
      \end{itemize}
    \item $\gF(0, 0)$ is the class of all cycles if $Q$ is either
      \begin{itemize}
          \item the orthogonal group $O_n$,
          \item the quantum orthogonal group $O_n^+$, or
          \item $G_n(E_o)$.
      \end{itemize}
  \end{enumerate}
  Moreover, every $Q$ falls into one of these cases.
\end{theorem}

  \section{A Characterisation of Orthogonal Easy Quantum Groups and Their Homomorphism Indistinguishability Relations}
  Up to this point, we have proved homomorphism indistinguishability results for all orthogonal easy quantum groups. We have also identified the corresponding classes of intertwiner graphs. Still, our theorem is stated in a rather abstract way in terms of intertwiners. In this final section, we categorise quantum matrices according to their intertwiners. This allows us to state several concrete corollaries of our theorem.

\subsection{Properties of Quantum Matrices and Intertwiners}

Our first observation is that the partitions used in the classification of easy quantum groups in fact correspond to very concrete properties of the fundamental representations.

 \begin{lemma}\label{lem:intertwinerprops}
  Let $\mathcal{Z}$ be a \Cstar-algebra and $\U \in \mathcal{Z}^{n \times n}$. Then $\U$ has intertwiner
  \begin{enumerate}
    \item \spe{} iff $\U\U\transpose = \mathbf{1}I$,
    \item \spa{} iff the entries of $\U$ commute,
    \item \spb{} iff $\U$ has row sum $\mathbf{1}$,
    \item \spbi{} iff $\U$ has column sum $\mathbf{1}$,
    \item $\spb \otimes \spb$ iff the product of any two row sums of $\U$ is $\mathbf{1}$,
    \item $\spbi \otimes \spbi$ iff the product of any two column sums of $\U$ is $\mathbf{1}$,
    \item \spf{} iff the entries $u_{ij}$ of $\U$ satisfy $u_{ik} u_{jk} = \delta_{ij}u_{ik}$ for all $i, j, k \in [n]$,
    \item \spc{} iff the entries $u_{ij}$ of $\U$ satisfy $u_{ki} u_{kj} = 0$ and $u_{ik} u_{jk} = 0$ for all $i, j, k \in [n]$ with $i \neq j$.
  \end{enumerate}
  \begin{proof} 
    We first show 1. Expanding $\Utp{2} M^{2, 0} = M^{2, 0} (\mathbf{1})$ yields
    \begin{equation}\label{eq:orthonormalbases}
      \sum_i \U e_i \otimes \U e_i = \sum_i \mathbf{1} (e_i \otimes e_i).
    \end{equation}
    We thus have
    \begin{align*}
      (\U\U\transpose)_{ij} &= \sum_k u_{ik} u_{jk} = \sum_k (\U e_k)_i \cdot (\U e_k)_j = \Big(\sum_k \U e_k \otimes \U e_k\Big)_{i , j} \\
      &= (M^{2, 0})_{i , j}\mathbf{1} = \delta_{ij}\mathbf{1}.
    \end{align*}

    The remaining parts can all be proven along the same lines, by expanding the intertwiner equation and then comparing the entries of the resulting matrices. We show this again for case 2. Let $u_{*i}$ and $u_{i*}$ denote the $i$-th column and row vector of $\U$ respectively. $\Utp{2} S = S \Utp{2}$ is equivalent to 
    \begin{equation*}
      \Utp{2}\Big(\sum_{i,j} (e_j \otimes e_i) \cdot (e_i \otimes e_j)\transpose \Big) = \Big(\sum_{i,j} (e_j \otimes e_i) \cdot (e_i \otimes e_j)\transpose \Big) \Utp{2}.
    \end{equation*}
    Expanding this yields
    \begin{equation*}
      \sum_{i,j} (u_{*j} \otimes u_{*i}) \cdot (e_i \otimes e_j)\transpose = \sum_{i,j} (e_j \otimes e_i) \cdot (u_{i*} \otimes u_{j*}),
    \end{equation*}
    and entrywise comparison gives
    \begin{align*}
      u_{i_1j_2} \cdot u_{i_2j_1} &= \sum_{i,j} u_{i_1j} \cdot u_{i_2 i} \cdot \delta_{ij_1} \cdot \delta_{jj_2}\\ 
                                  &= \Big(\sum_{i,j} (u_{*j} \otimes u_{*i}) \cdot (e_i \otimes e_j)\transpose\Big)_{i_1i_2,j_1j_2} \\           
                                  &= \Big(\sum_{i,j} (e_j \otimes e_i) \cdot (u_{i*} \otimes u_{j*}) \Big)_{i_1i_2,j_1j_2} \\
                                  &= \sum_{i,j} \delta_{i_1j} \cdot \delta_{i_2i} \cdot u_{ij_1} \cdot u_{jj_2}\\
                                  &= u_{i_2j_1} \cdot u_{i_1j_2}
    \end{align*}
    for all $i_1, i_2, j_1, j_2 \in [n]$, as desired.
  \end{proof}
\end{lemma} 

Now recall our definition of quantum permutation matrices as matrices with entries in some \Cstar-algebra, satisfying \cref{eq:symrep}, the definining relations of the quantum symmetric group. The previous lemma implies that we can rephrase this definition in terms of intertwiners.

\begin{corollary}
  Let $\mathcal{Z}$ be a \Cstar-algebra. A matrix $\U \in \mathcal{Z}^{n \times n}$ is a quantum permutation matrix if and only if it has intertwiners $\spb, \spf$.
  \begin{proof}
    We first note that if $\U$ has $\spb$ as an intertwiner then it also has $\spbi$ as an intertwiner, so the row and column sums are $\mathbf{1}$. It remains to show that the condition $u_{ik}u_{jk} = \delta_{ij} u_{ik}$ for all $i, j, k \in [n]$ is equivalent in this case to the weaker condition $u_{ij}^2 = u_{ij}^* = u_{ij}$ that we required in \cref{eq:symrep}.

    To see this, suppose that $u^2_{ij} = u^*_{ij} = u_{ij}$ and $\sum_\ell u_{\ell k} = \mathbf{1}$. Then 
    \begin{equation*}
      u_{jk} = u_{jk}\mathbf{1}u_{jk} = u_{jk} \Big(\sum_\ell u_{\ell k}\Big) u_{jk} = u_{jk} + \sum_{\ell \neq j} u_{jk}u_{\ell k}u_{jk},
    \end{equation*}
    and thus $\sum_{\ell \neq j} u_{jk}u_{\ell k}u_{jk} = 0$. Now note that all the terms $u_{jk} u_{\ell k} u_{jk} = (u_{\ell k}u_{jk})^*u_{\ell k}u_{jk}$ are positive. Since the sum of positive elements is positive again, we also have that $-u_{jk} u_{ik} u_{jk} = \sum_{\ell \neq i, j} u_{jk} u_{\ell k} u_{jk}$ is positive for all $k \in [n]$. Thus for $i \neq j$, both $u_{jk}u_{ik}u_{jk}$ and $-u_{jk}u_{ik}u_{jk}$ are positive, which is only possible if $u_{jk}u_{ik}u_{jk} = 0$. But we then have
    \begin{equation*}
      \norm{u_{ik}u_{jk}}^2 = \norm{(u_{ik}u_{jk})^*u_{ik}u_{jk}} = \norm{u_{jk}u_{ik}u_{jk}} = 0.
    \end{equation*}
    It follows that $u_{ik}u_{jk} = 0$ for $i \neq j$. Moreover, per assumption we have $u_{ik}u_{ik} = u_{ik}$, so $u_{ik}u_{jk} = \delta_{ij} u_{ik}$ as desired.
  \end{proof}
\end{corollary}

Through their intertwiners, we can also define quantum monomial, quantum signed permutation, quantum bistochastic, and quantum signed bistochastic matrices. 

\begin{definition}
  Let $\mathcal{Z}$ be a \Cstar-algebra. A non-zero matrix $\U \in \mathcal{Z}^{n \times n}$ whose entries $u_{ij}$ satisfy $u^*_{ij} = u_{ij}$ is called a
  \begin{enumerate}
    \item \emph{quantum monomial orthogonal matrix} if it has intertwiners \spe, \spc,
    \item \emph{quantum signed permutation matrix} if it has intertwiners $\spe, \spb \otimes \spb$, \spc,
    \item \emph{quantum bistochastic matrix} if it has intertwiners \spe, \spb,
    \item \emph{quantum signed bistochastic matrix} if it has intertwiners \spe, $\spb \otimes \spb$.
  \end{enumerate}
\end{definition}

In particular, the fundamental representations of $H_n^+$, $S_n'^+$, $B_n^+$, and $B_n^{\#+}$ are quantum monomial orthogonal-, quantum signed permutation-, quantum bistochastic-, and quantum signed bistochastic matrices respectively. In the same way as for quantum orthogonal and quantum permutation matrices, we recover the notions of monomial orthogonal, signed permutation, bistochastic, and signed bistochastic matrices when the \Cstar-algebra is $\C$. 

\begin{lemma}\label{lem:coincidingmatrices}
  Over $\C$, the following notions coincide
  \begin{enumerate}
    \item monomial orthogonal matrices and quantum monomial orthogonal matrices,
    \item signed permutation matrices and quantum signed permutation matrices,
    \item bistochastic matrices and quantum bistochastic matrices, 
    \item signed bistochastic matrices and quantum signed bistochastic matrices. 
  \end{enumerate}
  \begin{proof}
    1: Recall that monomial matrices are matrices with the same non-zero pattern as a permutation matrix, whose non-zero values are arbitrary complex numbers. Using \cref{lem:intertwinerprops}, it is straightforward to verify that every monomial orthogonal matrix has $\spe{}, \spc$ as intertwiners and is thus a quantum monomial orthogonal matrix. Suppose then $\U = (u_{ij})$ is quantum monomial orthogonal matrix over $\C$. Since $\U$ has intertwiner \spe{}, it is orthogonal and in particular has at least one non-zero entry per row and column. Now suppose $u_{ik} \neq 0, u_{jk} \neq 0$ for $i \neq j$. Since $u_{ik}, u_{jk} \in \C$, this implies $u_{ik}u_{jk} \neq 0$ contradicting---by \cref{lem:intertwinerprops}---the fact that $\U$ has \spc{} as an intertwiner. The case for $u_{ki} \neq 0, u_{kj} \neq 0$ is analogous. Consequently, $\U$ is orthogonal and has exactly one non-zero entry per row and column, making it a monomial orthogonal matrix.

    2: It is again straightforward to verify that every signed permutation matrix is a quantum signed permutation matrix. Let $\U$ be a quantum signed permutation matrix over $\C$. The same argument as for 1 shows that $\U$ is a monomial orthogonal matrix. It thus suffices to show that the non-zero entries are all $1$ or all $-1$.  Since $\spb \otimes \spb$ is an intertwiner, \cref{lem:intertwinerprops} implies that the product of any two row sums must equal $1$. This implies that 
    \begin{equation*}
      1 = \Big(\sum_\ell u_{i\ell} \Big)\Big(\sum_\ell u_{i\ell}\Big) = u_{ij}^2
    \end{equation*}
    for all $i$ and some $j$. In particular, all non-zero entries must be $\pm 1$. Moreover, if there existed entries $u_{ij} = 1$ and $u_{k\ell} = -1$, then 
    \begin{equation*}
      \Big(\sum_\ell u_{i\ell} \Big)\Big(\sum_\ell u_{k\ell}\Big) = -1,
    \end{equation*}
    contrary to our observation. It follows that $\U$ is a signed permutation matrix.

    3: Follows immediately from \cref{lem:intertwinerprops}.

    4: It is easy to see that every signed bistochastic matrix is a quantum signed bistochastic matrix. Suppose then $\U$ is quantum signed bistochastic matrix over $\C$. Then $\U$ is orthogonal because it has $\spe$ as an intertwiner. Moreover, the intertwiner $\spb \otimes \spb$ guarantees that the product of any two row sums equals $1$. Along similar lines as in 2, this implies that every row sum must be $\pm 1$, and in fact all row sums must have the same value. Analogously, the intertwiner $\spbi \otimes \spbi$ guarantees the same for the column sums. Finally, the rows and columns must sum to the same value, otherwise 
    \begin{equation*}
      \sum_{i,j} u_{ij} = \sum_i \sum_j u_{ij} = \sum_j \Big(-\sum_i u_{ij}\Big) = -\sum_{i,j} u_{ij},
    \end{equation*}
    contradicting the fact that the sum of entries of $\U \in \C^{n \times n}$ is $\pm n$.
    It follows that $\U$ is a signed bistochastic matrix.
  \end{proof}
\end{lemma}

An interesting situation again arises in the context of the quantum complexly modified bistochastic group, where the intertwiner spaces $\langle \spa, \spb \otimes \spb \rangle$ and $\langle \spa, \spj \rangle$ coincide in the classical/commutative, but not in the quantum case. For this reason, we define quantum complexly signed bistochastic matrices seperately.

\begin{definition}
  Let $\mathcal{Z}$ be a \Cstar-algebra. A matrix $\U \in \mathcal{Z}^{n \times n}$ is called a \emph{quantum complexly signed bistochastic matrix} if it has intertwiners $\spe, \spj$.
\end{definition}

In particular, the fundamental representation of the quantum complexly modified bistochastic group $B^{'+}_n$ is a quantum complexly signed bistochastic matrix.
Since $\C$ is commutative, we still recover signed bistochastic matrices when $\mathcal{Z} = \C$.

\begin{lemma}
  Over $\C$, the notions of quantum complexly signed bistochastic matrices, quantum signed bistochastic matrices, and signed bistochastic matrices coincide.
  \begin{proof}
    Since $\C$ is commutative, any quantum (complexly) signed bistochastic matrix over $\C$ has \spa{} as an intertwiner. Since $\spj \in \langle \spa, \spb \otimes \spb \rangle$, every quantum signed bistochastic matrix has \spj{} as an intertwiner and is thus a quantum complexly signed bistochastic matrix. Since $\spb \otimes \spb~ \in \langle \spa, \spj \rangle$, every quantum complexly signed bistochastic matrix has $\spb \otimes \spb$ as an intertwiner and is thus a quantum signed bistochastic matrix. The statement now follows from \cref{lem:coincidingmatrices}.
  \end{proof}
\end{lemma}

Finally, for each of the quantum matrices above we obtain a version with commuting entries by additionally requiring that \spa{} be an intertwiner. In the context of compact matrix quantum groups, this intertwiner is precisely what makes a quantum group classical, and we can prove a similar result for quantum matrices.

\begin{lemma}
  Let $\mathcal{Z}$ be a \Cstar-algebra, and $\U \in \mathcal{Z}^{n \times n}$ be a non-zero matrix with commuting entries. Then there exists a non-zero matrix $U \in \C^{n \times n}$, such that $U^{\otimes k} T = T' U^{\otimes \ell}$ for all $T, T' \in \C^{n^k \times n^\ell}$ with $\Utp{k} T = T' \Utp{\ell}$.
  \begin{proof}
    Let $\mathcal{Z}'$ be the \Cstar-subalgebra of $\mathcal{Z}$ generated by the entries $u_{ij}$ of $\U$. Then $\mathcal{Z}'$ is commutative and thus by Gelfand-Naimark isomorphic to the function algebra $X$ of some finite set $X$. More precisely, $X$ is the set of non-zero $^*$\nobreakdash-homomorphisms $\chi\colon \mathcal{Z}' \to \C$ and there exists an isomorphism $\phi\colon \mathcal{Z}' \to C(X)$ mapping $z \in \mathcal{Z}'$ to 
    \begin{equation*}
      \tilde{z}\colon X \to \C, \chi \mapsto \chi(z).
    \end{equation*}
    We now define a mapping $\rho \colon X \to \C^{n \times n}$ by $\rho(\chi)_{ij} = \phi(u_{ij})(\chi) = \chi(u_{ij})$. Since $\mathcal{Z}'$ is non-trivial, there exists a $\chi \in X$. Set $U = \rho(\chi)$ and let $T, T' \in \C^{n^k \times n^\ell}$ such that $\Utp{k} T = T' \Utp{\ell}$ for $k, \ell \geq 1$. Then since $\chi$ is an algebra homomorphism, we have
    \begin{align*}
      \left(U^{\otimes k} T\right)_{i_1\dots i_k,j_1\dots j_\ell} &= \sum_{r_1\dots r_k} \left(\prod_s U_{i_sr_s}\right) T_{r_1\dots r_k,j_1 \dots j_\ell} \\
                                                                  &= \sum_{r_1\dots r_k} \left(\prod_s \chi(u_{i_sr_s})\right) T_{r_1\dots r_k,j_1 \dots j_\ell} \\
                                                                  &= \chi\left(\sum_{r_1\dots r_k} \left(\prod_s u_{i_sr_s}\right) T_{r_1 \dots r_k,j_1 \dots j_\ell}\right) \\
&= \chi \left(\left(\Utp{k} T\right)_{i_1\dots i_k, j_1\dots j_\ell}\right)\\
                                                                  &= \chi \left(\left(T' \Utp{\ell}\right)_{i_1\dots i_k, j_1\dots j_\ell}\right)\\
                                                                  &= \chi\left(\sum_{r_1 \dots r_\ell} T'_{i_1 \dots i_k, r_1 \dots r_\ell} \prod_s u_{r_sj_s}\right) \\
                                                                  &= \sum_{r_1 \dots r_\ell} T'_{i_1 \dots i_k, r_1 \dots r_\ell} \prod_s U_{r_sj_s}\\
                                                                  &= \left(T' U^{\otimes \ell}\right)_{i_1 \dots r_k, j_1 \dots j_\ell}.
    \end{align*}

    The case for $k + \ell \leq 1$ follows from the observation that $\chi(\mathbf{1}) = 1$. Indeed, let $z \in \mathcal{Z}'$, $z \neq \mathbf{1}$, $z \neq 0$. Since $\chi$ is a non-zero $^*$-homomorphism, we have $\chi(z) = \chi(z \cdot \mathbf{1}) = \chi(z) \cdot \chi(\mathbf{1})$ where $\chi(z) \neq 0$. Since $\chi(z), \chi(\mathbf{1}) \in \C$, $\chi(\mathbf{1})$ must be $1$. An analogous argument as above then shows that for $T \in \C^{1 \times n^\ell}$, $\Utp{0}T = (\mathbf{1})T = T'\Utp{\ell}$ implies $U^{\otimes 0}T = (1)T = T'U^{\otimes \ell}$; for $T \in \C^{n^k \times 1}$, $\Utp{k}T = T'\Utp{0} = T'(\mathbf{1})$ implies $U^{\otimes k}T = T'(1) = T' U^{\otimes 0}$; and for $T \in \C^{1 \times 1}$, $\Utp{0}T = (\mathbf{1})T = T'(\mathbf{1}) = \Utp{0}$ implies $U^{\otimes 0}T = (1)T = T'(1) = T'U^{\otimes 0}$. This concludes the proof.
  \end{proof}
\end{lemma}

This means that we can essentially identify quantum matrices with commuting entries with classical matrices over $\C$ that have the same intertwiners. In particular, \cref{lem:intertwinerprops,lem:coincidingmatrices} apply: For instance, if there exists a quantum orthogonal matrix $\U$ with commuting entries and $\U A_G = A_H \U$, then there exists a corresponding orthogonal matrix $U$ with $U A_G = A_H U$.

\subsection{Homomorphism Indistinguishability}

We have seen that in the cases of classical- and liberated quantum groups, the matrices in \cref{thm:mainresult} either correspond to classical matrices or can be given meaningful names. Together with our results of \cref{sec:constructing}, where we investigated the intertwiner graphs of the different orthogonal easy quantum groups, we can state various corollaries of \cref{thm:mainresult} in a more readable form. 

The first result is a new characterisation of homomorphism indistinguishability over the class of all graphs, which according to Lovász \cite{lovasz1967operations} is equivalent to graph isomorphism.

\begin{corollary}\label{cor:homind-allgraphs}
	For graphs $G$ and $H$, the following are equivalent:
  \begin{enumerate}
    \item $G$ and $H$ are homomorphism indistinguishable over the class of all graphs.
    \item There exists a permutation matrix $U$ such that $U A_G = A_H U$.
    \item There exists a signed permutation matrix $U$ such that $U A_G = A_H U$.
    \item There exists an orthogonal monomial matrix $U$ such that $U A_G = A_H U$.
  \end{enumerate}
\end{corollary} 

Moreover, we recover Man\v{c}inska and Roberson's characterisation of quantum isomorphism, avoiding the detour through quantum orbits inherent to the original proof. We also obtain two additional, equivalent characterisations that mirror the classical case.

\begin{corollary}\label{cor:homind-planar}
	For graphs $G$ and $H$, the following are equivalent:
  \begin{enumerate}
    \item $G$ and $H$ are homomorphism indistinguishable over the class of all planar graphs.
    \item There exists a quantum permutation matrix $\U$ such that $\U A_G = A_H \U$.
    \item There exists a quantum signed permutation matrix $\U$ such that $\U A_G = A_H \U$.
    \item There exists a quantum orthogonal monomial matrix $\U$ such that $\U A_G = A_H \U$.
  \end{enumerate}
\end{corollary}

Homomorphism indistinguishability over the class of all cycles is well-known to be equivalent to cospectrality, and hence the existence of an orthogonal matrix $U$ with $U A_G = A_H U$. We recover this result and extend it by a quantum version.

\begin{corollary}\label{cor:homind-cycles}
	For graphs $G$ and $H$, the following are equivalent:
  \begin{enumerate}
    \item $G$ and $H$ are homomorphism indistinguishable over the class of all cycles.
    \item There exists an orthogonal matrix $U$ such that $U A_G = A_H U$.
    \item There exists a quantum orthogonal matrix $\U$ such that $\U A_G = A_H\U$.
  \end{enumerate}
\end{corollary}

Finally, we recover and extend the characterisation of homomorphism indistinguishability over the class of all paths and cycles---noted by Seppelt \cite[Corollary 3.3.3]{seppelt_homomorphism_2024} and \cite[Theorem 3]{van2007cospectral} to correspond to the existence of a bistochastic matrix $U$ with $U A_G = A_H U$---by the following criteria.

\pagebreak
\begin{corollary}\label{cor:homind-pathscycles}
  	For graphs $G$ and $H$, the following are equivalent:
  \begin{enumerate}
    \item $G$ and $H$ are homomorphism indistinguishable over the class of all paths and cycles. 
    \item There exists a bistochastic matrix $U$ such that $U A_G = A_H U$.
    \item There exists a signed bistochastic matrix $U$ such that $U A_G = A_H U$.
    \item There exists a quantum bistochastic matrix $\U$ such that $\U A_G = A_H \U$.
    \item There exists a quantum signed bistochastic matrix $\U$ such that $\U A_G = _H \U$.
    \item There exists a quantum complexly signed bistochastic matrix $\U$ such that $\U A_G = A_H \U$.
  \end{enumerate}
\end{corollary}

We finally  note that the assertions of \cref{cor:homind-allgraphs,cor:homind-cycles,cor:homind-pathscycles,cor:homind-planar} are mutually not equivalent.
That is, homomorphism indistinguishability over cycles, cycles and paths, planar graphs, and all graphs
forms a descending chain of four pairwise distinct graph isomorphism relaxations.
This is implied by the stronger assertion that all (the disjoint union closures of) these graph classes are homomorphism distinguishing closed, cf.\ \cite[Theorems~4.7, 8.3]{roberson_oddomorphisms_2022} and \cite[Theorem~7.1.4]{seppelt_homomorphism_2024}.
A graph class $\mathcal{F}$ is \emph{homomorphism distinguishing closed} \cite{roberson_oddomorphisms_2022}
if, for all $F \not\in \mathcal{F}$, there exist graphs $G$ and $H$ such that $G \cong_\gF H$ and $\hom(F, G) \neq \hom(F, H)$.

More generally, the characterisations in  \cref{cor:homind-allgraphs,cor:homind-cycles,cor:homind-pathscycles,cor:homind-planar} make the graph isomorphism relaxations considered here subject to the emerging theory of homomorphism indistinguishability.
In particular, they imply that the graph isomorphism relaxations are preserved under certain operations \cite{seppelt2024logical}
and that the notions in \cref{cor:homind-cycles,cor:homind-pathscycles} can be decided in polynomial time \cite{seppelt_algorithmic_2024},
while those in \cref{cor:homind-planar} are undecidable \cite{atserias2019quantum,lupini2020nonlocal}
and those in \cref{cor:homind-allgraphs} are in quasi-polynomial time \cite{babai_graph_2016}.

  \section{Conclusion}
  We studied the connection between orthogonal easy quantum groups and homomorphism indistinguishability.
We proved \cref{thm:mainresult}, which characterises the feasability of the matrix equation $\U A_G = A_H \U$, where $A_G, A_H$ are adjacency matrices of graphs $G$ and $H$, and $\U$ is a matrix indeterminate with entries in some \Cstar-algebra satisfying the relations $\Utp{k} M_i = M_i \Utp{\ell}$ for intertwiners $M_i$ of some easy quantum group, as being equivalent to homomorphism indistinguishability over the unlabelled intertwiner graphs generated by the corresponding bilabelled graphs $\bM_i$ and the bilabelled edge $\mathbf{A}$. In particular, the theorem generalises Man\v{c}inska and Roberson's characterisation of quantum isomorphism, in which case $\U$ is a quantum permutation matrix, and the generated unlabelled intertwiner graphs are precisely the planar graphs.

The result is particularly nice because it joins a now quite extensive list of homomorphism indistinguishability relations characterising the feasability of matrix equations of the form $X A_G = A_H X$, cf.\ \cite[Table 1.2]{seppelt_homomorphism_2024}.
Salient examples of such relations include the existence of a pseudo-stochastic $X$, being equivalent to homomorphism indistinguishability over all paths, and the existence of doubly stochastic matrix $X$, being equivalent to homomorphism indistinguishability over all trees \cite{dell2018lov}.
In fact, we have shown that the conditions $\Utp{k} M_i = M_i \Utp{\ell}$ very naturally define well-known classes of matrices and corresponding quantum relaxations. 
This allowed us to state some consequences of \cref{thm:mainresult} in a form more closely resembling these existing homomorphism indistinguishability results. 

\subsection{The Limits of Quantum Groups for Homomorphism Indistinguishability}
Interestingly, we reprove certain characterisations, namely those of the existence of orthogonal, permutation, and bistochastic solutions to the matrix equation $X A_G = A_H X$. Moreover, some homomorphism indistinguishability relations were already known for all the intertwiner graph classes of orthogonal quantum groups.

This suggests the question to what extent our results are exhaustive. The short answer is that we derived all homomorphism indistinguishability relations induced by orthogonal easy quantum groups. However, as Man\v{c}inska and Roberson show in \cite{manvcinska2020quantum}, orthogonal easy quantum groups are special instances of the more general graph-theoretic quantum groups. These are quantum groups whose intertwiners are given by homomorphism tensors of arbitrary bilabelled graphs. It turns out that our proof generalises to these more general quantum groups without much difficulty. The resulting homomorphism indistinguishability relations, however, characterise the existence of a solution $\U$ not only to $\U A_G = A_H \U$, but simultaneously for $\Utp{k_i} \mathbf{K}^{(i)}_G = \mathbf{K}^{(i)}_H \Utp{\ell_i}$---the $\mathbf{K}^{(i)}$ being the $(k_i, \ell_i)$-bilabelled graphs whose homomorphism tensors span the intertwiners of the graph-theoretic quantum group in question. This seems like a rather arbitrary relation from a general point of view. 

A more interesting class of quantum groups have been introduced by Tarrago and Weber in \cite{tarrago2018classification} and \cite{tarrago2017unitary}. They introduce the \emph{unitary easy quantum groups} which generalise the orthogonal easy quantum groups. 
The unitary easy quantum groups are the quantum groups whose intertwiners can be characterised in terms of \emph{coloured} partitions, where the elements of the underlying set are either black or white. 
These coloured partitions correspond to linear maps in exactly the same way as uncoloured partitions, and hence can also be seen as homomorphism tensors of edgeless bilabelled graphs. Moreover, the operations of composition, tensor product, and taking adjoints are generally defined as for their uncoloured counterparts. Merely the composition requires that the bottom row of the first and the top row of the second partition additionally match in colours. In particular, the operations on bilabelled graphs and on coloured partitions coincide whenever they are defined on partitions. We can get rid of this caveat by considering bilabelled graphs with coloured labels, allowing composition only when label colours match. This, in turn, potentially restricts the set of intertwiner graphs we are able to construct. Under these definitions, our proofs go through with some obvious minor changes. The remaining task would be to investigate which concrete graph classes result from these definitions.

As the name suggests, the fundamental representations of unitary easy quantum groups do not have to be orthogonal, which makes them a promising candidate to obtain homomorphism indistinguishability relations coarser than the ones derived in this paper. 

\subsection{Multigraphs}
An indispensible tool in our construction of the intertwiner graph classes of orthogonal easy quantum groups is the suppression of parallel edges. This, however, has the drawback that the resulting graph classes collapse into essentially four cases. Therefore, it might be enlightening to consider the classes of multigraphs generated when keeping duplicate edges. For the results to fit into our framework of homomorphism indistinguishability, however, we have to count homomorphisms from multigraphs.  One option to do this is to define
\begin{equation*}
  \hom(F, G) = \sum_{f \colon V(F) \to V(G)} \prod_{uv \in E(F)} m_{E(G)}(f(u)f(v)),
\end{equation*}
where $m_{E(G)} \colon E(G) \to \N$ is the multiplicity function of the multiset of edges $E(G)$. For bilabelled graphs $(F, \tup{a}, \tup{b})$ and $(G, \tup{a'}, \tup{b'})$, we instead sum only over functions $f\colon V(F) \to V(G)$ with $f(\tup{a}) = \tup{a}'$ and $f(\tup{b}) = \tup{b'}$.
Under this definition, the $uv$-entry of $\mathbf{A}_G$ is exactly equal to the number of edges between $u$ and $v$, which fits the usual definition of the adjacency matrix of a multigraph. Moreover, as Man\v{c}inska and Roberson note in the conclusion of \cite{manvcinska2020quantum}, allowing multiedges does not break any of the correspondence results between homomorphism tensors, linear maps, and partitions.

An interesting caveat with respect to multigraphs is also noted by Man\v{c}inska and Roberson. Namely, there is no established definition of the quantum automorphism group of a multigraph $G$. Proceeding as for simple graphs, by simply requiring that $\U A_G = A_H \U$, would mean that all of our proofs go through nearly unchanged. However, this might not be the natural choice. In the classical case, we usually want to capture the permutation of parallel edges in our automorphisms, while the above definition only captures the action on vertices. 

On the other hand, it is not clear that trying to mirror the classical automorphism group of multigraphs is the correct approach in the first place. To illustrate this point, recall that a central step in our proof is to add the adjacency matrix $A_G$ to the intertwiners of an easy quantum group. For classical groups, this is equivalent to requiring that $U A_G = A_G U$ for all its elements $U$, which for the symmetric group happens to yield the automorphism group $\Aut(G)$ of $G$. However, it is
generally not clear what it means to require that the elements $U$ of a matrix group satisfy $U A_G = A_G U$. The symmetric group is the exception, since it acts on graphs by conjugating the adjacency matrix. $U A_G = A_G U$, being equivalent to $U A_G U\inverse = A_G$, thus simply says that the adjacency matrix has to be invariant under the group action, which leaves precisely the permutations that are automorphisms. For other matrix groups, this conjugation does not define a group action, since $U A_G U\inverse$ is generally not an adjacency matrix anymore. Depending on whether there is a natural interpretation of the subgroup of a matrix group defined by the relation $U A_G = A_G U$, simply adding $A_G$ as an intertwiner might ultimately prove to be a more natural choice. In fact, whether such an interpretation exists is an interesting question in its own right.

  \printbibliography
\end{document}